\documentclass[%
prx,
twocolumn,
 reprint,
 letterpaper,
 superscriptaddress,
 amsmath,amssymb,
 aps,
floatfix,
]{revtex4-1}
\usepackage{graphicx}
\usepackage[dvipsnames]{xcolor}
\usepackage{threeparttable}
\usepackage{physics}

\begin{document}

\title{Gradient Ascent Pulse Engineering with Feedback}

\author{Riccardo Porotti}
\email{riccardo.porotti@mpl.mpg.de}
\affiliation{Max Planck Institute for the Science of Light, Erlangen, Germany}
\affiliation{Department of Physics, Friedrich-Alexander Universit{\"a}t Erlangen-N{\"u}rnberg,
Germany}

\author{Vittorio Peano}
\affiliation{Max Planck Institute for the Science of Light, Erlangen, Germany}

\author{Florian Marquardt}
\affiliation{Max Planck Institute for the Science of Light, Erlangen, Germany}
\affiliation{Department of Physics, Friedrich-Alexander Universit{\"a}t Erlangen-N{\"u}rnberg,
Germany}

\date{\today}

\begin{abstract}
Efficient approaches to quantum control and feedback are essential for quantum technologies, from sensing to quantum computation. Open-loop control tasks have been successfully solved using optimization techniques, including methods like gradient-ascent pulse engineering (GRAPE) , relying on a differentiable model of the quantum dynamics. For feedback tasks, such methods are not directly applicable, since the aim is to discover strategies conditioned on measurement outcomes. In this work, we introduce feedback-GRAPE, which borrows some concepts from model-free reinforcement learning to incorporate the response to strong stochastic (discrete or continuous) measurements, while still performing direct gradient ascent through the quantum dynamics. We illustrate its power considering various scenarios based on cavity QED setups. Our method yields interpretable feedback strategies for state preparation and stabilization in the presence of noise. Our approach could be employed for discovering strategies in a wide range of feedback tasks, from calibration of multi-qubit devices to linear-optics quantum computation strategies,  quantum-enhanced sensing with adaptive measurements, and quantum error correction.
\end{abstract}

\maketitle
\section{Introduction}
The application of optimal-control techniques to quantum systems \cite{glaser_training_2015,koch_controlling_2016} forms a cornerstone of modern quantum technologies, ranging from the tailoring of laser pulses acting on molecules to the synthesis of unitaries in multi-qubit systems as part of the ``compilation" of quantum algorithms for specific hardware platforms. Since the equations of quantum dynamics are explicitly known and even differentiable, one can exploit this knowledge and specifically make use of powerful gradient-based techniques. The most prominent approach is ``gradient-ascent pulse engineering" (GRAPE)\cite{khaneja_optimal_2005, machnes_comparing_2011}, with its efficient evaluation of gradients, together with its variants. GRAPE is the state of the art method for quantum optimal control and is extremely widely employed. In fact, it has been used to find optimal control sequences for spin systems \cite{khaneja_optimal_2005,dolde_high-fidelity_2014,yang_silicon_2019}, coupled qubits \cite{schulte-herbruggen_optimal_2005,sporl_optimal_2007}, an implementation of the Jaynes-Cummings model \cite{heeres_implementing_2017}, and qubit-cavity lattices \cite{fisher_optimal_2010}, among many other examples. It has also been used to optimize open dynamics \cite{schulte-herbruggen_optimal_2011,boutin_resonator_2017}, has been turned into an adaptive approach to cope with parameter uncertainties \cite{egger_adaptive_2014,wu_data-driven_2018,wu_learning_2019,ding_collaborative_2021},  and has been extended to second-order optimization techniques \cite{de_fouquieres_second_2011}. Other efficient gradient-based optimal control approaches have also been presented recently (e.g. \cite{machnes_tunable_2018}). 

However, there is one crucial extension that is not easily addressed by such gradient-based  techniques: feedback. Conditioning the control sequence based on the stochastic outcomes of quantum measurements is an important component of many more challenging tasks \cite{zhang_quantum_2017}. It allows to remove entropy from the system and is therefore essential in applications like state preparation and stabilization in the presence of noise \cite{handel_modelling_2005,sayrin_real-time_2011,vijay_stabilizing_2012,hirose_coherent_2016,andersen_entanglement_2019}, adaptive measurements \cite{hentschel_efficient_2011}, or quantum error correction with its syndrome extraction (e.g. \cite{ahn_continuous_2002,cramer_repeated_2016,ryan-anderson_realization_2021,krinner_realizing_2021}). Unfortunately, discovering feedback strategies is a formidable challenge. These strategies live in a space that is combinatorially larger than that of open-loop control strategies, since every sequence of measurement outcomes may require a different response. Beyond that general difficulty, it is unclear a priori how to take gradients through strong stochastic quantum measurement events.

 In principle, there is a set of techniques from machine learning that can discover feedback strategies without taking gradients through quantum dynamics: so-called model-free reinforcement learning  (RL)\cite{sutton_reinforcement_2018} approaches (for a brief review, see Appendix \ref{appendix-recap-RL}). During the last few years, a number of groups have demonstrated numerically the promise of model-free RL for quantum physics. This included both open-loop control tasks (e.g. \cite{bukov_reinforcement_2018,august_taking_2018,niu_universal_2019,porotti_coherent_2019}, also in an experiment \cite{baum_experimental_2021}), but in particular the more challenging quantum real-time feedback tasks that rely on adaptive responses to measurement outcomes \cite{fosel_reinforcement_2018,borah_measurement-based_2021,sivak_model-free_2021,porotti_deep_2021}, recently showcased in first experiments \cite{reuer2022realizing,sivak2023breakeven}.  In model-free RL, the quantum device is treated as a black box, which can be an advantage in applications to  experimental setups with unknown parameters \cite{sivak_model-free_2021,baum_experimental_2021, reuer2022realizing, sivak2023breakeven}.
On the other hand, much of the training time is therefore spent in learning implicitly a model of the dynamics while simultaneously attempting to find good feedback strategies. This can make learning  inefficient, leading to longer training times and/or suboptimal strategies.

It would therefore seem desirable to find a way to incorporate feedback based on arbitrary quantum measurements into a direct gradient-based optimal control technique like GRAPE, making efficient use of our knowledge of the differentiable quantum dynamics. 

In this work, we present such a technique, which we refer to as 'feedback-GRAPE'.  In the language of RL, it would be classified as a model-based technique \cite{moerland_model-based_2022}. On the one hand, it keeps the ability of GRAPE to exploit gradients through the quantum dynamics, making the learning more efficient. On the the other hand, similar to model-free RL, it provides a flexible approach to incorporate feedback, even in the presence of strong stochastic measurements, and to do this efficiently using Monte-Carlo simulations.

Overall, the technique we introduce here, feedback-GRAPE, is conceptually simple: GRAPE-type gradient ascent for the continuous control parts, possibly implemented using automatic differentiation for convenience \cite{leung_speedup_2017,abdelhafez_universal_2020,schafer_differentiable_2020,liao_differentiable_2019,coopmans_protocol_2021,abdelhafez_gradient-based_2019}, and in any case exploiting modern gradient optimizers, supplemented with stochastic sampling of measurement outcomes. 

We show that introducing Monte-Carlo sampling in the framework of gradient-based feedback optimization requires the addition of an important  correction term to the overall reward function, for discrete measurement outcomes, or a ``reparametrization" of the measurement probability density, for continuous measurement outcomes. This  innovation allows us to optimize any differentiable reward over long sequences of  measurements, both discrete and continuous. 

In this way, feedback-GRAPE is able to go significantly beyond existing gradient-based optimization methods for feedback. These are typically limited to  greedy optimization over one or, at most, a few measurements \cite{sayrin_real-time_2011}. Otherwise, for non-greedy optimization, they are limited  to optimal control problems that can be mapped onto so-called classical linear-quadratic-Gaussian control problems, or to a special linear ansatz for the feedback protocol as in so-called  Markovian quantum feedback methods \cite{wiseman_quantum_2009, zhang_quantum_2017}.  The special limiting case of weak Gaussian-distributed measurements, which does not yet require the mathematical treatment that we will introduce, has recently been considered by \cite{schafer_control_2021}, which can thus be considered an important first step towards the general method we are going to discuss here. Another important aspect of our method is that, in contrast to so-called Bayesian quantum feedback approaches \cite{doherty_feedback_1999,vijay_stabilizing_2012, zhang_quantum_2017}, it does not  require to simulate the system dynamics during deployment in an experiment, as the controller  is only provided with the measurement outcomes. This feature is important both for real-time control at fast time scales and the scalability  to more complex systems.

 We illustrate the power of feedback-GRAPE in a series of different tasks, considering different experimental setups  relevant for modern quantum computing employing cavity modes \cite{blais_circuit_2021}.  Although our method is general,  we have focused on   feedback sequences with a modular structure, i.e. where building blocks like unitaries and measurements are combined in discrete time steps. These are useful scenarios, since they can make it easier to interpret the resulting strategies.

In the following, we will first present the general method, then analyze the numerical examples,  discuss aspects of the optimization landscape and scalability, and finally present further extensions.

\begin{figure}
    \centering
    \includegraphics[width=\columnwidth]{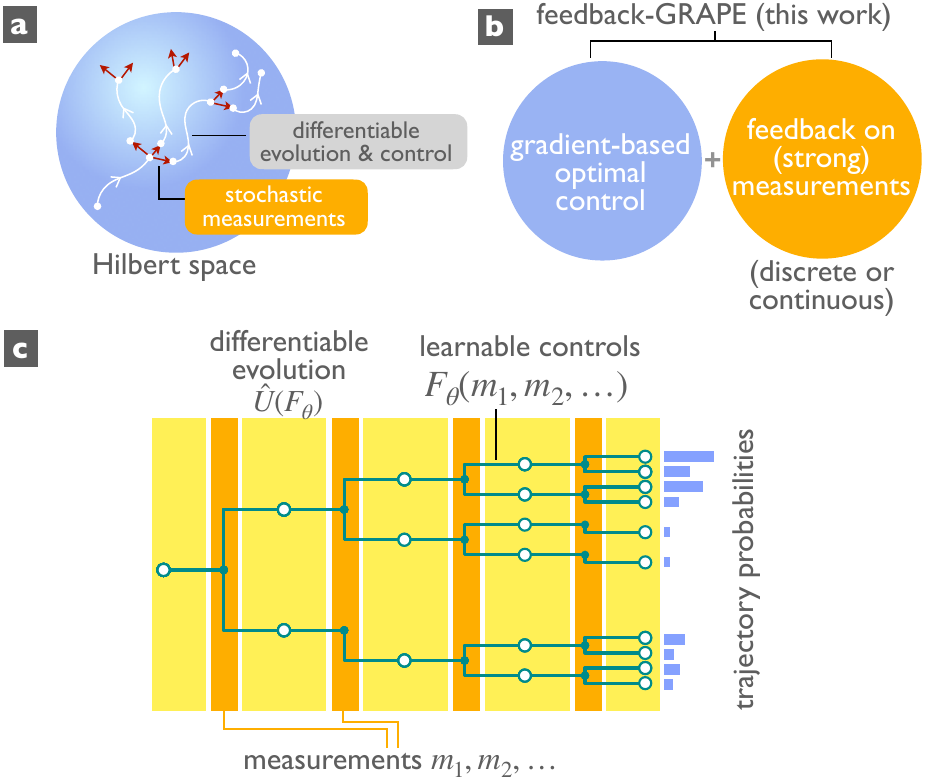}
    \caption{\label{Fig1_Schematic}Quantum feedback control with strong measurements. (a) Such feedback tasks combine smooth, differentiable dynamics in Hilbert space with measurement-induced jumps. (b) This work integrates model-based techniques relying on gradients and feedback to strong stochastic measurements. (c) Schematic decision-tree representation of a feedback strategy for discrete measurement outcomes. Intervals of differentiable evolution with optimizable control functions $F^j_{\theta}$ depend on the sequence of outcomes $m_1,\ldots,m_j$. In general, the evolution can be dissipative.}
\end{figure}

\section{Feedback-GRAPE Method}

\label{Section:Method}

We consider a general dissipative quantum system with feedback (for an overview of the scheme, see Fig.~\ref{Fig1_Schematic}). Suppose measurements are performed at times $t_1, t_2, \ldots, t_N$, and the evolution is controlled -- in a manner to be optimized -- based on the corresponding measurement outcomes $m_j$. Specifically, the control parameter $F^j_\theta$ (which might be a vector) applied during the time interval $[t_j,t_{j+1}]$ is a function of all previous measurement results $\mathbf{m}_j=(m_1,\ldots, m_{j-1},m_{j})$.   Below, we refer to  the set of controls $\{F^{j}_{\theta}(\mathbf{m}_j)\}$  for all possible measurement outcomes as a feedback strategy or simply strategy, see Fig.~\ref{Fig1_Schematic}(c).
Here, we anticipated that the feedback-control functions $F^{j}_{\theta}(\mathbf{m}_j)$ are parametrized, depending on trainable parameters $\theta$ that will be optimized via gradient ascent ($\theta$ is typically a high-dimensional vector).  We assume $F^{j}_{\theta}$ to be differentiable with respect to $\theta$. Ultimately, the value of $F^j_{\theta}$ will be provided by a neural network, or, alternatively, a lookup table: we comment on these different approaches further below, but the present considerations are independent of this aspect. In practice, the control vector $F^j_{\theta}$ might enter a Hamiltonian or directly a parametrized unitary gate. On a minor note, in some scenarios, during the first time interval $[0,t_1]$, one might apply a control $F^0$ that does not depend on any previous measurement outcomes but can still be optimized. 

With this notation in place, the time evolution of the system's density matrix, for a particular measurement sequence $\mathbf{m}=(m_1,m_2,\ldots )$, can be written in the general form

\begin{equation}
{\hat \rho}(T | \mathbf{m}) = \Phi^{(\mathbf{m})}(F^N_{\theta}, F^{N-1}_{\theta}, \ldots, F^0_{\theta})[{\hat \rho}(0)]\, \label{eq-CPmap}
\end{equation}
where $\Phi^{(\mathbf{m})}$ is the map that depends on the control parameters and implements the quantum-dissipative time evolution throughout the whole time interval $[0,T]$, {\em conditioned on the given fixed sequence $\mathbf{m}$ of measurement outcomes}. Note that our definition implies that $\Phi^{(\mathbf{m})}$ itself is not a completely positive (CP) map, because it contains the renormalization of the quantum state required after each measurement (it implements a ``quantum instrument"), which introduces a nonlinear dependence on the initial state. To obtain the unconditional average quantum state, the average $\left\langle \ldots \right\rangle_\mathbf{m}$ of this expression may be taken over all possible measurement sequences, weighted with their respective probabilities. 

Eq.\,(\ref{eq-CPmap}) is valid formally 
even if the overall evolution is non-Markovian. It can be simplified in the important Markovian case. Then, evolution proceeds step-wise. Let us denote by $\Phi_j$ the CP map for the continuous evolution during the time interval $[t^{+}_{j},t^{-}_{j+1}]$, where $t^{-}$ is shorthand for a time point just prior to the measurement at $t$, and correspondingly $t^+$ is right after the measurement. Then we have ${\hat \rho}(t^{-}_{j+1})=\Phi_{j}(F^j_{\theta}(\mathbf{m}_j))[{\hat \rho}(t^{+}_{j})]$. In the special case of unitary dynamics, the evolution itself simplifies further to ${\hat \rho}(t^{-}_{j+1})={\hat U}_j(F^j_{\theta}(\mathbf{m}_j)) {\hat \rho}(t^{+}_{j}) {\hat U}_j(F^j_{\theta}(\mathbf{m}_j))^{\dagger}$. Here ${\hat \rho}(t^{-}_{j+1})$ is understood to be the quantum state at time $t_{j+1}$ for a fixed sequence $m_1,\ldots,m_j$ of previous measurement outcomes, just prior to the next positive-operator-valued measure (POVM) measurement implemented at $t_{j+1}$. 

This measurement is described by some POVM element that can be written in the form ${\hat M}(m')^{\dagger} {\hat M}(m')$, with the POVM normalization condition $\sum_{m'} {\hat M}(m')^{\dagger} {\hat M}(m') = 1$ and ${\hat M}\equiv{\hat M}_{j+1}$ depending on the physics of the measurement. It will yield a particular outcome $m_{j+1}\equiv m'$ with probability $P(m')={\rm tr} [{\hat M}(m')^{\dagger} {\hat M}(m') {\hat \rho}(t^{-}_{j+1})]$ and an updated state ${\hat \rho}(t^{+}_{j+1})= {\hat M}(m') {\hat \rho}(t^{-}_{j+1}) {\hat M}(m')^{\dagger} / P(m')$. 

Our goal is to maximize some overall cumulative reward $R$, which is called ``return" in the nomenclature of reinforcement learning. For example, in a state-preparation task this might be the final fidelity with respect to some target state ${\hat \sigma}$. For a given sequence $\mathbf{m}$ of outcomes, we would define  $R(\mathbf{m})=\left( {\rm tr} \sqrt{\sqrt{\hat \sigma} {\hat\rho}(T|\mathbf{m}) \sqrt{\hat \sigma}} \right)^2$. This would be averaged eventually over all possible measurement outcome sequences to yield $\bar{R}=\left\langle {R}(\mathbf{m}) \right\rangle_\mathbf{m}$. The return $R$ could also involve penalties for suppressing larger control amplitudes etc. These additional contributions depend on the specific sequence $\mathbf{m}$ as well, via the controls $F^j_{\theta}(\mathbf{m}_j)$.

It might now seem straightforward to employ automatic differentiation for optimizing $\bar{R}$ via gradient ascent, updating $\delta \theta = \eta \frac{\partial {\bar{R}}}{\partial \theta}$, with some learning rate $\eta$ and with all the trainable parameters combined in a vector $\theta$.

The crucial observation to be made at this stage is that the introduction of stochastic measurement results into this scheme requires some extra care. The following considerations constitute the main conceptual steps needed to enable the discovery of feedback-based quantum control strategies based on gradient ascent.

We have to distinguish between discrete and continuous measurement outcomes, which require substantially different treatment. 

For the particularly interesting discrete case (e.g. strong projective qubit measurements), the essential insight is that the probabilities $P$ for obtaining the different measurement outcomes themselves depend on all the controls $F^j_{\theta}$ applied during previous time intervals, simply because the quantum state itself carries this dependence. This has to be taken care of during the evaluation of gradients with respect to  $\theta$. Illustrating this in the case of a single measurement $m$ at time $t_1\in[0,T]$, we have

\begin{equation}
\langle R \rangle_m=\sum_{m} P(m| {\hat \rho}(t_1^{-})) R(m) \label{eq-one-discrete-msmt}
\end{equation}
Here $P(m|{\hat \rho})$ is the probability for measurement outcome $m$ given state ${\hat \rho}$. As we take the gradient with respect to the trainable parameters $\theta$,  we observe that the derivative acts not only on the return $R$ based on the time-evolved state $\Phi_1(F^1_{\theta}(m))[{\hat \rho}(t_1^{+})]$ (the second factor inside the sum) but also on the probability $P(m)$ itself, due to its dependence on the initial control, ${\hat \rho}(t_1^{-}) = \Phi_0(F^0_{\theta})[{\hat \rho}(0)]$.

Generalizing this observation, we cannot simply implement gradients of the measurement-averaged return $\bar{R}=\langle R(\mathbf{m}) \rangle_\mathbf{m}$ by averaging the gradient of the sequence-specific return, $\langle {\partial R(\mathbf{m}) / \partial \theta} \rangle_\mathbf{m}$. Rather, observe $ \langle R(\mathbf{m}) \rangle_\mathbf{m} = \sum_\mathbf{m} P(\mathbf{m}) R(\mathbf{m})$. Thus, when evaluating $ \partial \langle R(\mathbf{m}) \rangle_\mathbf{m} / \partial \theta$, we will get two contributions: $\partial  [R(\mathbf{m}) P(\mathbf{m})] / \partial \theta =  P(\mathbf{m}) \partial R(\mathbf{m}) / \partial \theta + R(\mathbf{m})  \partial P(m) / \partial \theta$. To enable stochastic sampling of the second term, we rewrite it using $ \partial P(\mathbf{m}) / \partial \theta = P(\mathbf{m}) \partial \ln P(\mathbf{m}) / \partial \theta $. This then leads to:

\begin{equation}
    \frac{\partial \left\langle R(\mathbf{m}) \right\rangle_\mathbf{m}}{\partial \theta} = \left\langle \frac{\partial R(\mathbf{m}) } {\partial \theta} \right\rangle_\mathbf{m} + \left\langle R(\mathbf{m}) \frac{\partial \ln P_{\theta}(\mathbf{m})}  {\partial \theta} \right\rangle_\mathbf{m} \label{eq-extra-term} .
\end{equation}
Here we displayed explicitly the parameter-dependence of $P_{\theta}(\mathbf{m})$, which represents the probability of the full sequence of outcomes $\mathbf{m}=(m_1,m_2,\ldots)$, given the parameters $\theta$ that determined the shape of the control functions $F^j_{\theta}$. 

The mathematics for the extra term appearing here, with the gradient of the log-likelihood, is well known from policy-gradient-based approaches in {\em model-free} reinforcement learning. However, there this term appears for a different reason. It arises due to the deliberate choice of implementing stochastic controls, in order to avoid any need to take gradients through the possibly unknown dynamics of the system to be controlled. For more details see Appendix \ref{appendix-recap-RL}. In our case, by contrast, we do take gradients through the known dynamics and the controls themselves are deterministic when conditioned on a fixed sequence of measurements. The randomness enters via the stochastic measurement outcomes (these are observations of the ``environment" in RL language). 

Due to the sequential nature of the control procedure, the log-likelihood term can be rewritten as a sum of contributions, $\ln P_{\theta}(\mathbf{m}) = \sum_j \ln P_{\theta}(m_j|\mathbf{m}_{j-1})$. Thus, during the individual time evolution trajectory, this term may be easily accumulated step by step, since the conditional probabilities are known (these are just the POVM measurement probabilities). The gradients of Eq.~(\ref{eq-extra-term}) can then be taken for such an individual trajectory or a batch, substituting stochastic sampling for an exact average over $\mathbf{m}$. The whole approach, with its calculational pipeline, is illustrated schematically in Fig.~\ref{fig:Figure2_Pipeline}.  Additionally,  a more detailed algorithmic flow-chart representation is given in Appendix \ref{App:flow_chart}.

\begin{figure}
    \centering
    \includegraphics[width=\columnwidth]{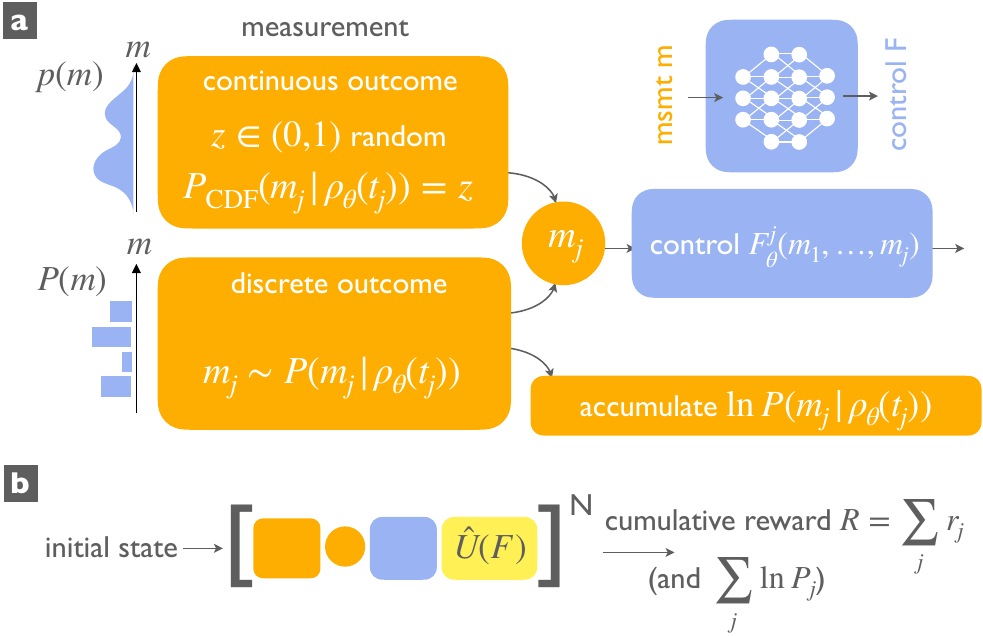}
    \caption{Quantum feedback sequences considered within feedback-GRAPE, set up for automatic differentiation. (a) The measurement samples a stochastic outcome $m_t$, adopting a different method depending on whether the outcome is continuous or discrete. In both cases, the probability distribution depends in a differentiable way on the trainable parameters $\theta$, via the preceding unitary controls that have generated the present quantum state $\rho_{\theta}$. Depending on the measurement outcome, a learnable control $F$ is applied that may be implemented either via a neural network or a lookup table.
    (b) Full sequence. This consists of repeated application of the blocks depicted in (a), plus subsequent implementation of unitary controls depending on $F$, potentially with decay and decoherence included in the model of the system's evolution. For discrete measurement outcomes, the logarithmic term (in brackets) has to be accumulated and is used to evaluate a log-likelihood correction term  when optimizing the overall return $R$ using gradient ascent,  cf Eq.~(\ref{eq-extra-term}).  An algorithmic flow chart representation of the learning pipeline for the case of discrete measurements is provided  in Appendix \ref{App:flow_chart}.
    }
    \label{fig:Figure2_Pipeline}
\end{figure}

The Monte-Carlo evaluation of the average reward $ \langle R(\mathbf{m}) \rangle_\mathbf{m} = \sum_\mathbf{m} P(\mathbf{m}) R(\mathbf{m})$  is a crucial ingredient to tackle long sequences of stochastic measurements, as it allows to focus only on the most likely measurement outcomes among the exponentially large set of such outcomes. Only for very short sequences would it be feasible to instead explicitly evaluate the sum over all possible measurement outcomes, producing less noisy gradients.

The evaluation of the gradients of the return with respect to the trainable parameters $\theta$ can proceed in two different ways, using either automatic differentiation (see below) or exploiting analytical approaches to obtain explicit expressions for the gradients that can then be evaluated numerically. In the latter case, one can either set up evolution equations for the parameter-gradient of the quantum state, ${\partial_{\theta}}{\hat \rho}$ or, in the suitable scenario, directly apply a modified version of the original GRAPE technique to efficiently evaluate the gradients. We describe both of these procedures in detail in Appendix \ref{appendix-GRAPE-evol-eqs}. In the language of current machine learning concepts, taking the gradient through the continuous-evolution intervals would be generally speaking an example of the concept of neural ordinary differential equations, a rather recent development \cite{chen_neural_2018}.

Alternatively, and sometimes more conveniently, the whole evolution pipeline described above can straightforwardly be implemented in an automatic differentiation framework, such as TensorFlow \cite{tensorflow2015-whitepaper}, PyTorch, JAX, or others. Gradients of the resulting overall return and of the log-probability can then be obtained using that framework without extra effort. The sequence of discrete measurement outcomes of a given trajectory is considered fixed when taking the gradient in this manner. The automatic-differentiation approach is particularly helpful and efficient in cases where the whole time evolution can be split into many building blocks (parametrized gates, i.e. unitaries, acting during fixed time intervals), as is common practice for many quantum control tasks in present quantum computing platforms. Whenever this latter situation is encountered, it also aids interpretability, as we will see in the numerical examples.

As we remarked at the beginning, the central quantity of our approach are the measurement-dependent controls $F^j_{\theta}(\mathbf{m}_j)$. For accessing those, one can simply adopt a lookup table, at least for the case of discrete measurements discussed up to now and when the total number of measurements during the full time evolution is not too large. The table for $F^j_{\theta}$ needs $ M^j $ entries, if there are $M$ possible outcomes for each measurement, corresponding to the exponentially many possible sequences. In that case, the entries of this table would directly represent the trainable parameters $\theta$.  Alternatively, the controls $F^j_{\theta}(\mathbf{m}_j)$ can be implemented via a neural network that takes measurement results as input and maps those to the current control vector. Since the number of available measurement results is different for each time step $j$, one may choose to set up a different network at each $j$. However, training efficiency and generalization ability can be improved by constructing a single recurrent network, i.e. a network with memory that is employed in sequence processing tasks \cite{hochreiter_long_1997}. It takes the temporal sequence of measurements as input, one step at a time, producing a control vector at each such time step. This approach can possibly generalize to infinitely long feedback control sequences, for example during state stabilization tasks. In the course of our numerical experiments, to be detailed later, we observed both scenarios where the neural network outperformed the lookup table but also the reverse.

We note in passing  that both the look-up table and recurrent NN approaches do not require a Bayesian estimate of the state during deployment in an experiment: they operate purely on the measurement outcomes.  For the sake of comparison, however, we also considered a Bayesian quantum feedback approach \cite{doherty_feedback_1999} in which a fully connected NN is provided with the quantum state before each time step. This approach suffers problems with scalability, because it can be deployed in
an experiment only in combination with real-time simulations of a stochastic master equation \cite{doherty_feedback_1999} to update the quantum state based on the measurement outcomes. We will see later that our numerical results indicate that the knowledge of the full quantum state does not appear to provide any substantial learning advantage compared to our non-Bayesian approaches. 

Continuous measurement outcomes can be treated in exactly the same way as discrete ones. However, for that scenario there also exists an alternative, which obviates the need for the logarithmic-likelihood correction term: we can adopt a general version of what is known as the 'reparametrization trick' in stochastic neural networks (e.g. in variational autoencoders). The idea is that we can generate a stochastic variable $z$ according to some {\em fixed} probability density and then transform this into the required measurement probability density $p(m|{\hat \rho})$, which does depend on control parameters (via the state ${\hat \rho}$, as explained above) and must be subjected to gradients. This parameter-dependent transformation can be implemented in a differentiable way, as we now show. We first obtain the cumulative distribution function $f(m)=\int_{-\infty}^m p(m'|{\hat \rho}) dm'$, by discretizing $p$ as a vector on a lattice and using a cumulative sum for an Euler approximation of the integral (this operation exists in frameworks like TensorFlow). We then draw a random uniformly distributed $z\in[0,1]$ and {\em invert} $f(m)$. The last step also needs to be performed in a differentiable way. One option is to set $m=f^{-1}(z)\approx\sum_n {\tilde m}_n H(z-z_n) H(z_{n+1}-z)$. Here $z_n=f(m_n)$ defines the lattice version of $f$, $H$ is the Heaviside step function, the sum ranges over the lattice points, and ${\tilde m}_n$ solves the piecewise linearized approximation of $m=f^{-1}(z)$ associated with the interval $n$:  ${\tilde m}_n=(m_{n+1}-m_n)(z-z_n)/(z_{n+1}-z_n)+m_n $. The set of measure zero where the gradient is undefined can be ignored, as is common practice in using activation functions like rectified linear units in neural networks.

In this way, one can implement, within the automatic differentiation framework, for example measurements of discrete variables with continuous outcomes. A typical case would be a qubit measurement with $m=\sigma + \xi$, where $\sigma=\pm1$ is the qubit state and $\xi$ some measurement noise of density $q(\xi)$. Formally, $p(m|{\hat \rho})=\sum_{\sigma} q(m-\sigma) \rho_{\sigma \sigma} $, and ${\hat M}(m)=\sum_{\sigma} \sqrt{q(m-\sigma)} \left|\sigma\right\rangle \left\langle \sigma\right|$. One can also perform measurements on continuous variables, e.g. a weak measurement of position, $p(m|{\hat \rho})=\int dx q(m-x) \rho(x,x)$, with ${\hat M}(m)=\int dx \sqrt{q(m-x)} \left|x\right\rangle \left\langle x\right| $. The dependence of the probability density $p$ in each case on the parameters determining the control functions at earlier times will be correctly taken into account, and one can now use the straightforward formula ${\partial \bar{R} / \partial \theta}=\left\langle {\partial R(\mathbf{m}) / \partial \theta}\right\rangle_\mathbf{m}$ for stochastic sampling of the gradient. Note that the discrete-outcome case (above) and the continuous-outcome case can also be easily combined in our approach.

Our reparametrization trick allows to switch from any arbitrary state-dependent probability density $p(m|\hat{\rho})$ to an easy-to-sample fixed probability density, allowing to obviate the need for the log-likelihood term that was required in the case of discrete outcomes discussed earlier. There is a special limiting case in which an even simpler linear  reparametrization  achieves the same goal. We are referring to the case in which $p(m|\hat{\rho})$ is a Gaussian which depends on the state of the system only via its mean value $\bar{m}$, $p(m|\hat{\rho})={\cal N}(\bar{m}(\hat{\sigma}),1)$. In this case,  one can switch to a fixed (easy-to-sample) Gaussian probability density with the reparametrization  $z=m-\bar{m}(\hat{\sigma})\sim {\cal N}(0,1)$.   This reparametrization has the advantage that it can be trivially inverted. This kind of description automatically arises in the well-known quantum trajectories approach applied to homodyne detection \cite{wiseman_quantum_2009} of light emerging from a cavity.  In that setting, the  observable $m$ of interest is the observed homodyne detection current, appropriately rescaled and integrated over a time window $\Delta t$ much smaller than the typical system decay time. Its distribution depends on the state of the system only via its  mean  value  $\bar{m}=\epsilon \tr (\hat{O}\hat{\rho})$ where  $\hat{O}$ is the system observable that couple to the bath and $\epsilon\propto \sqrt{\Delta t}$. In the framework of this quantum trajectory approach, one can, thus, find feedback strategies conditioned on the homodyne current by differentiating through the system dynamics  without introducing our more flexible reparametrization of $p(m|\hat{\rho})$  or  our additional log-likelihood term. This approach has been pursued recently by Sch\"afer et al. \cite{schafer_control_2021}.

So far, controls have been continuous and represented via  functions (differentiable with respect to parameters) depending on previous measurement results. However, sometimes one might want to {\em also} take discrete actions, e.g. deciding whether some measurement should be performed at all or not, or whether some fixed qubit gate should be applied. This can be incorporated without any substantial changes to the approach discussed here,  borrowing from policy-gradient model-free reinforcement learning, by introducing stochastic actions $a$, in contrast to the deterministic continuous actions discussed so far:  use a network or a lookup-table to calculate the probability $P_\theta(a_j|\mathbf{m}_j,\boldsymbol{a}_{j-1})$ of taking a discrete action $a_j$ at step $j$ given the previous measurement record $\mathbf{m}_j$ and actions $\boldsymbol{a}_{j-1}=(a_{j-1},\ldots,a_2, a_1)$ and then sample from all actions accordingly. Then, the same form for the gradient for the reward applies as in Eq.~(\ref{eq-extra-term}) but now with both the reward $R$ and the probability $P_\theta$  depending not only on the measurement history $\mathbf{m}$ but also on the history of all the actions taken throughout the trajectory, $\boldsymbol{a}=(a_1,a_2,\ldots)$. As before, the probability in Eq.~(\ref{eq-extra-term}) can be replaced by Monte Carlo sampling while the log-likelihood term can be accumulated according to
$\ln P_\theta(\mathbf{m},\boldsymbol{a})=\sum_j \ln P_\theta(m_j|\mathbf{m}_{j-1}, \boldsymbol{a}_{j-1})+\ln P_\theta(a_j|\mathbf{m}_{j},\boldsymbol{a}_{j-1})$.

\begin{figure}
\centering
\includegraphics[width=\columnwidth]{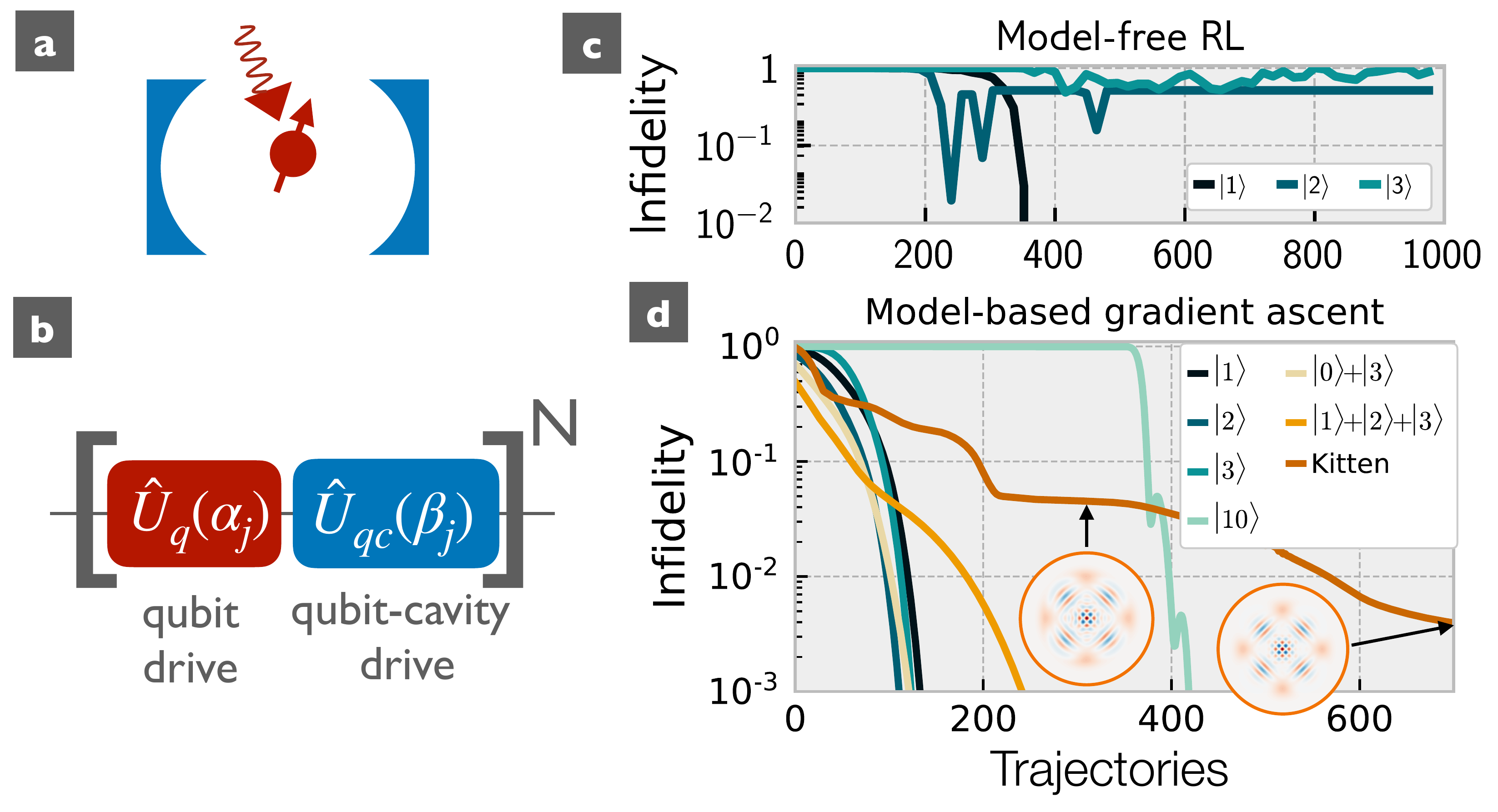}
\caption{State preparation from the ground state (open-loop control). a) Schematic of the Jaynes-Cummings system. b)  Sequence of parametrized controls during a  time-evolution trajectory. Each trajectory comprises $N$ time steps.  A time step consists in a qubit drive gate followed by a qubit-cavity interaction gate.  c,d) Fidelity as a function of the number of trajectories sampled during training (here, that number is equal to the number of gradient ascent iterations). The target state is indicated.  Model-free RL performs very poorly, while the direct gradient-based approach used as the basis for our method converges well even when long sequences are required ($N=20$ for the four-components kitten state).
}
\label{fig:Figure3_a_open-control}
\end{figure}

\section{Numerical examples}

We now turn to an illustration of the feedback-GRAPE method by solving several different challenging quantum feedback control tasks. We will consider five separate tasks of increasing difficulty: starting with noiseless state preparation (an open-loop control task) as a baseline benchmark for GRAPE-type control in this scenario, then moving to purification (a task that already benefits from feedback, i.e. adaptive measurements), to feedback-based state preparation in the presence of noisy control parameters or out of a thermal state, and feedback-based state stabilization.
Along the way, we will explore a handful of different experimental scenarios.

\subsection{State preparation with Jaynes-Cummings controls}
\label{sec:open_loop}
 As a preliminary step, we consider state preparation of a target state starting from a pure state. In addition, we assume that any coupling to an external environment is negligible  and that the parametrized controls can be implemented perfectly. In this setting, the preparation of a quantum state  does not require any feedback and, thus, we will not yet be able to test our feedback extension of GRAPE. Instead, the purpose of this section is to provide a compelling motivation for our   approach, showing that, even before feedback is introduced, (GRAPE-type) model-based optimal-control approaches outperform, some times dramatically, their model-free counterparts. 

As a first example, we consider the state preparation of a cavity resonantly coupled to an externally driven qubit, cf Fig.~\ref{fig:Figure3_a_open-control}a. This scenario is modelled by the well-known Jaynes-Cummings Hamiltonian \cite{jaynes_comparison_1963}. It is the first and simplest light-matter coupling scenario that emerged in quantum optics \cite{jaynes_comparison_1963,shore_jaynes-cummings_1993} but is nowadays of practical relevance for modern quantum-computing platforms \cite{blais_circuit_2021}.  In those, it is employed both for qubit readout and for qubit-enabled nonlinear manipulation of cavity states.  Here, we will consider a particular sequence of parametrized unitary gates originally introduced by Law and Eberly \cite{law_arbitrary_1996}. The sequence consists in a series of two interleaved gates, cf Fig.\ref{fig:Figure3_a_open-control}b.    In the first gate, the qubit is driven externally to implement an arbitrary rotation about an equatorial axis, implementing the unitary gate ${\hat U}_q(\alpha)=\exp[-i\left( \alpha \hat \sigma^c_+ + \alpha^* \hat \sigma^c_- \right)/2 ]$. Here we introduced the qubit raising (lowering) operator $\hat{\sigma}^c_+$ ($\hat{\sigma}^c_-$), and  $|\alpha|$ is the rotation angle while ${\rm arg}(\alpha/|\alpha|)$ is the azimuthal angle of the rotation axis. In the second gate, the qubit and cavity mode with ladder operator ${\hat a}$ can be coupled for a variable duration, exchanging excitations, ${\hat U}_{qc}(\beta)=\exp[-i \left( \beta \hat a \hat \sigma^c_+ + \beta^* \hat a^{\dagger} \hat \sigma^c_- \right)/2]$. Here, $|\beta|$ is proportional to the interaction time, see Appendix \ref{App:Law_and_Eberly}. Depending on the target state, the control parameters $\alpha$ and $\beta$ can be chosen to be complex or real. In the latter case,  the control vector $F^j$ defined in Section II is simply $F^j=(\alpha_j,\beta_j)$. 

In their groundbreaking work \cite{law_arbitrary_1996}, Law and Eberly showed that  any arbitrary superposition of Fock states with maximal excitation number $N$ can be prepared out of the ground state in a sequence of $N$ such  interleaved gates, providing also an algorithm to find the correct  angles and interaction durations (see Appendix \ref{App:Law_and_Eberly}).  This solution has been used to remarkable effect in experiments with superconducting qubits \cite{hofheinz_synthesizing_2009}. Here, we use it as a benchmark to test different RL approaches. 

With the goal of recovering the strategies predicted by Law and Eberly, we set the return $R$ equal to the state fidelity at the final time step, prescribing a fixed number of time steps equal to the maximum number of excitations in the target state, e.g. for the state $\propto |1\rangle+|3\rangle$ we set $N=3$. 

Somewhat surprisingly, state-of-the-art model-free reinforcement learning is not able to cope well with this challenge. We employed proximal-policy optimization (PPO) \cite{PPO}, a powerful and widely used modern general-purpose advantage actor-critic approach to optimize the continuous controls. It performs well only for the very simple task of preparing Fock state $\left|1\right\rangle$, while getting stuck at bad final overlaps for higher Fock states, cf Fig.~\ref{fig:Figure3_a_open-control}c. This statement holds even after training for many episodes and varying the hyperparameters, and even for other modern general model-free RL algorithms that we tried, see Appendix \ref{appendix-model-free-RL}. 

In contrast, direct gradient ascent through the unitary evolution, using the control parameters   as learning parameters,  $\theta=\{F^j=(\alpha_j,\beta_j)\}$, allows to find optimal state preparation strategies performing as well as the known Law-Eberly algorithm for a broad range of states. As examples, we have prepared Fock states with excitations numbers up to $n=10$, and superpositions of two Fock states, cf Fig.~\ref{fig:Figure3_a_open-control}d. For all these states, convergence of the training protocol has been obtained in a single run and the infidelity can be decreased up to the numerical precision of the algorithm.   

In addition, we have also considered a much more challenging four-component kitten state built from four coherent states,  $|\psi^{\rm Kit4}_{\tilde{\alpha}}\rangle\propto\sum_{j=0}^3 |i^j\tilde{\alpha}\rangle$. Here, we consider $\tilde{\alpha}=3$, corresponding to the average photon number $\bar{n}\approx |\tilde{\alpha} |^2=9$.   We attempt to prepare this state using a long sequence of $20$ time steps.  In this case, we see that the infidelity tends to decrease step-wise during the training, cf Fig.~\ref{fig:Figure3_a_open-control}d. In Appendix \ref{App:State_prep_kitten}, we show that the detailed evolution of the fidelity  during training depends strongly on the initialization. On the other hand, the height of the steps is an intrinsic feature of the target state. Interestingly,  each step  can be associated to a particular intermediate  state that can be reached using a large number of different strategies. Each such strategy corresponds to a saddle point of the optimization landscape. The training becomes particularly slow close to these saddle points because  the curvature of the optimization landscape  in the direction of increasing fidelity is zero, giving rise to a narrow valley. This can cause the training to stall on a suboptimal solution. Nevertheless, if the preparation sequences are made longer (larger $N$), good solutions can be found in any training run. For more details see Appendix \ref{App:State_prep_kitten}.

Physically, the presence of a large number of narrow valleys and plateaus in the optimization landscape  is due to the fact that in the Law and Eberly protocol the excitations can only be added  one by one, first exciting the qubit and then transferring them to the oscillator.
They are present even for simple  tasks such as the preparation of a simple Fock state with excitation number $n$ prepared in $N=n$ time steps (as in Fig.~\ref{fig:Figure3_a_open-control}c,d). In this case, it is enough that a single control parameter ($\beta_j$ or $\alpha_j$ for any $j$) is zero for the fidelity and its gradient to vanish. If two or more parameters are small also the Hessian is zero, leading to a plateau. In the model-based approach this leads to a slow convergence for large $n$, cf Fig.~\ref{fig:Figure3_a_open-control}d). On the other hand, lack of direct access  to the gradient in a model-free approach prevents convergence even for small values of $n$, cf Fig.~\ref{fig:Figure3_a_open-control}d.

In order to further substantiate that  model-based GRAPE-type  optimal control approaches are more efficient than their model-free counterparts, we analyze the state preparation of a so-called Gottesman-Kitaev-Preskill state  \cite{Gottesman_encoding_2001} using the same set of universal controls recently adopted by Sivak et al. \cite{sivak_model-free_2021}   to demonstrate  their model-free quantum control approach. Direct comparison of our results, reported in Appendix \ref{App:grid_state}, with those of Ref.~\cite{sivak_model-free_2021} shows the following: in this setting with more powerful controls, in which model-free RL  performs already well, a GRAPE-type approach performs even better. In particular, it allows to explore much
larger parameter spaces 
and, thus, obtain better quality solutions using only a small fraction (about
one percent) of simulated training trajectories.

Regardless of these detailed observations, these examples indicate that model-based gradient ascent approaches can outperform model-free generic methods for optimizing quantum control in settings relevant for quantum technologies. Given the large performance difference already in the open-loop control scenario we focused entirely on the feedback-GRAPE approach in the subsequent exploration of the more advanced challenges that do include feedback.

\begin{figure}
    \centering
    \includegraphics[width=\columnwidth]{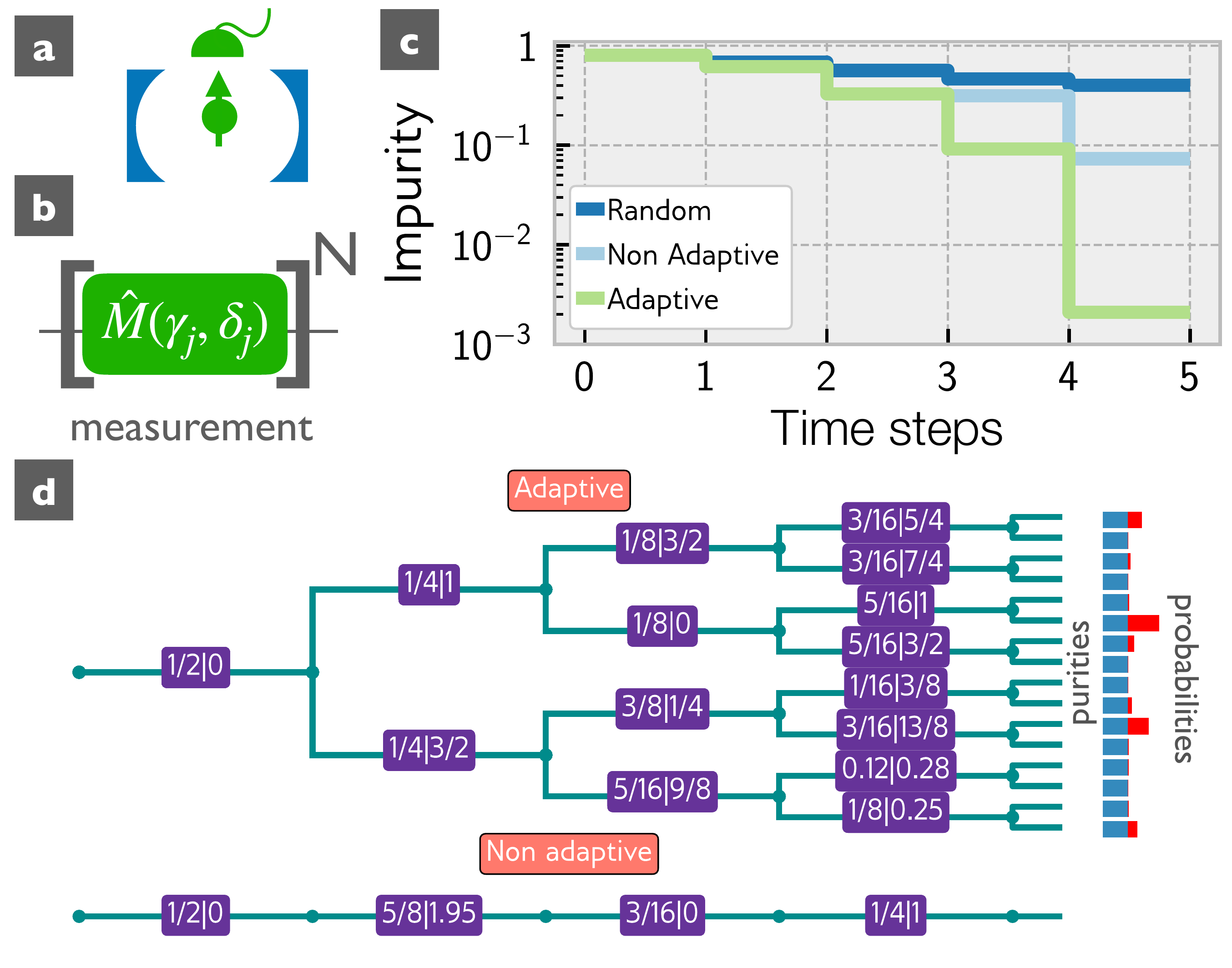}
    \caption{ Purification of a thermal state. a) Sketch of the experimental set up. The cavity is coupled dispersively to an ancilla qubit. b) Sequence of parameterized controls during a time-evolution trajectory. Each time step consists in a parameterized measurement. c) Purification of a thermal state (here, with ${\bar n} =2$). An adaptive strategy (found using feedback-GRAPE) clearly outperforms other approaches. e) Extracted purification strategy visualized in a decision tree. The purple boxes display the measurement parameters $(\gamma/\pi|\delta/\pi)$.}
    \label{fig:Figure3_b_purification}
\end{figure}

\subsection{State purification with qubit-mediated  measurement}
\label{sec:state_pur}
Next, we move  to a first example of a situation that requires feedback. We will now imagine that the cavity is initially in a mixed state. The goal will be to purify the cavity's state, i.e. the reward is determined by the purity ${\rm tr}{\hat \rho_{\rm cav}}^2$ of the cavity state at the final time. Purification can be achieved by applying repeated quantum measurements, which remove entropy from the quantum system.

In the following, we consider an adaptive  measurement scheme originally proposed in \cite{brune_quantum_1990} and demonstrated in a series of experiments on Rydberg atoms interacting with microwave cavities  \cite{brune_observing_1996,guerlin_progressive_2007,  gleyzes_quantum_2007, sayrin_real-time_2011}.
In this scheme,  the cavity is coupled to an ancilla qubit, which can then be read out to update our knowledge of the cavity's quantum state, 
cf sketch in Fig.~\ref{fig:Figure3_b_purification}a. 

The measurement comprises several steps, which we will list individually before summarizing their combined effect on the cavity state. In a first step, the ancilla  qubit with Pauli operators $\hat{\sigma}^a_{i=x,y,z}$ is prepared in the $+x$ eigenstate. Subsequently it is coupled dispersively to the cavity for a variable amount of time. The dispersive coupling in experiments is linear in the photon number to a very good approximation in the low-photon regime, and  is described by a unitary of the form:  ${\hat U}(\gamma)=\exp(-i \gamma {\hat \sigma^a_z}{\hat a}^{\dagger} {\hat a})$ with parameter $\gamma$ depending linearly on the interaction time. This means the qubit precesses by an angle that depends linearly on the number of photons inside the cavity. In the next, final step, the ancilla qubit is projected along some selected axis $\hat \sigma^a_x\cos \delta +\hat \sigma^a_y\sin \delta $, yielding a discrete result $m\in \{-1,+1\}$.  The combined effect of these operations is to perform a POVM on the cavity, with outcome probability $P(m)={\rm tr}[{\hat M}(m)^{\dagger} {\hat M}(m) {\hat \rho}]$ and an updated state ${\hat M}(m){\hat \rho}{\hat M}(m)^{\dagger}/P(m)$. Here ${\hat \rho}$ is the state of the qubit-cavity system, excluding the measurement qubit which has been eliminated in this description. The measurement operator ${\hat M}(m)$ is given by 
\begin{equation}
\label{eq:POVM_Haroche} {\hat M}(m=+1)={\cos}(\gamma {\hat a}^{\dagger} {\hat a} + \delta/2)\, , 
\end{equation}
and likewise for $m=-1$, with $\cos$ replaced by $\sin$. This formula indicates that after the measurement the probabilities of the different cavity Fock states $\left|n\right>$ will be multiplied by a sinusoidal ``mask", where the period is determined by $1/\gamma$ and the phase shift is set by both $\delta$ and the measurement outcome $m$. This helps to pinpoint the state of the cavity, especially when multiple such measurements are carried out with suitably chosen periodicities \cite{brune_quantum_1990} and phase shifts \cite{gleyzes_quantum_2007}.

\begin{figure*}
    \centering
    \includegraphics[width=\textwidth]{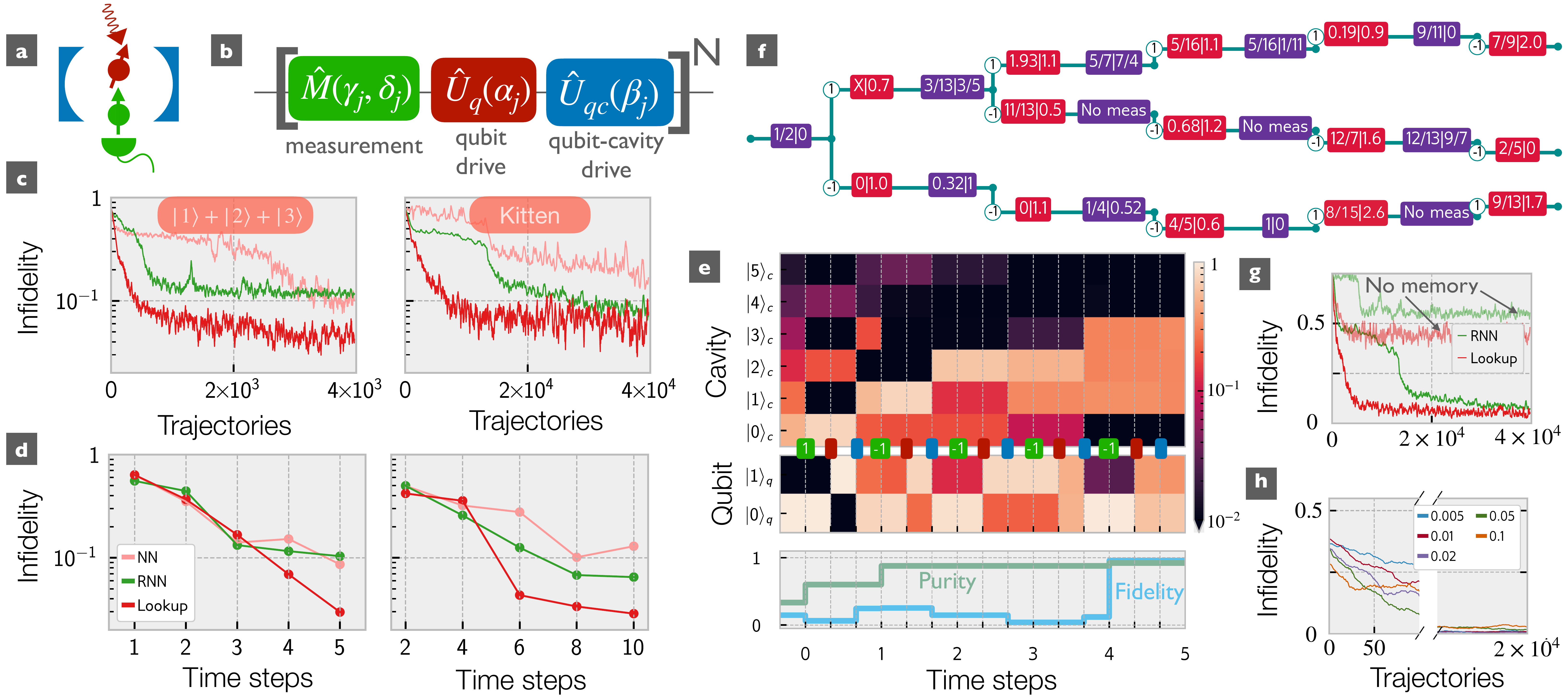}
    \caption{State preparation from a thermal state (with average occupation ${\bar n}=1$), employing feedback. a) Schematic of the Jaynes-Cummings system with an additional ancilla qubit used for the measurement. b) Sequence of parameterized controls. Each trajectory corresponds to $N$ time steps. Each time step consists in a parameterized measurement, followed by two unitary gates.  c) Gradient-ascent progress for two target states (the curves are smoothed with a moving average), and d) final infidelity vs total number of time steps. In (d), each point is the best out of 30 training runs. The statistics of the final infidelity for random training initialization is analyzed in Appendix I.  is analyzed in Appendix \ref{app-further-num-res}. e) Evolution of reduced qubit and cavity state (probability as color) for one trajectory of the converged strategy (target $\left| 1\right\rangle + \left| 2\right\rangle + \left| 3\right\rangle $); time points of measurements (with results) and controls are indicated as in (b). f) Corresponding decision tree, for the most probable sequences of measurement outcomes. The red boxes show $(\alpha/\pi|\beta/\pi)$, and ``no meas" means the parameters $\gamma,\delta$ are such that no measurement takes place. g,h) Gradient-ascent progress (for the same state and number of steps as in e,f) with and without memory and for different values  of the learning rate (see legend).
        \label{fig:Figure3_JCResults}}
\end{figure*}

Fig.~\ref{fig:Figure3_b_purification}c shows the results of applying the feedback-GRAPE method to this problem (labeled 'Adaptive'). We employ a recurrent neural network to produce the controls $F^j=(\gamma_j,\delta_j)$ when provided with the measurement outcome sequence (more details on numerical parameters can be found in appendices D,E). As we see, the impurity quickly decreases with the number of allowed measurements, and it does so significantly better than in a non-adaptive scheme, where the sequence of measurement controls $\delta_j$ and $\gamma_j$ is still optimized, but where these controls are not allowed to depend on previous measurement outcomes. To visualize and analyze the numerically obtained strategy, we introduce in Fig.~\ref{fig:Figure3_b_purification}d a decision tree. This is extracted via an automated numerical procedure, by running many trajectories and noting in each case the controls suggested by the adaptive strategy. The controls are a deterministic function of previous measurement outcomes. 

Such a decision tree will  contain all information about the adaptive strategy learned by the NN and can possibly allow the user to give it a physical interpretation and extrapolate analytical solutions for larger numbers of control steps. This might require to leverage  any available physical understanding of the control operations, e.g. identifying physically significant values of the control parameters.  Using our understanding of the model's physics, we can choose to (analytically) interpret the controls, e.g. trying to represent them in terms of fractional multiples of $\pi$. 
This kind of analysis is optional, and independent of our method, but it nicely demonstrates what can be usefully done in settings with discrete measurements, generating additional insights after running the general-purpose algorithm. For example, here, we were able to take inspiration from the decision tree for four measurements and a specific value of the temperature to extrapolate the optimal purification  strategy for any temperature and any number of measurements, see appendix \ref{app-analysis-strategies}. 

\subsection{State preparation from a thermal state with Jaynes-Cummings  controls} 
We now turn to a task that involves both feedback and control simultaneously. Specifically, we consider state preparation out of a thermal state, for target states that are selected as arbitrary superpositions of the first few Fock states.  For this purpose, we 
consider the setup shown in Fig.~\ref{fig:Figure3_JCResults}a, comprising both an ancilla and a control qubit to combine
 the parameterized  measurements introduced in Section \ref{sec:state_pur} with the Jaynes-Cummings control gates introduced in Section \ref{sec:open_loop}. The resulting sequence of parameterized controls is shown in Fig.~\ref{fig:Figure3_JCResults}b.

Results  for the state $(|1\rangle+|2\rangle+|3\rangle)/\sqrt{3}$ and a four-component kitten state $|\psi^{\rm Kit4}_{\tilde{\alpha}}\rangle$ with $\tilde{\alpha}=\sqrt{2}$, are shown in Fig.~\ref{fig:Figure3_JCResults}c,d. Feedback-GRAPE converges in about 1000 gradient-ascent steps (each operating on a batch of 10 sampled trajectories).  We ran the method several times, starting with different initial random configurations of the trainable  parameters $\theta$, demonstrating that convergence is robust, despite the usual absence of a guarantee for such a non-convex optimization problem. For more details see Appendix \ref{app-further-num-res}. 

It is interesting to analyze in some more detail the convergence behaviour. As one noteworthy observation, despite the overall very good performance, we sometimes find that the algorithm may get stuck at suboptimal solutions if we increase the total number of time steps available for the feedback sequence (Fig.~\ref{fig:Figure3_JCResults}d). Ideally, an increased number of steps should always lead to an improvement (in the present scenario), but apparently the larger space of control variables then becomes challenging. This can be mitigated to some extent by running the gradient ascent repeatedly from random starting conditions.   We discuss other possible solutions to this general problem in Section \ref{section-scalability}.

One motivation for the use of a neural network instead of a lookup table is that the number of parameters needed for a tree-type table grows exponentially, while a neural network could in principle make use of a much smaller number of parameters. Also, it may be expected that the strategy of a network generalizes to situations with a number of time steps larger than the one it was trained on. Despite these obvious advantages of neural networks, we found (to our surprise) that lookup tables often converge to better fidelities than recurrent NNs, in the present example scenario with feedback, cf the red and green lines in Fig.~\ref{fig:Figure3_JCResults}c,d.  Another important observation is that  both look-up table and recurrent NN methods fare at least equally well if not better than a quantum-state-aware NN controller (pink line in Fig.~\ref{fig:Figure3_JCResults}c,d). As mentioned above, these methods are preferrable as they do not require any real-time simulations during deployment in an experiment. The reasons for these observations are still unclear and merit future investigation.

What is the nature of the feedback strategies that the algorithm discovers? Naively, we might expect the following strategy: an optimized adaptive purification phase, of the kind discussed above, leading to some Fock state $\left|n \right\rangle$, followed by state preparation that is derived from the Law-Eberly protocol (e.g. going back down to the ground state and then building up the arbitrary target state from there). However, the actual strategies discovered by feedback-GRAPE are significantly more efficient. They interleave adaptive measurements and controls already in the first stage of the process. This can be seen in Fig.~\ref{fig:Figure3_JCResults}e,f, where the goal was to prepare the equal superposition $(\left| 1 \right\rangle + \left| 2 \right\rangle + \left| 3 \right\rangle)/\sqrt{3}$. Again, it is possible to obtain more information about the full strategy (as opposed to a single trajectory), by extracting a decision tree (Fig.~\ref{fig:Figure3_JCResults}f). There, we observe that measurements are sometimes deliberately performed in such a way that certain Fock states are completely ruled out (their probability is set to 0), which requires certain choices of measurement control parameters. Simultaneously, qubit-cavity interaction cycles are employed to reduce the excitation number of the cavity.

\begin{figure}
    \centering
    \includegraphics[width=\columnwidth]{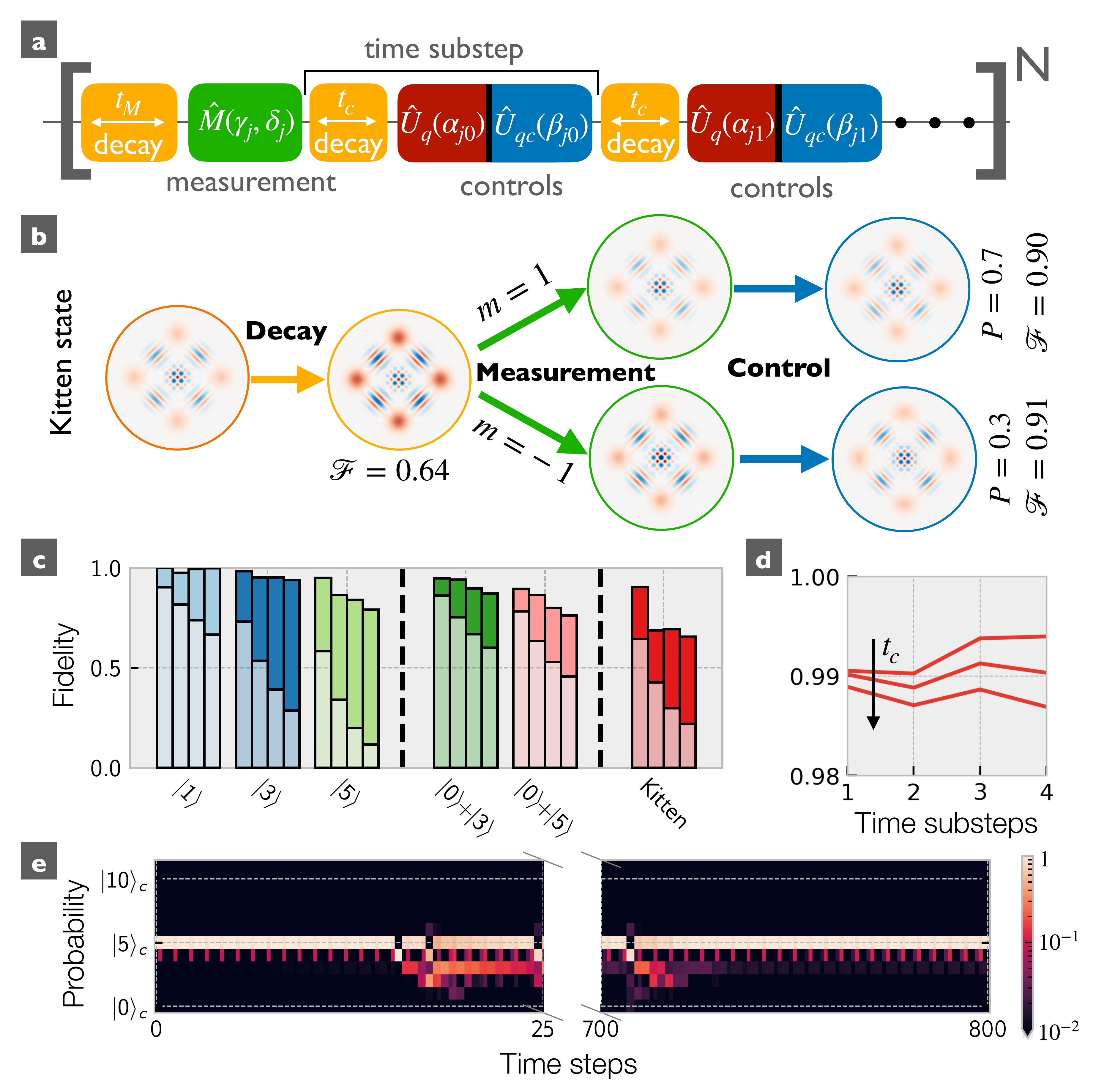}
    \caption{State stabilization with Jaynes-Cummings controls. a) Sketch of the feedback control sequence, including also physical decay of the cavity. We assume that multiple control substeps can be applied after each   measurement. Each substep comprises a qubit driving gate followed by a qubit-cavity interaction gate.   The decay with rate $\kappa$ is incorporated by interleaving the parameterized controls with dissipative evolution of fixed durations $t_M$  and $t_c$ before a measurement and control substep, respectively. See Appendix \ref{appendix:NN_lookup} for more details.  b) Trajectories after optimization. We show the evolution of the oscillator Wigner function  during one step of the feedback control sequence stabilizing  a four-legged kitten state with average photon number $\bar{n}=9$. Indicated are also the probabilities $P$ of each measurement outcome as well as the fidelity ${\cal F}$ after the decay and control steps. c) Performance of the strategy found by feedback-GRAPE.  We show the final  fidelity for various target states and number of steps $N$. After each decay and measurement, a single control sequence (i.e. only one choice of $\alpha$ and $\beta$) is applied.  The four columns represents different numbers of decay steps experienced by the state (here: N=1, 2, 3 and 4). The bars with lower values show the bare decay of the fidelity (here for $\kappa t_M=0.05$ and $t_c=0$), when no feedback strategy is employed. d) For sufficiently low dissipation (small $t_c$), the fidelity can be increased by applying more control substeps.   We show the  fidelity as a function of the number of substeps for $N=1$, $\kappa t_M=0.1$ and three different decay durations  $t_c$, $\kappa t_c=0,
    1\times 10^{-4}, 2\times 10^{-4}$. e) Stabilization of a Fock state (here $\ket{5}$) for an arbitrarily long time, employing the generalization ability of a recurrent neural network (RNN).}
    \label{fig:Figure4_stabilization}
\end{figure}
 
\subsection{State stabilization in a noisy environment with Jaynes-Cummings controls}
\label{sec:state_stabilization}
Quantum state stabilization in a noisy environment  represents another challenging task that can be solved using feedback-GRAPE. 

In this scenario,  the interaction with the environment induces  decay and decoherence of the quantum state.  
Both these effects can   be suppressed  by probing the system with an appropriate stream of quantum measurements interleaved with unitary gates, leading to the long-term stabilization of the quantum state.

We use the same Jaynes-Cummings  feedback control scheme as in the previous section, allowing for multiple control substeps. Each substep comprises  a qubit drive and a qubit-oscillator interaction gate. In addition,
we  model physical decay of the cavity with decay rate $\kappa$, interleaving the substeps of the feedback control sequence  with intervals of dissipative evolution of fixed durations  $t_M$ and $t_c$, cf Fig.~\ref{fig:Figure4_stabilization}(a). These could be interpreted as waiting times before applying instantaneous measurement and control gates, respectively, but more realistically they can effectively incorporate the decay and decoherence  that has occured during finite-duration operations. For this approximation to work the decay needs to be weak, which is the case for our scenario.

As an illustrative example,  we show the two  possible trajectories for a circuit comprising a single step of the feedback control sequence and optimized to stabilized a four-legged kitten state  $|\psi^{\rm Kit4}_{\tilde{\alpha}}\rangle$  with average photon number $\bar{n}\approx 9$, cf Fig.~\ref{fig:Figure4_stabilization}(b). The learning algorithm selects the measurement parameters  $\delta=0$, and $\gamma=\pi/2$. This seems a natural choice  because in this case $\hat{M}(m=1)|\psi^{\rm Kit4}_{\tilde{\alpha}}\rangle=|\psi^{\rm Kit4}_{\tilde{\alpha}}\rangle$. This implies that the measurement leaves invariant the target Kitten state. In addition, a measurement outcome $m=-1$ postselects an orthogonal state. The latter outcome is verified, for example, if a single-photon leaks out of the cavity and occurs with $30\%$ probability after the dissipative evolution of duration $\kappa t_M=0.05$. The fidelity recovers up to $91\%$  after a single Jaynes-Cummings control sequence is applied. This is a good result considering the limited expressivity of our control scheme. The fidelity can be moderately increased by allowing more steps provided that the decay during the control protocol is not too large, cf. Fig.~\ref{fig:Figure4_stabilization}(d). 

The fidelity for several quantum states, including Fock states and superpositions thereof, for a varying number of time steps (up to 4) is shown in Fig.~\ref{fig:Figure4_stabilization}(c). As a comparison,  the bare decay in the absence of any control is also displayed. These  results demonstrate the ability of feedback GRAPE to discover strategies to mitigate the effect of dissipation for a variety of quantum states.

As we explained above, lookup tables often perform surprisingly well. We now briefly demonstrate, in the context of state stabilization, one example where the power of a neural network is clearly helpful (Fig.~\ref{fig:Figure4_stabilization}e). We first train a RNN on sequences of 20 steps, with the goal to stabilize a given Fock state for an arbitrarily long time. For this example, the cumulative reward of a trajectory is not only the final fidelity, but the sum of fidelities at all time steps. After training, we test on a 40 times longer simulation, and we see that the strategy learned by the RNN generalizes well even for longer sequences. We note how the strategy can recover, even when some ``unlucky" measurement outcomes significantly perturb the quantum state.

\begin{figure*}[t]
    \centering
    
    \includegraphics[width=\textwidth]{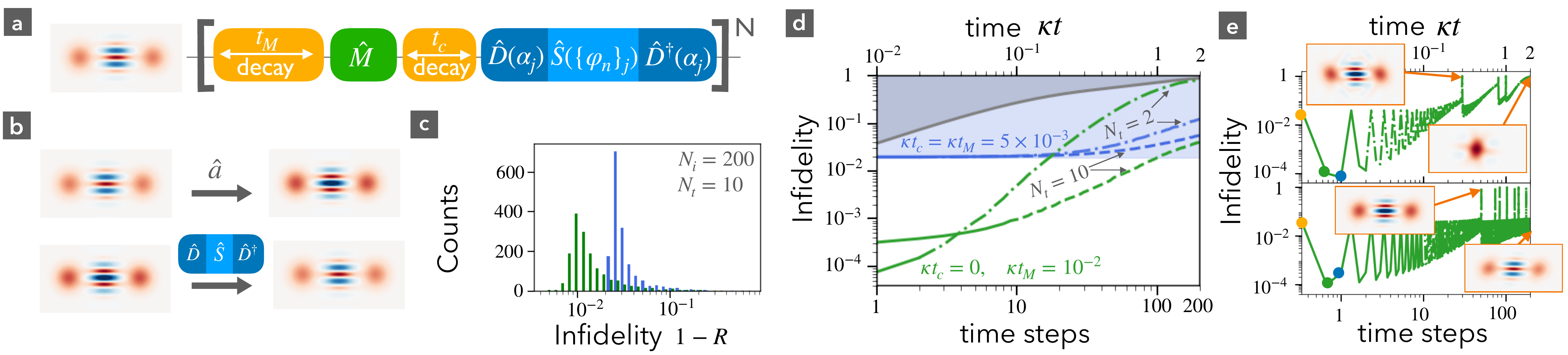}
    \caption{\label{fig:figure_5}State stabilization with more expressive controls. a) Sketch of the feedback control sequence. A time-evolution step consists in a parity measurement, interleaved between two dissipative evolution substeps  and a block of unitary gates  comprising two phase-space displacement gates interleaved with a SNAP gate.  b) Evolution for an even kitten state. The decay of a single excitation induces the transition to an odd kitten state.  An optimized feedback sequence can transform this state back into the original kitten state with high fidelity. c) Histograms showing the  probability distribution for the time-averaged infidelities of single quantum trajectories for $N_i=200$ time steps during inference, while training runs only had  $N_t=10$ time steps.     The histogram colors identify two different dissipation scenarios. The durations $t_c$ and $t_m$ of the dissipative evolution substeps    are displayed in  d. The underlying feedback strategies are predicted  by two different NNs, chosen as the best performing out of 5 training runs. d) Infidelities as a function of time, averaged over a representative set of  trajectories. We consider two dissipation scenarios (color) and two  values of $N_t$. The curves are plotted as solid lines in the time interval seen during training. The NNs used to predict the feedback strategies are the best performing out of 5 training runs. Also shown is the infidelity in the absence of any control (gray) and  after    dissipative evolution for a duration  $(t_c+t_M)/2$ (edge of blue-shaded region) e) Time evolution of the infidelity for two typical inference trajectories, predicted using NNs trained on trajectories of duration $N_t=2$ (top) and $N_t=10$ (bottom). Here, we show also the infidelity after each substep. The first  dissipative substep/measurement is marked as a yellow/green dot.  
    Also shown are the Wigner functions after the first odd parity measurement and at the end of the time evolution.}
 \end{figure*}

\subsection{State stabilization with SNAP gates and displacement gates}

Using feedback-GRAPE applied to the Jaynes-Cummings scenario has allowed us to discover strategies extending the lifetime of a range of quantum states. However, for more complex quantum states like  kitten states the infidelity becomes significant after just  a few dissipative evolution  steps in spite  of the feedback, cf.~Fig.~\ref{fig:Figure4_stabilization}(c). This raises the question whether the limited quality of the stabilization is to be attributed to a failure of our feedback-GRAPE learning algorithm to properly explore the control parameters landscape or rather due to the limited expressivity of the controls. With the goal of addressing this question, we test our  method on the state stabilization task using a more expressive control scheme.
Specifically, we use a universal control scheme originally proposed in \cite{Krasnatov_Universal_2015}.  This  consists in a sequence of interleaved \textit{Selective Number-dependent Arbitrary Phase} (SNAP) gates $\hat{S}(\{\varphi_n\}))=\sum_{n}e^{i\varphi_n}|n\rangle \langle n|$  and displacement gates $\hat{D}(\alpha)=\exp[\alpha\hat{a}^\dagger-\alpha^*\hat{a}]$. This is the same control scheme adopted by Sivak et al \cite{sivak_model-free_2021} to demonstrate their model-free optimal-control approach  for state preparation and, as mentioned above, we  have also used it to demonstrate the   preparation of a so-called Gottesman-Kitaev-Preskill state \cite{Gottesman_encoding_2001} with open-loop controls, see Appendix \ref{App:grid_state}. We now go one step further and employ these powerful controls inside a feedback-based state stabilization scheme, optimized via feedback-GRAPE.

As a test example we consider the feedback-based stabilization of a  kitten state built from two coherent states, $|\psi^{\rm Kit2}_{\tilde{\alpha}}\rangle\propto |{\tilde{\alpha}}\rangle+|-{\tilde{\alpha}}\rangle$ with ${\tilde{\alpha}}=2$, corresponding to an average photon number of $\bar{n}\approx |{\tilde{\alpha}}|^2=4$. This state has even parity 
$\hat{P}|\psi^{\rm Kit2}_{\tilde{\alpha}}\rangle=|\psi^{\rm Kit2}_{\tilde{\alpha}}\rangle$ with $\hat{P}=\exp[i\pi\hat{a}^\dagger\hat{a}]$. After an excitation leaks out of the cavity, it decays into an odd cat state with the same ${\tilde{\alpha}}$, cf Fig.~\ref{fig:figure_5}(b). Thus, we can  detect  these  decay processes using repeated  parity measurements.
After such a process, an optimized  control step  can transform the odd kitten  state back into the target kitten state with  high fidelity, Fig.~\ref{fig:figure_5}(b). These considerations motivate us to use  the feedback control sequence  shown in 
Fig.~\ref{fig:figure_5}(a). We use as  control parameters the real and imaginary part of the phase-space displacements $\alpha^j$ together with the phases $\varphi^j_n$ for the first $N_{\rm SNAP}$ Fock states (the remaining phases are set to zero).

We have trained several  NNs  using as a reward the  time-averaged   fidelitiy,   $R=\sum^N_j {\cal F}_j/N$ with  ${\cal F}_j$ calculated  after applying the block of unitary gates. We have considered  two different durations $N_t$ of the quantum trajectories  seen during training. We have also considered two different scenarios for our description of the dissipative evolution. In a first scenario, the dissipative evolution is concentrated before the measurement, $t_c=0$ and $\kappa t_M=0.01$. In a second scenario, the dissipative evolution is subdivided into two intervals before and after the measurement, $\kappa t_M=\kappa t_c=0.005$.  Finally, we have tested the NNs performance in stabilizing the kitten state for $N_i=200$ time-steps. This is much larger than the number of time-steps seen during training, $N_t=2$ or $N_t=10$, and the typical decoherence time $t_d$ for our kitten state, $\kappa t_d=\bar{n}^{-1}\approx 1/4$ corresponding to $\approx 25$ time steps, Fig.~\ref{fig:figure_5}(c-d).

Our  results obtained using the best performing NNs for each of the four scenarios discussed above
are summarized in Fig.~\ref{fig:figure_5}(c-d). Panel (d) shows the infidelity  as a function of time, averaged over a  representative set of trajectories. The infidelity is plotted as a solid line in the time interval seen during training (and dashed thereafter). As a comparison,  the infidelity without any stabilization (gray line) and the infidelity after a single interval of dissipative evolution of duration $(t_c+t_M)/2$ (the lower-edge of the blue-shaded region) are also shown. The latter represents a theoretical lower bound for the infidelity in the scenario with $t_c=t_d$ (as the effects of dissipation after the last measurement can not be corrected).  Panel (c) shows a histogram for the distribution of the time-averaged infidelity  $1-R$   for single quantum trajectories.

From these results, we can generally conclude that the feedback strategies discovered using feedback-GRAPE allow to maintain a low infidelity for measurement sequences much longer than those seen during training, demonstrating a remarkable power of generalization. For the scenario with dissipation injected after the measurement (blue lines), the fidelity  remains just above the theoretical lower bound for much of the time-evolution. In both dissipation scenarios, the NNs trained on longer measurement sequences perform better. This tendency is most evident in the dissipative scenario with $t_c=0$. In this case, the NN trained on sequences of just two measurements performs very well on a similar time horizon. However, its ability to generalize the feedback strategy to a longer time evolution is poorer.

In order to better understand this behavior, we plot the time evolution of the infidelity 
 for two typical trajectories (one for each NN), showing also the infidelity after each substep (decay, measurement, or control), cf Fig.~\ref{fig:figure_5}(e). From these results, one can see that the strategies learned by the two NNs are qualitatively different: The NN trained on  shorter measurement sequences pursues a greedy strategy that decreases the infidelity after each block of unitary gates (top). In contrast, the NN trained on a longer measurement sequence  learns a non-greedy strategy which increases the infidelity after even parity measurements (bottom). Another obvious difference between the two strategies is that the latter triggers more often odd parity measurements, which are imprinted in the infidelity as peaks of unit height cf Fig.~\ref{fig:figure_5}(e). This is a good sign because it indicates that the parity measurements are able to extract more of the entropy injected during the dissipative part of the dynamics or, equivalently, better suppress decoherence. This is also reflected in the clear interference fringes in the Wigner function of the state after $200$ time steps, $8$ times larger than the typical decoherence time, cf inset of Fig.~\ref{fig:figure_5}(e).

 This example highlights the importance of training on long measurement sequences to develop more robust stabilization strategies.
 
 Overall, we conclude that expressive controls like the well-established SNAP gate allow feedback-GRAPE to discover excellent feedback strategies in an efficient manner, and that strategies discovered for shorter training sequences generalize nicely to much longer sequences. To the best of our knowledge, the level of state stabilization performance demonstrated here has not been achieved with any other method, despite this being an area of active research in the context of quantum devices and quantum error correction.

\begin{figure*}[t]
    \centering
    
    \includegraphics[width=\textwidth]{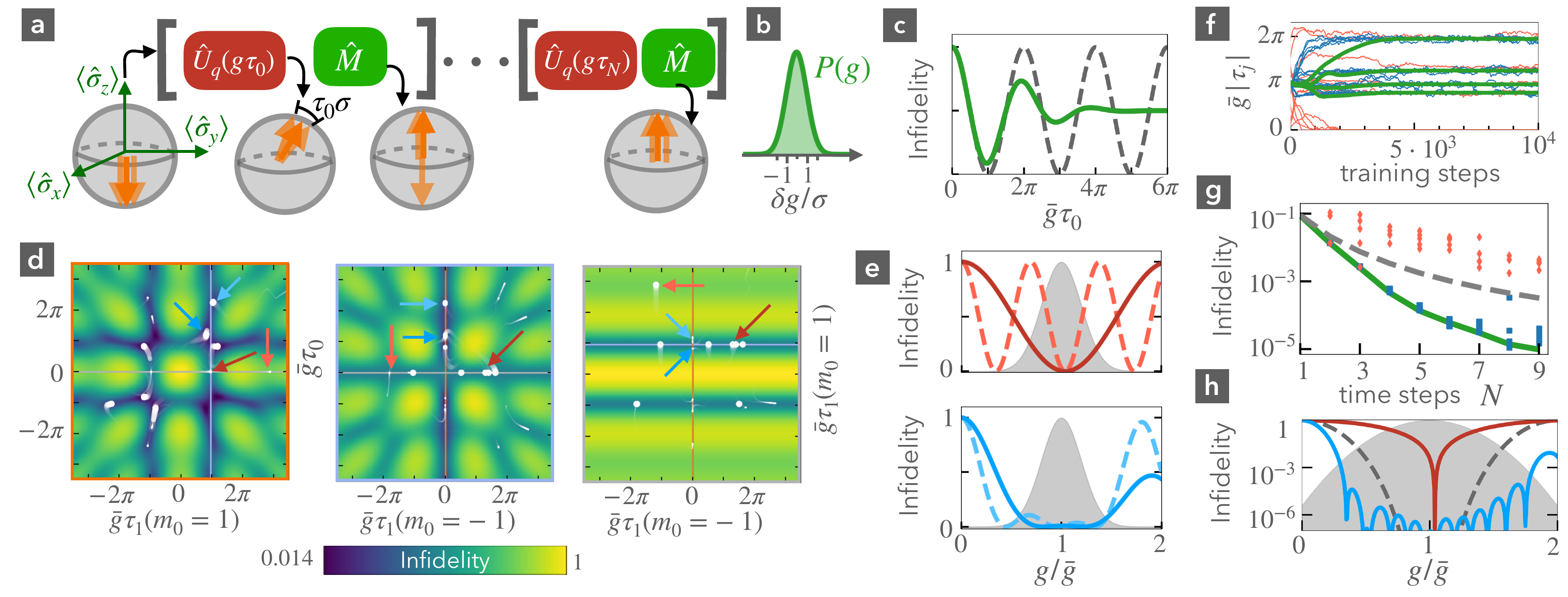}
    \caption{\label{fig:figure_6}State preparation with parameter uncertainty. (a) Sketch of the feedback control sequence consisting of $N$ pulses of duration $\tau_i$ interleaved with projective measurements. We also show the corresponding dynamics of an (inhomogeneous) ensemble of qubits represented as spins on the Bloch sphere. The goal of the feedback strategy is to prepare as many spins as possible in the ``up" state. (b) The coupling $g$ to the driving field is assumed to be a random Gaussian variable of width $\sigma$. (c) Optimization landscape for the case $N=1$ and  $\sigma/\bar{g}=0.2$ (solid line) and $\sigma=0$ (dashed line). (d) Three 2D cuts of the 3D optimization landscape for the case  $N=2$ and $\sigma/\bar{g}=0.2$. The 1D projection of each 2D cut into the other cuts is marked as a line of the same color as the frame.   Also shown (on all cuts) are the evolutions of the three learning parameters during training for 26 different training runs. Each run is represented by $200$ snapshots which in turn are represented by dots of varying size and transparency, to represent the distance from the 2D plane and the time at which they where recorded. (e) Infidelity $1-\langle{\cal F}_N(g)\rangle_{m}$ as a function of the coupling $g$ for four strategies.      Also shown  is the distribution of the coupling $g$. The underlying control parameters are marked by the arrows of the same color in (d). f) Evolution of the control parameters during training.  Shown are the durations $\tau_j$  of the $j$-th pulse, conditional on the qubit being in the groundstate for all previous measurements, $\tau_j\equiv\tau_j(\mathbf{m}_{j-1})$ with $\mathbf{m}_{j-1}=(1,\ldots,1)$, for $0\leq j\leq N-1$. Several training runs are depicted, running feedback-GRAPE   with random initialization (red lines, one run with $N=7$ and one run with $N=6$), an initialization close to the intuitive feedback strategy with durations $\bar{g}\tau_j\approx\pi$ and zero otherwise   (blue lines; five runs with $N=4$), or the constrained optimization protocol (green thick lines, 1 run with $N=4$). g) Infidelity $1-\langle\langle{\cal F}_N(g)\rangle_{m}\rangle_g$ as a function of the number of time steps $N$ for the three different optimization protocols (for the first two protocols, five runs  for each $N$ are shown). Also shown is the infidelity for the intuitive feedback strategy $\bar{g}\tau_j\approx\pi$. h) Same as (e), here, for the optimal strategy without feedback (red), the optimal (blue) and intuitive (gray dashed) feedback strategies with $N=8$.}
   
\end{figure*}

\subsection{State preparation in the presence of model uncertainty}
\label{state_prep_uncertain_model}
Until now we have assumed that a model for the stochastic dynamics of our quantum system is known without any uncertainty. However, in practice, the model parameters are known only with a finite precision, they might deviate from theoretical predictions because of disorder and be difficult to measure precisely. Moreover,  they might even be subject to slow drifts because of environment-induced changes in the quantum device. For all these reasons, an important direction of research in optimal control focusses on improving model-based methods to better perform in the presence of model uncertainties \cite{egger_adaptive_2014,wu_data-driven_2018,wu_learning_2019,ding_collaborative_2021}. Our goal in the present section is to explore how feedback-GRAPE can contribute to this challenge. At the same time, this will also allow us to study the optimization landscape in a manageable example.

There are two fundamentally different approaches to deal with model uncertainties in a coherent control setting. In the first, ``data-driven" approach experimental data are used during training. In the second ``fluctuation-model-based" approach,  the  model parameters are sampled from a probability distribution during training. The resulting parameter fluctuations reflect an imperfect knowledge of the model parameters, and the strategy is optimized for being resilient against these fluctuations. 

Several extensions of GRAPE have been proposed to incorporate uncertainty in the model parameters following either the ``data-driven" approach \cite{egger_adaptive_2014, wu_data-driven_2018,ding_collaborative_2021} or the ``fluctuation-model-based" approach \cite{wu_learning_2019}. However, we emphasize that none of these extensions include feedback.

Our feedback-GRAPE method can also be extended to account for parameter uncertainties,  both by using a ``data-driven" or a ``fluctuation-model-based"  approach. For the ``data-driven" approach, one could follow a similar approach as d-GRAPE  \cite{wu_data-driven_2018} and c-GRAPE \cite{ding_collaborative_2021},  modifying the analytical formula for the  learning gradient (see Appendix \ref{appendix-GRAPE-evol-eqs}) to incorporate also operators  estimated in quantum tomography experiments.   Alternatively, feedback-GRAPE could be used in combination  with model-free RL with both methods sharing the same controller in the form of a recurrent NN. In this setting, feedback-GRAPE would be used for the initial training of the controller allowing to explore higher-dimensional  control parameter spaces. Afterwards,  the controller will be  trained on experimental data using a model-free approach to obtain a more accurate feedback strategy. 

In the following, we demonstrate instead in more detail the  ``fluctuation-model-based" approach. This approach is not only more straightforward to implement. It is also best suited to a scenario with feedback: Since the measurement statistics depends on the model parameters,  the measurement outcomes  carry  information about the underlying model parameters. At the same time,  the strategies are conditional on the measurement outcomes and, thus,  can be adapted to the most likely underlying model parameters. This approach  is so powerful that it is sometimes worth  to adopt it  even in  a model-free RL approach to forgo costly training on experimental data. A prominent example of this  is the control of tokamak plasma where  zero-shot transfer from simulations to hardware with imprecisely known parameters has been demonstrated \cite{degrave_magnetic_2022}. 

We consider a simple toy model in which an inhomogeneous ensemble of qubits initially in the ground state are subject to a series of $N$ pulses interleaved with projective measurements on their computational basis, cf Fig.~\ref{fig:figure_6}a. The duration of the pulses $\{\tau_j\}$ can be controlled but the coupling $g$ of a qubit to the driving field is a random variable distributed  according to a Gaussian distribution of average $\bar{g}$ and standard deviation $\sigma$, cf Fig.~\ref{fig:figure_6}b. As a consequence,  the Bloch sphere rotation angles $\alpha_j=g \tau_j$  will also be  Gaussian random variables, now with standard deviations $\tau_j\sigma$, cf Fig.~\ref{fig:figure_6}a.

Our goal is to maximize the number of qubits correctly flipped to their excited state or, equivalently, the fidelity averaged over the measurement outcomes and the coupling $g$, $\langle\langle{\cal F}_N\rangle_{\mathbf{m}\sim P(\mathbf{m}|g)}\rangle_{g\sim P(g)}$. In this case, we are interested in the optimal solution for a fixed number of pulses $N$.  

Before discussing this problem in general, let us consider the limiting case of only one time step, $N=1$. In this case, the pulse duration $\tau_0$ is the only control parameter and there is no feedback by any previous measurement.  The average fidelity as a function of this parameter is shown in Fig.~\ref{fig:figure_6}c. In the limiting case without fluctuations, a single $\pi$-pulse of duration $\tau_0=\pi/g$ or any of its odd integer multiples  will achieve zero infidelity, cf dashed line in Fig.~\ref{fig:figure_6}c. Once parameter fluctuations are introduced, one might still expect   $\tau_0=\pi/\bar{g}$ and any of its odd integer multiple to be optimal. In this way, the spins with coupling $g=\bar{g}$, corresponding to the peak of the parameter distribution $P(g)$, would be flipped with unit probability. However, we observe that in reality shorter pulses are favored  because they give rise to a narrower distribution $P(\alpha_0)$ of the rotation angles $\alpha_0=g\tau_0$. This physics leads to a single optimal pulse of duration slightly shorter than $\tau=\pi/\bar{g}$ and a series of suboptimal pulse durations corresponding to local minima of the average fidelity, see Fig.~\ref{fig:figure_6}c. 

Next, we consider the  simplest scenario  with feedback, which corresponds to $N=2$ time steps, i.e. feedback on a single measurement. In this case, the control parameters are the first pulse duration $\tau_0$ and the two conditional durations of the second pulse,  $\tau_1(m_0=1)$ and $\tau_1(m_0=-1)$. If we use our look-up table approach to directly optimize these control parameters,
 the underlying optimization landscape is a 3D function.  In Fig.~\ref{fig:figure_6}d, we show three 2D cuts of the optimization landscape along with the evolution of the training parameters for 26 feedback-GRAPE training runs, starting from random initial conditions. If we view as equivalent  strategies that are connected by  transformations of the control parameters that leave invariant the fidelity $\langle {\cal F}_N(g)\rangle_{\mathbf{m}}$, we can associate most runs to just four final feedback strategies, see Appendix \ref{Appendix_opt_landscape} for more details. Their coupling-dependent infidelity  $1-\langle {\cal F}_N(g)\rangle_{\mathbf{m}}$ is displayed in Fig.~\ref{fig:figure_6}e. A plurality of the training runs (10 out of 26 runs) converges to the optimal strategy which consists in the combination of a pulse of duration slightly shorter than $\pi/\bar{g}$ followed or preceded by a slightly longer pulse. This leads to a small infidelity  over the full width of the coupling distribution $P(g)$, cf  blue continuous line in Fig.~\ref{fig:figure_6}e.  However,  for a significant number of runs (9 out of 26)   the duration  $\tau_0$ of the first pulse converged asymptotically to zero during training, cf rightmost cut in Fig.~\ref{fig:figure_6}d. In other words, the first pulse is  switched off, effectively reducing the number of time steps to $N=1$. The resulting infidelity is small on a much narrower band, cf top of Fig.~\ref{fig:figure_6}e. The training can lead to this type of solutions because  isolated attractors for  $N=1$ optimization landscape are promoted  into 1D manifolds of attractors in the $N=2$ optimization landscape, cf the middle and right-hand cuts in  Fig.~\ref{fig:figure_6}d with Fig.~\ref{fig:figure_6}c.

The insight on the learning dynamics gained for the case with $N=2$ time steps can be transferred to the general case of
 an arbitrary number of time steps $N$. For a typical run, one or more pulses are switched off, effectively reducing the number of time steps, cf red lines in Fig.~\ref{fig:figure_6}f.  This leads to many runs ending up in suboptimal solutions,  cf the red diamonds in Fig.~\ref{fig:figure_6}g.  We note in passing that we have observed a similar learning  dynamics  for the state preparation of complex superposition of Fock states using the Jaynes-Cummings controls. Also in that case, a large number of local extrema for an optimization task with $N$ time steps could be constructed adding idle time steps to the optimal solution for a smaller number of time steps, see Appendix \ref{App:State_prep_kitten}.  We expect the same type of local extrema to appear in many  optimal control  tasks that take as ansatz a quantum circuit comprising a finite sequence of parametrized building blocks, irrespective of whether measurements are  present or not. 
 
The challenge posed by the local minima can be addressed by taking a physically motivated initialization of the control parameters, e.g.   $\tau_j(\mathbf{m}_{j-1}) \approx \pi/\bar{g}$  if $\mathbf{m}_{j-1}=(1,\ldots,1)$ and $\tau_j(\mathbf{m}_{j-1})=0$ otherwise. This is close to a $\pi$-pulse for $g\approx\bar{g}$ if no previous measurement with outcome $m=-1$ indicated  that the spin has  flipped. With this approach, we consistently reach the optimal solution for $N\leq 6$, cf blue squared in Fig.~\ref{fig:figure_6} g. For even larger $N$, we reach a regime in which for a typical batch of trajectories used to calculate the gradient all spins have flipped. This leads to  a very noisy gradient making it difficult to distinguish between many available feedback strategies with low infidelity. We eliminate this problem taking the ansatz $\tau_j(\mathbf{m}_{j-1})=0$ if $\mathbf{m}_{j-1}\neq(1,\ldots,1)$ for the feedback protocol. In this way, the feedback task is reduced  to the optimization of  $N$ control parameters  $\tau_j\equiv\tau_j(\mathbf{m}_{j-1})$ with $\mathbf{m}_{j-1}=(1,\ldots,1)$. At the same time,  the number of possible measurement outcomes is also reduced  to the same value. This reduction of the decision tree to only $N$ branches makes it efficient to evaluate the sum in Eq.~(\ref{eq-one-discrete-msmt})  without resorting to measurement sampling, reducing the gradient fluctuations. In addition, it eliminates the basin of attraction of the suboptimal strategies with a reduced effective number of pulses. Overall, it leads to very robust learning even for large values of $N$, cf green lines in Fig.~\ref{fig:figure_6}fg.

Finally, we comment on the robustness to parameter uncertainty for the feedback strategies obtained using feedback-GRAPE. The average infidelity $\langle\langle{\cal F}_N\rangle_{\mathbf{m}}\rangle_{g}$ obtained using the feedback strategy with $N=8$ time steps is of the order  $\sim 10^{-5}$, cf Fig.~\ref{fig:figure_6}g. This means that, in spite of the broad distribution of couplings $g$, only one qubit out of every $10^5$ would remain in the ground state. This compares to approximately $50$ with the  intuitive feedback strategy and $10^4$ using the optimal strategy without feedback. We note that  the infidelity $1-\langle {\cal F}_N(g)\rangle_{m}$ is suppressed in a broad range of couplings $g$, remaining  below the threshold $10^{-3}$ in a broad band of width $\approx 1.5\bar{g}$, cf Fig.~\ref{fig:figure_6}h. From this, we can conclude  that the robustness of our strategy to parameter uncertainty will  extend beyond the particular coupling distribution used during training. 

This example thus has convincingly shown the ability of feedback-GRAPE to deal with parameter uncertainties, both by finding strategies that properly take into account the size of the fluctuations and, on top of that, by exploiting the extra information obtained via measurements.

\section{Scalability and Optimization Landscape in Feedback-Grape}
\label{section-scalability}

As we have seen in the numerous examples presented so far, feedback-GRAPE performs very well for quantum feedback tasks in physically relevant scenarios, including the preparation and stabilization of rather complex states. Even though it occasionally got stuck in local optima, in our examples this could often be remedied very simply by re-running from different random starting points a number of times. Nevertheless, in the present section we want to address the aspects of the non-convex optimization landscape and scaling towards larger quantum systems, like those consisting of many qubits, in a more general fashion. These challenges are of course by no means unique to feedback-GRAPE, and we will consequently rely a lot on observations that have been made in the literature, starting from the original GRAPE and going towards recent results on variational quantum circuits.

Already in the original GRAPE article \cite{khaneja_optimal_2005} it was recognized that generally speaking GRAPE is a non-convex optimization problem, and it was suggested that adding stochasticity to the gradient update step could help to jump out of local minima. More refined approaches would perform simulated annealing, i.e. slowly reducing the noise strength over time. We note that, in contrast to GRAPE itself, some form of stochasticity is automatically generated in our case by the random outcomes of measurements. The noise strength   can effectively be reduced over time  using a  learning rate schedule. 

Two common approaches to work around local minima are also demonstrated in our own numerical experiments, reported in the present article. The first approach consists in avoiding the local minima by starting from a smart initialization of the training parameters.
 This could be physically motivated as in Section \ref{state_prep_uncertain_model} or, as it is sometimes done in RL, obtained using some form of   pre-training based on supervised learning  \cite{xie_pretraining_2022} of an existing approximate strategy. The second approach  consists in modifying the ansatz for the feedback protocol which will also modifies the optimization landscape, possibly eliminating or reducing the problematic local minima.  A better ansatz could be found using physical insight, as in Section \ref{state_prep_uncertain_model}, or using some form of derivative-free optimization, e.g. model-free RL as in \cite{Yao_reinforcement_2021}.

Another option for addressing the issue of local optima is to employ the natural gradient. In \cite{Wierichs_avoiding_2020} it was shown for variational quantum circuits in systems of up to 40 qubits that this technique, though computationally more expensive, most successfully avoids local minima when compared to the more well-known techniques, i.e. direct adaptive gradient descent or quasi-Newton methods like the Broyden–Fletcher–Goldfarb–Shanno algorithm. Natural gradient can directly be applied to state preparation problems by computing the Fubini-Study metric of the final state based on its dependence on the control parameters. Thus, it could be employed to help convergence in feedback-GRAPE when the technique is applied to larger qubit numbers.

The examples of feedback-controlled quantum dynamics we have focussed on in this article can be viewed as a combination of parameterized quantum circuits with classical measurements and feedback. When dealing with the question of scalability towards larger qubit numbers, 
the recent literature on variational quantum circuits (without feedback) suggests that there may arise another
challenge that goes beyond the generic problem of getting stuck in local optima for non-convex optimization tasks. As it has been recognized first in \cite{mcclean_barren_2018}  and subsequently discussed at length in the literature, one may be stuck in
parts of the parameter landscape with essentially zero gradients, i.e. gradients that are 
exponentially small in the number of qubits; this is the infamous problem of "barren plateaus".
Fortunately, the importance of this problem for the quantum computing community has
led to a succession of possible suggested solutions, all of which could be transferred
to an application like feedback-GRAPE in case the issue arises when applying it eventually
to systems with larger numbers of qubits. The proposed solutions comprise (i) smart parameter
initialization \cite{grant_initialization_2019,volkoff_large_2021}, e.g. choosing parameters values that initially lead to unitary blocks equal to the identity , (ii) being smart in the choice of circuit ansatz but avoiding overparametrization (too expressible ansatz structures) \cite{holmes_connecting_2022}, (iii) constructing a cost function from local observables instead of a global cost function like the fidelity \cite{khatri_quantum-assisted_2019,cerezo_cost_2021}. Last and most relevant to our work, it has been recently shown that incorporating in the quantum circuit the same type  of stochastic local measurements that are also  used in feedback-GRAPE  could by itself induce a phase transition to a regime without barren plateaus  \cite{wiersema_measurement-induced_2021}. While in their work the measurements only introduce decoherence, it would be worthwhile to explore whether they also help to avoid such plateaus in true feedback scenarios.

Beyond these aspects of the optimization landscape, the overall performance of feedback-GRAPE, and hence its scalability, is also governed by the computational cost associated with single trajectories. Just like any other model-based method, our method can deal only with systems whose time-evolution  can be efficiently simulated on a classical computer. The computational cost of calculating the gradient through the system dynamics grows linearly with the number of time  steps, just like the cost of the direct time evolution itself. Furthermore, when addressing the scalability for multi-qubit systems, it is true that the cost of a simulation will increase with the size of the Hilbert space and thus rise exponentially with the number of qubits. However, this scaling is no worse than in the original GRAPE or for any other model-based RL method, including those methods that are based on simulations interacting with a model-free approach. From experience with numerical simulations of multi-qubit systems, we deem feedback-GRAPE to be feasible still up to about 10 qubits when simulating master equations and maybe 20 qubits when resorting to quantum jump approaches, evolving pure states. This already covers a lot of unexplored territory for quantum feedback.

Another important aspect is the number of trajectories needed for convergence towards the optimal strategy. This is extremely scenario-dependent and therefore hard to predict in general. However, empirically we have seen that typically thousands of time-evolution trajectories are to be evaluated to converge to an optimum. It is here that the model-based approach of feedback-GRAPE has a big advantage over model-free approaches, since the latter have to employ a lot more trajectories just to implicitly learn the expected behaviour of the quantum system. Indeed, among our examples we have briefly discussed the superior performance of GRAPE vs a model-free RL approach in the case of SNAP-gate-based cavity state preparation.

We mention in passing another advantage of feedback-GRAPE: It does not require a real-time Bayesian estimate of the quantum state during deployment in an experiment. In this sense, it is more scalable than other existing model-based quantum feedback approaches based on so-called Bayesian quantum feedback.

Finally, another aspect that may affect scaling is generalizability. Our method, for certain feedback tasks, allows a generalization of the feedback strategy. This is exemplified by our stabilization of a kitten state for a long sequence of $200$ measurements, $20$ times longer than the sequences seen during training. This generalization power decreases substantially the computational cost expended during training.

\section{Extensions}

Before concluding our discussion, we outline possible extensions of the general feedback-GRAPE technique introduced above.

\subsection{Reducing sampling noise by using a value function}
\label{section:Q_function}
The average of the return over different measurement outcome sequences is obtained by sampling, which introduces noise into the estimate of the gradients. We can help suppress the noise by adopting value function approaches that are known as a general technique in reinforcement learning \cite{sutton_reinforcement_2018}. 

To start, we need to discuss the structure of the rewards more carefully. Above, we introduced the overall return (cumulative reward) as the quantity to be optimized. We can also assign the rewards more specifically to individual time steps. For example, during state stabilization we can evaluate the fidelity at each time step and sum it over time to obtain the return. Likewise, it is customary in some optimal control settings to punish large control amplitudes at any given time step. In all these cases, the return is a sum $R=\sum_{j=1}^N r_j$ of individual rewards.

More precisely, in the original approach, we had simply set $R=r_1(m_1|\theta) + r_2(m_2,m_1|\theta) + \ldots $. Here $r_j(m_j,m_{j-1},\ldots | \theta)$ is the instantaneous reward obtained after time step $j$ (which consisted of some control, some measurement yielding $m_j$, and possibly a further control step before assigning the reward). For any time step $j$, this then yields two contributions to the overall gradient ascent update. For example, at $j=2$ we obtain, in a given trajectory with randomly sampled $m_1, m_2, \ldots$ the following contributions:

\begin{eqnarray}
 \partial_{\theta} r_2(m_2,m_1|\theta)  + \partial_{\theta} \ln P(m_2|m_1,\theta) & \cdot & \nonumber \\
\left\{ r_1(m_1|\theta) + r_2(m_2,m_1|\theta) + r_3(m_3,m_2,m_1|\theta) + \ldots \right\} \label{eqn-naive-reward-sum}
\end{eqnarray}
Adding up these contributions for all $j$ and averaging over trajectories yields precisely Eq.\,(\ref{eq-extra-term}).

This is a Monte-Carlo sampling approach. One concern in any such approach is the sampling noise, i.e. in our case the fluctuations of the quantity shown above between different trajectories. We can now take inspiration from the domain of model-free reinforcement learning and the general theory of reinforcement learning \cite{sutton_reinforcement_2018}, where approaches have been invented to reduce the variance in estimations of the gradient update. Recall that in our case, the variance stems from the stochasticity of measurements, whereas in model-free RL it stems from the stochasticity of policy action choices that is encountered in policy-gradient and actor-critic approaches, plus any stochasticity of the environment dynamics. Even though the following steps follow very closely the corresponding tricks known in the model-free RL community, we display them explicitly here, for our modified scenario. This should help avoid any confusion and make this presentation self-contained.

First, when evaluating the gradient above, we need only include the sum of future rewards, since only those can be influenced by the present measurement result. In the example of Eq.\,(\ref{eqn-naive-reward-sum}), this means the term $r_1(m_1|\theta)$ on the second line may be dropped, as it is independent of $m_2$, i.e. the new measurement result. Mathematically, this follows because when we eventually perform the average over trajectories, we have to multiply Eq.\,(\ref{eqn-naive-reward-sum}) by $P(m_2,m_1|\theta)=P(m_2|m_1,\theta) P(m_1|\theta)$. Collecting terms, the $m_2$-dependency for the $r_1$ contribution ends up in a sum $\sum_{m_2} \partial_{\theta} P(m_2|m_1,\theta)$. This sum turns out to be zero due to the normalization of the conditional probability for any value of $\theta$. This insight holds for any $j$, where it is used to drop all $r_k$ $(k<j)$ when they multiply $\partial_\theta \ln P(m_j|m_{j-1},\ldots,\theta)$.

Second, to further suppress stochastic fluctuations one can learn a value function $V$, which is a function of the current state and represents the expected future cumulative reward, averaged over all possible future measurement outcomes. Thus $V(m_j,m_{j-1},\ldots|\theta)$ is defined to be

$$E(r_{j+1}+r_{j+2}+\ldots |m_j,m_{j-1},\ldots,\theta)\,,$$
where the label $E$ stands for the expectation value over future rewards, conditioned on the preceding measurement results.

Typically, $V$ would be expressed as a neural network, though a lookup table can also be used in the case of a (modest) number of discrete measurements. The input to the value network would be some representation of the current ``state" $s$. This state could be identified directly with the sequence of previous measurement results, as indicated in our notation above, $s_j=m_j,m_{j-1},\ldots$ (which uniquely determines the current state). Alternatively, this state could also be represented by some version of the current quantum state (e.g. the density matrix), if that proves easier to handle for the network. The value network would be trained to output the expected (averaged) future cumulative reward, counted from this state onwards. The value training would proceed in the fashion known from general reinforcement learning, i.e. using the Bellman update equation \cite{sutton_reinforcement_2018} $V^{\rm new}(s_j) = V(s_j)+\alpha (r_{j}+\gamma V(s_{j+1}) - V(s_j))$, with $\alpha<1$ some update factor and $\gamma\leq1$ some discount factor to reduce the weight of long-term rewards ($\gamma\rightarrow 1$ in the ideal case discussed up to now). When using a neural network, $V^{\rm new}$ would be the new target value for the value network during a supervised-learning update. Once an approximation to the value function has been learned in this manner, we can proceed as in advantage actor-critic approaches to model-free RL. This means that in the gradient ascent procedure of the feedback-GRAPE approach, one would replace the (future) return by the {\em advantage} $A_j = r_j + \gamma V(s_{j+1}) - V(s_j)$, which expresses the improvement over the currently expected future return. In effect, this reduces the variance of the gradient estimates by subtracting a convenient baseline, without changing the average gradient update.

Concretely, Eq.\,(\ref{eqn-naive-reward-sum}), the gradient contribution from time step $j=2$, would be replaced by the following:

\begin{eqnarray}
 \partial_{\theta} r_2(m_2,m_1|\theta)  + \partial_{\theta} \ln P(m_2|m_1,\theta) & \cdot & \nonumber \\
\left\{ r_2(m_2,m_1|\theta) + \gamma V(m_2,m_1|\theta) - V(m_1|\theta) \right\} \label{eqn-enhanced-reward-sum}
\end{eqnarray}
The first line is unchanged, but in the second line $r_1$ was dropped, as explained before. Moreover, the sum of $r_3 + r_4+\ldots $ has been replaced by $\gamma V(m_2,m_1|\theta)$, which is the expectation of the future return (such that averaging over $m_3, m_4,\ldots$ has already been carried out, reducing sampling noise). Finally, $V(m_1|\theta)$ was subtracted, to reduce further the variance by canceling the expected value, given $m_1$. This is possible for the same reason that we could drop $r_1(m_1)$, as explained above. The extension to arbitrary $j\neq2$ is obvious.

In summary, such an enhanced feedback-GRAPE method would run trajectories with deterministic continuous controls and stochastic discrete quantum measurements just as before. However, it would learn a value function to represent expected future returns, and it would use that value function to modify the gradient ascent procedure and reduce fluctuations.

\subsection{Multi-target quantum feedback control}

Whenever we are employing neural networks to represent the feedback-based controls, a straightforward but powerful extension of feedback-GRAPE suggests itself. We may feed a representation of a variable target state $\Psi$ (or, in general, the target task, however it is defined) into the network: $F_j(\theta_j, m_j, \ldots; \Psi)$. The whole feedback-control strategy is then trained on many different randomly chosen tasks (e.g. many possible target states). 

Such approaches have been successful recently for other control challenges, e.g. they are being investigated in robotic navigation and the general field of multi-target reinforcement learning \cite{mousavian_visual_2019,kim_goal-aware_2021}. Multi-target schemes have also been recently suggested to improve variational quantum circuits \cite{self_variational_2021}. The benefit is data-efficiency: the network learns to generalize from the training tasks to other similar tasks, which requires less overall effort than to retrain a freshly initialized network for each task.

\section{Conclusions and Outlook}

In this work, we have presented a general scheme for the direct gradient-based discovery of quantum feedback strategies. This scheme, which we have labeled feedback-GRAPE, works for arbitrarily strong (discrete or continuous) nonlinear stochastic measurements, which so far had been possible only using the less data-efficient approaches of model-free reinforcement learning.

We observed very good performance, significantly beyond the state of the art, when testing the method on a challenging set of feedback tasks in an important, practically relevant quantum-optical scenario. Overall, our method opens a new route towards solving challenging feedback-based control tasks, including tasks in quantum communication and quantum error correction on multi-qubit or qubit-cavity systems. Besides presenting and analyzing the basic approach, we have also discussed extensions such as advantage functions (for reducing sampling noise) and training on multiple targets (to increase data efficiency and exploit transfer learning).

\section*{Acknowledgments}

The research is part of the Munich Quantum Valley, which is supported by the Bavarian state government with funds from the Hightech Agenda Bayern Plus.

\appendix

\section{Brief recap of reinforcement learning}
\label{appendix-recap-RL}

For the reader with a physics or optimal control background, we very briefly recall some key concepts in reinforcement learning (RL). However, these remarks serve only to provide additional context and are not necessary to understand the feedback-GRAPE technique introduced in the main text.

The term RL covers a set of techniques for discovering optimal control strategies, typically involving feedback \cite{sutton_reinforcement_2018}. The setting can always be phrased as an agent, i.e. a controller, interacting with an environment, where the latter may represent e.g. a device to be controlled. The goal is always to discover a good strategy for the agent, by optimizing some reward -- e.g. a fidelity, in the quantum setting. A first distinction is between model-based approaches, which require and exploit a simulation of the environment, and model-free approaches, where the environment is treated as a black box and the agent only learns implicitly about the environment's behaviour through repeated training runs. Feedback-GRAPE would be classified under the domain of model-based approaches, while most general RL algorithms used in machine learning applications nowadays are model-free. A sub-category, sometimes leading to confusion, are those cases where model-free algorithms are used to train an agent in silico, i.e. on simulated environments.

While we explain feedback-GRAPE in depth in the main text, here we briefly comment on one of the two main classes of model-free algorithms, namely policy gradient approaches, since we briefly compare and contrast some aspects of those against feedback-GRAPE in the main text. In such approaches, one represents the policy as a conditional probability to choose an action $a$ given an observed state $s$ of the environment: $\pi_{\theta}(a|s)$ in standard notation of the field. Here $\theta$ is a set of parameters that will be updated during training. Training proceeds by performing gradient ascent on the cumulative reward $R$, in the form
\begin{equation}
    \delta \theta = \eta \frac{\partial E[ R ]}{ \partial \theta} = \eta \sum_t E[ R \frac{\partial \ln \pi_{\theta}(a_t|s_t)}{\partial \theta}].
\end{equation}
Here $t$ is the time step, $s_t$ and $a_t$ are the sequence of states and actions in a particular trajectory, $R$ is the cumulative reward for that trajectory, and $E$ denotes the expectation value over many trajectories. We note in the main text that the logarithmic derivative appearing here relates to the probability of the agent's actions whereas a superficially similar logarithmic derivative appearing in feedback-GRAPE relates to quantum measurement probabilities, i.e. a property of the environment and not the agent.
\begin{figure}
    \centering
    \includegraphics[width=\columnwidth]{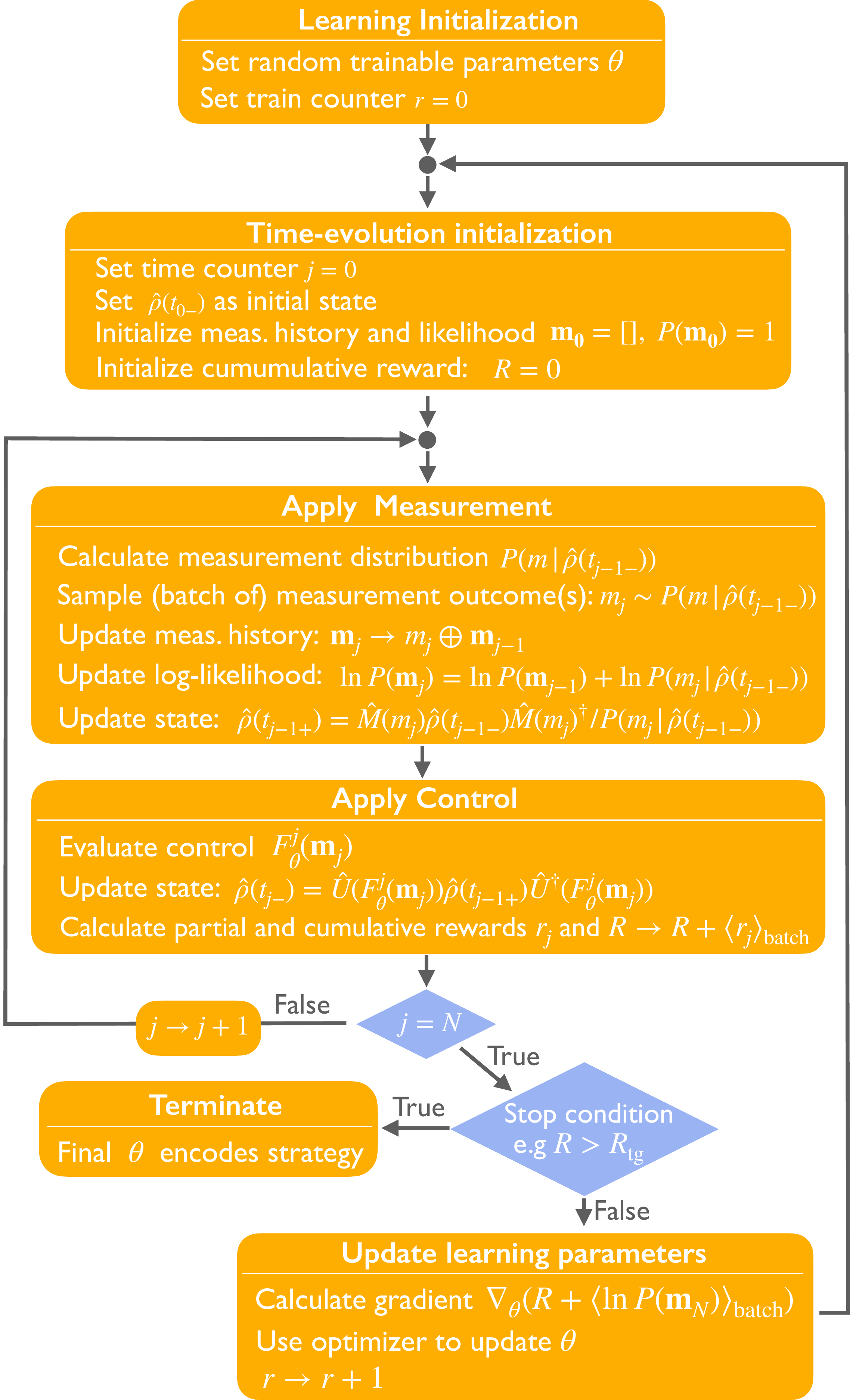}
   \caption{\label{fig:flow_chart} Algorithmic flow chart representation of Feedback-GRAPE. The measurement outcome $m_j$ at time step $j$ and all other quantities that depend on it  have an additional batch dimension that is not explicitly indicated. The symbol $\oplus$ denotes concatenation. Examples of stop conditions include the reward $R$ to be larger of a target value $R_{\rm tg}$,  $R>R_{\rm tg}$, or the number of training iterations $r$ having reached a maximum value $r_{\rm Max}$, $r=r_{\rm Max}$. The controls $F_\theta(\mathbf{m}_j)$ can be evaluated using a recurrent NN or directly read out from a look-up table.}
\end{figure}
\section{Algorithmic flow chart representation of feedback GRAPE}
\label{App:flow_chart}

In Figure \ref{fig:flow_chart}, we represent the working flow of feedback-GRAPE for the special case of discrete measurement outcomes as  an algorithmic flow chart. This representation provides more details than the conceptual representations in the main text.

\section{Evaluation of the parameter gradients of the time-evolving quantum state}
\label{appendix-GRAPE-evol-eqs}

In the numerical results in the main text, we have employed automatic differentiation to evaluate gradients with respect to the trainable parameter vector $\theta$. This approach is very convenient using modern machine learning tools. However, alternatively, it is also possible to directly work out analytical formulas to evaluate such gradients, based on our knowledge of the evolution equations. In a particular scenario, where the entries of the  vector of trainable parameters $\theta$ directly correspond to the controls at different time points, this then produces a suitable extension of the approach advocated in the original GRAPE manuscript \cite{khaneja_optimal_2005}.

In the following formulas, we will assume for simplicity unitary evolution outside the measurements, but the extension to (Markovian) dissipative dynamics is comparatively straightforward (using a Liouvillian superoperator instead of the Hamiltonian). 

We will first describe a general approach which works for any arbitrary choice of the parametrization $\theta$.  Further below, we will then specialize to a scenario where the original GRAPE idea for efficient gradient evaluation can be applied.

The general task is to obtain the gradient of the quantum state with respect to the trainable parameters $\theta$ that enter the controls (and, likewise, the gradient of the final probability $P(\mathbf{m})$ of a measurement sequence $\mathbf{m}$). 

In modern machine learning language, tracking the evolution of parameter-gradients in the manner described in the following is connected to the recent developments of neural ordinary differential equations \cite{chen_neural_2018}, where efficiency is obtained by not using automatic differentiation as a black box but rather evaluating analytically the form of the equations of motion for the gradients (and then solving those equations numerically with any efficient solver available). We can obtain the parameter gradient of the quantum state by solving the following evolution equation during measurement-free time intervals:
\begin{equation}
i \partial_t \partial_{\theta} {\hat \rho} = [{\partial_{\theta}}{\hat H}, {\hat \rho}] + [{\hat H}, \partial_{\theta} {\hat \rho}], \label{eq:dthetarho}
\end{equation}
where ${\hat \rho}$ is the solution to the original equation of motion, $ i \partial_t {\hat \rho} = [{\hat H}, {\hat \rho}] $, and the initial condition at time 0 would be $\partial_{\theta} {\hat \rho}=0$ (we have set $\hbar\equiv1$ for brevity). The interesting step now happens at a measurement, where $ {\hat \rho}(t^+) = {\hat M}(m){\hat \rho}(t^-) {\hat M}^{\dagger}(m) / P_m  $, with the probability for the measurement outcome, $P_m = {\rm tr}[{\hat M}(m){\hat \rho} {\hat M}^{\dagger}(m)]$. For brevity we suppress the index $j$ (used in the main text) that would indicate the number of the measurement in the sequence. It now follows that we have
\begin{eqnarray}
  {\partial_{\theta}}{\hat \rho}(t^+) = {\hat M}(m) {\partial_{\theta}}{\hat \rho}(t^-) {\hat M}^{\dagger}(m) / P_m - & & \nonumber \\
{\hat \rho}(t^+) {\rm tr}[{\hat M}(m){\partial_{\theta}}{\hat \rho}(t^-) {\hat M}^{\dagger}(m)] / P_m  . & &
\end{eqnarray}
Here the required $ {\partial_{\theta}}{\hat \rho}(t^-) $ is the outcome of solving the previous continuous evolution equation up until time $t$. After this update, the continuous evolution of $ {\partial_{\theta}}{\hat \rho}(t) $ will proceed. We note, however, that the controls (embedded inside ${\hat H}$ in the present setup) will now depend on the measurement outcome $m$ that was selected. Likewise for later time intervals, they will depend on the whole previous sequence, as described in the main text.

At the end, we also need the gradient of the extra term, the log-likelihood of the whole measurement sequence, $ \ln P(m_1,m_2,\ldots) $. One way to obtain this is to evolve an unnormalized version of the quantum state, ${\tilde {\hat \rho}}$, whose trace will give $P$, which follows the same evolution as the quantum state itself, but without the normalization factors that are the probabilities for the individual measurement outcomes. The $\theta$-gradient of this unnormalized state again follows an evolution equation of the form like Eq.~\ref{eq:dthetarho}, just with ${\tilde {\hat \rho}}$ substituted for $ {\hat \rho}$, during the unitary evolution intervals. However, at a measurement-induced update, we obtain the simpler rule $ {\tilde {\hat \rho}}(t^+) = {\hat M}(m){\tilde {\hat \rho}}(t^-) {\hat M}^{\dagger}(m)  $ and consequently $ {\partial_{\theta}} {\tilde {\hat \rho}}(t^+) = {\hat M}(m){\partial_{\theta}} {\tilde {\hat \rho}}(t^-) {\hat M}^{\dagger}(m) $.

What we have described here so far uses less assumptions than GRAPE, because the  vector of trainable parameters $\theta$ can enter the controls in an arbitrary manner. In GRAPE \cite{khaneja_optimal_2005}, an additional assumption was used to simplify the gradients further and gain efficiency: The components of the vector of trainable parameters $\theta$ were supposed to directly correspond to the control values applied at different time steps. That is, schematically speaking, we would have $\theta_1,\theta_2,\ldots$ associated with the controls at time steps $j=1,2,\ldots$. This then leads to a further simplification in the evaluation of the gradients. Importantly, if the number of parameters scales with the number of time steps $N$, then this approach has a runtime growing only linearly in $N$, while the general approach outlined above would need $N^2$ operations.

Let us briefly recall the GRAPE approach to gradient evaluation \cite{khaneja_optimal_2005}, before extending it. In the simplest possible version, with unitary evolution, let us consider the fidelity ${\rm tr}({\hat \sigma}(T) {\hat U}(T,0) {\hat \rho}(0) {\hat U}(0,T))$. The derivative with respect to parameters $\theta$ entering the Hamiltonian will produce a contribution for each time $t\in(0,T)$ in the evolution. Specifically, the contribution from time $t$ will be an expression of the type ${\rm tr}({\hat \sigma} {\hat U}(T,t) [-i\frac{\partial {\hat H}}{  \partial \theta}, {\hat \rho}(t) ] {\hat U}(t,T))$. Using the cyclic property of the trace, this can be reordered to obtain ${\rm tr}({\hat U}(t,T) {\hat \sigma}(T) {\hat U}(T,t) [-i\frac{\partial {\hat H}}{  \partial \theta}, {\hat \rho}(t) ] )$. This can now be re-interpreted, namely as the overlap between a backward-evolved target state ${\hat \sigma}(t)={\hat U}(t,T){\hat \sigma}(T){\hat U}(T,t)$ and the perturbation of the forward-evolved state at time $t$: ${\rm tr}( {\hat \sigma}(t) [-i\frac{\partial {\hat H}}{  \partial \theta}, {\hat \rho}(t) ] )$. 

In machine learning language, the GRAPE procedure of obtaining gradients in this way can essentially be viewed as an analytically derived version of backpropagation for this specific case of a quantum-physical evolution. It is very efficient, since the effort scales only linearly in the number of time steps, even if there is a different, independently optimizable parameter $\theta(t)$ for each time step.

The question is how this procedure needs to be modified in the presence of measurements. Let us imagine we have a particular trajectory with a given fixed sequence of measurement outcomes. We find that we can perform the temporal backpropagation (starting from the final time $T$) in the same manner as reviewed above, until a point in time ${\tilde t}$ where a measurement has happened (unless of course we talk about a time point $t$ later than the last measurement).  At that point ${\tilde t}$, we need to replace ${\hat \sigma}({\tilde t}) = {\hat U}({\tilde t},T){\hat \sigma}{\hat U}(T,{\tilde t})$ by the following expression:
\begin{equation} {\hat \sigma}'({\tilde t}) = \frac{1}{P} {\hat M}^{\dagger} {\hat \sigma}({\tilde t}) {\hat M} - \frac{1}{P^2} {\hat M}^{\dagger} {\hat M} {\rm tr}( {\hat M}^{\dagger} {\hat \sigma}({\tilde t}) {\hat M} {\hat \rho}({\tilde t}) ) \label{app-eq-msmt-update-GRAPE} \end{equation}
Here we have defined, for brevity, the measurement operator $ {\hat M} \equiv {\hat M}_{{\tilde m}} $ at time point ${\tilde t}$, with measurement outcome ${\tilde m}$, and the associated probability $P \equiv P_{{\tilde m}} = {\rm tr}( {\hat M}_{{\tilde m}} {\hat \rho({\tilde t})} {\hat M}_{{\tilde m}}^{\dagger} )$, where ${\hat \rho({\tilde t})}$ is already conditioned on previous measurement outcomes, for times less than ${\tilde t}$ and has been obtained by the forward evolution starting from time $0$ (with measurements and re-normalization of the state after each measurement).

After this procedure has been implemented for the measurement at ${\tilde t}$, we would proceed with the backward evolution of ${\hat \sigma}$ until point $t$, where the derivative is to be evaluated. There, we employ the same formula as in the usual GRAPE approach, i.e. we would evaluate ${\rm tr}( {\hat \sigma}(t)^{\dagger} [-i\frac{\partial {\hat H}}{\partial \theta}, {\hat \rho}(t) ] )$.

If there are multiple measurements between $t$ and $T$, the backward evolution will proceed by alternating unitary evolution and applying the formula in Eq.~(\ref{app-eq-msmt-update-GRAPE}).

If we want to treat the unnormalized quantum state in the same manner, e.g. for obtaining the log-likelihood term, we will only need the trace of that unnormalized state ${\tilde {\hat \rho}}$ at the end of the time evolution (see our discussion above). Formally, this is as if we were to calculate the fidelity against a state ${\hat \sigma}(T)=1$, which is given by the identity matrix. We can now evolve this state backwards in the manner discussed above, but in addition,  Eq.~(\ref{app-eq-msmt-update-GRAPE}) simplifies: One needs to drop the second term and also formally set $P=1$ in the first term. 

Finally, we briefly remark how the procedure will change if we are dealing with continuous measurement outcomes (strong continuous measurements, as briefly discussed in the main text, using the 'reparametrization trick'). In that case, we do not need the log-likelihood term. However, we now do need to differentiate the measurement outcome $m=f^{-1}_{\hat \rho}(z)$ which depends on some random variable $z$ (of a fixed distribution, not dependent on $\theta$) and the quantum state $\hat \rho$ (that does depend on $\theta$). As a consequence, Eq.~(\ref{app-eq-msmt-update-GRAPE}) needs to be modified. We have to add the following terms to the right-hand-side:
\begin{equation}
    \frac{1}{P} {\partial_{\theta}} ({\hat M}^{\dagger} {\hat \sigma}({\tilde t}) {\hat M}) - \frac{1}{P^2}  {\rm tr}( {\hat M}^{\dagger} {\hat \sigma}({\tilde t}) {\hat M} {\hat \rho}({\tilde t}) ) {\partial_{\theta}} ({\hat M}^{\dagger} {\hat M})
\end{equation}
Here ${\partial_{\theta}}$ in both parts of this expression is supposed to act only on the ${\hat M}^{\dagger}$ and ${\hat M}$ terms. This derivative is to be applied in the way ${\partial_{\theta}} {\hat M}(m) = ({\partial_m} {\hat M}(m)) ({\partial_{\theta}} m)$, where the derivative of $m$ with respect to $\theta$ must be evaluated using the dependence of the inverse cumulative distribution function on the $\theta$-dependent quantum state at that time-point.

\section{Law-Eberly algorithm}
\label{App:Law_and_Eberly}
\begin{figure}
    \centering
    \includegraphics[width=\columnwidth]{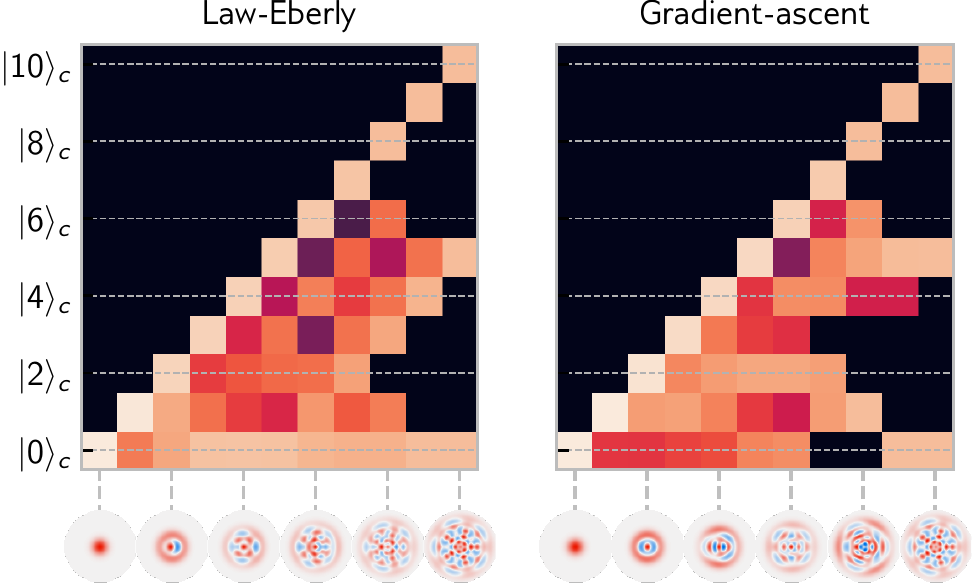}
   \caption{\label{fig:Figure1_supplementary}Comparison between one solution obtained analytically from the Law-Eberly protocol and a strategy found by using gradient ascent. The target state is $\ket{\psi} = (\ket{0}+\ket{5}+\ket{10})/\sqrt{3}$. Even though some details look different, we have verified that the gradient ascent strategy is a valid alternative solution for the Law-Eberly equations (which do not determine the controls uniquely). }
\end{figure}
As a benchmark with an analytical solution (but still without feedback), we consider the task of preparing an arbitrary pure cavity state in a cavity-qubit system . This can be achieved by exploiting the well-known Law-Eberly protocol \cite{law_arbitrary_1996}. 
This algorithm relies on the essential assumption that we start from the ground state. We briefly review it below.\\
The Hamiltonian that describes the system is a Jaynes-Cummings model with controllable couplings:
\begin{equation}
    \hat H(t) = (A(t) \hat \sigma_+ +  A^*(t) \hat \sigma_-) + (B(t) \hat a \hat \sigma_+  + B^*(t) \hat a^{\dagger} \hat \sigma_-)
\end{equation}
where the first term corresponds to the qubit drive and the second  to the cavity-qubit interaction. The two complex controls $A(t)$ and $B(t)$ can assume continuous values. 

Law and Eberly uses the particular  ansatz  $A(t)=0$ if $B(t)\neq 0$ and vice versa. In this scenario, the dynamics can be viewed as being subdivided into a discrete number of steps $N$, with each  step consisting in one qubit excitation  gate, $\hat{U}_q(\alpha_j)=\exp{-i (\alpha_j \hat \sigma_+  + \alpha^*_j \hat \sigma_-) }$, followed by one cavity-qubit interaction gate, $\hat{U}_{qc}(\beta_j)=\exp{-i (\beta_j \hat a\hat \sigma_+ + \beta^*_j \hat a^{\dagger}\hat \sigma_-) }$. [For  given  $A(t)$ and $B(t)$, $\alpha_j$ and $\beta_j$ can be easily obtained by integrating over the relevant time-interval.] Since the excitations  can be added only one by one via the qubit drive, one can further refine the ansatz assuming that the number of steps $N$ is equal to the maximum number of excitations in the target state,  $\ket{\psi}_{target} = \sum_{n=0}^N c_n\ket{n, g}$. Summarizing, the goal is to find the parameters $\{\alpha_j\}$ and $\{\beta_j\}$ that solve
\begin{equation}
    \ket{\psi}_{target} = \hat{U}\ket{0, g},
\end{equation}
with 
\begin{eqnarray}
    \hat{U} = \hat{U}_{qc}(\beta_N) \hat{U}_{q}(\alpha_N) \hat{U}_{qc}(\beta_{N-1}) \hat{U}_{q}(\alpha_{N-1}) \nonumber\\ \times \ldots\hat{U}_{qc}(\beta_1) \hat{U}_{q}(\alpha_1)
\end{eqnarray}

The Law-Eberly idea is to start from the target state and progressively remove excitations from the cavity until it becomes empty. In other words, one focuses on the time-reversed time evolution
\begin{equation}
    \ket{0, g} = \hat{U}^\dagger\ket{\psi}_{target},
\end{equation}
with 
\begin{eqnarray}
    \hat{U}^\dagger =\hat{U}^\dagger_{q}(\alpha_1) \hat{U}^\dagger_{qc}(\beta_1) \ldots\hat{U}^\dagger_{q}(\alpha_N) \hat{U}^\dagger_{qc}(\beta_N). 
\end{eqnarray}
and recursively (for decreasing $j$ starting from $j=N$) 
find the $\beta_j$ and $\alpha_j$ imposing the conditions $\langle j, g|\psi_j\rangle=0$, and $\langle j-1, e|\psi_j\rangle=0$ with
$$
\left| \psi_j \right\rangle =\hat{U}^\dagger_{q}(\alpha_j) \hat{U}^\dagger_{qc}(\beta_j) \ldots\hat{U}^\dagger_{q}(\alpha_N) \hat{U}^\dagger_{qc}(\beta_N) \ket{\psi_{target}}
$$
 being the state after  $N+1-j$ time steps of the time-reversed evolution. These conditions are enforced by the complex nonlinear equations 
\begin{equation}
\begin{split}
    \braket{j, g}{\psi_{j+1}}\cos(|\beta_j|\sqrt{j})(\beta/|\beta|)\\+i\braket{j-1,e}{\psi_{j+1}}\sin(|\beta_j|\sqrt{j})=0,\\
    \bra{j-1,e} \hat{U}^{\dagger}_{qc}(\beta_j)\ket{\psi_{j+1}}\cos(|\alpha_j|)(\alpha^*/|\alpha|)\\+i\bra{j-1,g}\hat{U}^{\dagger}_{qc}(\beta_j)\ket{\psi_{j+1}}\sin(|\alpha_j|)=0,
\end{split}
\label{eq:law_and_eberly}
\end{equation}
with $\left| \psi_{N+1} \right\rangle \equiv\ket{\psi_{target}}$ for the first  iterative step (corresponding to  $j=N$).

It should be noted that the solution of these equations is not unique. This is why Fig.~\ref{fig:Figure1_supplementary} shows two different strategies for the same task, although both of them fulfill the Law-Eberly ansatz.

\section{Model-free reinforcement learning for the Jaynes-Cummings scenario}
\label{appendix-model-free-RL}

It turns out that state-of-the-art model-free RL has surprising difficulties in addressing a physical scenario as important and conceptually simple as the Jaynes-Cummings model. In this subsection we provide some more details. 

We will only consider the (simpler) no-feedback case, meaning only the  controls $\alpha_j$ and $\beta_j$ (see main text) are available. Since model-free RL already has severe problems in this case, we did not explore further the more challenging cases.

In our numerical experiments, we relied on the RL library Stable Baselines \cite{stable-baselines}, which implements many of the most well-known optimized state-of-the-art RL algorithms. The RL environment (not to be confused with a ``physical" environment) has been implemented in the following way:
\begin{itemize}
    \item Action $a_j$: The two continuous controls, $\alpha_j$ and $\beta_j$.
    \item State $s_j$ (i.e. input to the agent): In principle, the no-feedback task requires no state input. However, we chose to make it easier for the agent, by supplying the full current quantum state of the system at time $t_j$. Since the state is pure and the system is closed, we simplify the observation by only using the state vector $\ket{\psi_j}$  (instead of the density matrix). Since it is complex-valued, we split its real and imaginary part and so we have a vector of length $2N$, where $N$ is the size of the Hilbert space. 
    \item Reward $r_j$: the fidelity at step $t_j$ (in various versions, see below).
\end{itemize}

We have used a variety of different approaches  to solve the task of pure state preparation. These included: using either a sparse final reward (i.e. $r_j\neq0$ only if $j=N$) or else a reward based on the fidelity at each time step, either discrete (discretized) actions or  continuous actions, and several different optimization algorithms (PPO\cite{PPO}, A2C\cite{A2C}, HER\cite{HER}, TRPO\cite{TRPO}, DDPG\cite{DDPG}). The results shown in \ref{fig:Figure3_a_open-control}c) are the best results we could manage to produce among all these approaches. They were obtained with PPO, continuous actions and sparse rewards and using the  hyperparameters in Table \ref{tab:PPO}. 

\begin{table}[h]
\centering

\begin{tabular}{|c|c|}

\hline
Parameter&Value\\
\hline
gamma & 0.99\\
n\_steps & 128\\
ent\_coef & 0.01\\
learning\_rate & 0.00025\\
vf\_coef & 0.5\\
max\_grad\_norm & 0.5\\
lam & 0.95\\
nminibatches & 4\\
noptepochs & 4\\
cliprange & 0.2\\
\hline
\end{tabular}
\caption{\label{tab:PPO} Hyperparameters table for the PPO model-free RL optimization algotithm used in Fig.3(c). See also the Stable Baselines PPO documentation.}
\end{table}

\begin{figure}[t]
    \centering
    \includegraphics[width=\columnwidth]{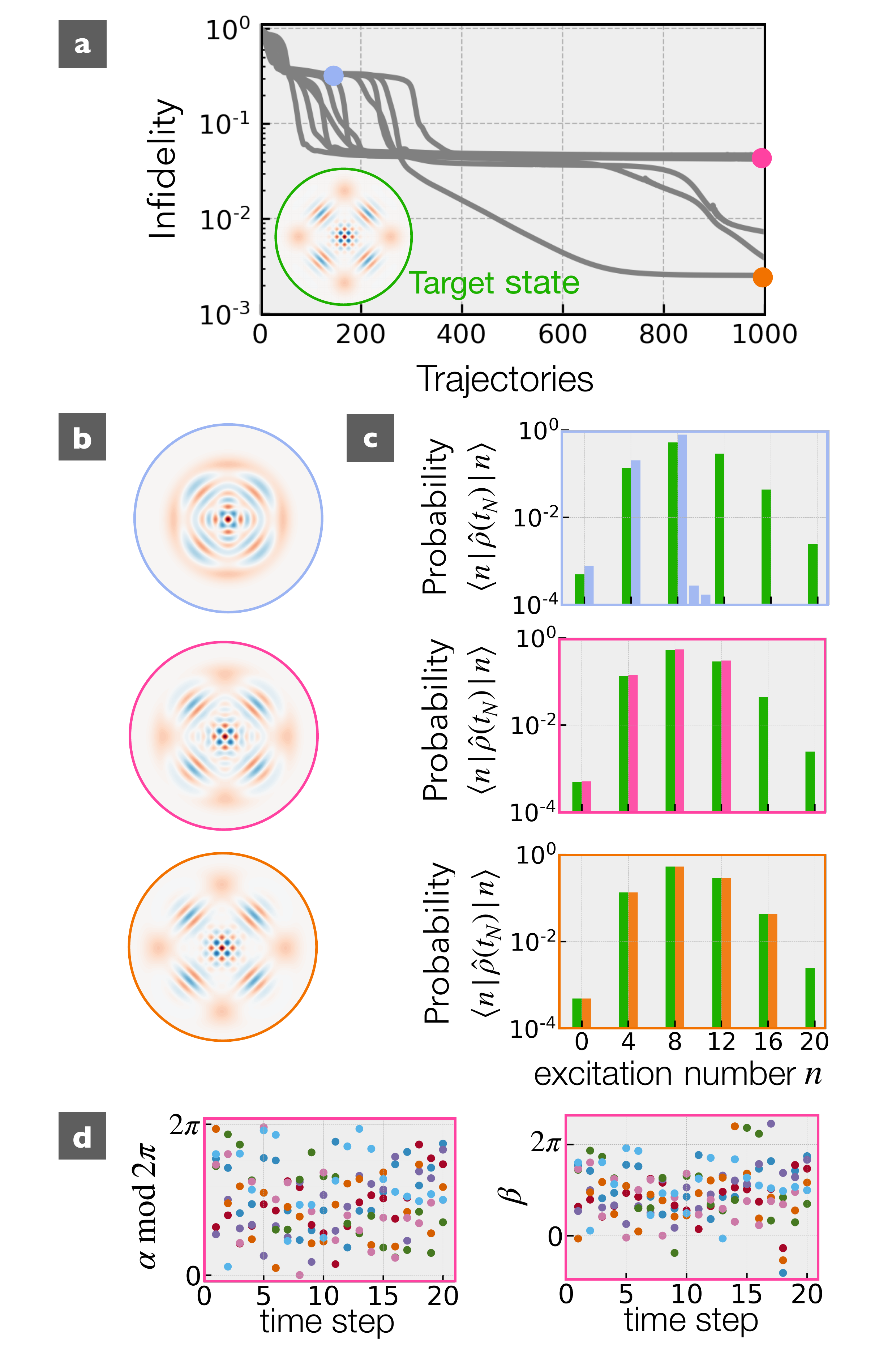}
    \caption{Preparation of a four-components kitten state with average excitation number $\bar{n}=9$ like in Fig.~\ref{fig:Figure3_a_open-control}(d) of the main text. (a) Shows the infidelity as a function of the number of trajectories seen during training for  10 training runs with different initial conditions. The Wigner function of the target state is shown as an inset. (b) Wigner function and (c) excitation-number distribution after state preparation for three different control strategies discovered during training. The fidelity and  number of trajectory used during training are marked in (a) as dots of the same color as the frames in (b) and (c). In (c),  the target distribution is also shown in green.   Seven different training runs leads to approximately the same fidelity after 1000 training iterations, cf the pink dot in (a). The corresponding underlying strategies are displayed in (d).  For each strategy, the controls as a function of the time step are shown as dots of the same color. Each of these  strategies leads to the preparation of approximately the same state,  displayed in the second row of b and c. \label{fig:kitten_state_preparation_supplementary}}
\end{figure}
\section{Details on learning dynamics using Jaynes-Cummings open-loop controls}
\label{App:State_prep_kitten}

In this Appendix, we give more details on the learning dynamics of the open-loop control strategy to prepare a four-component Kitten state,  $|\psi^{\rm Kit4}_{\tilde{\alpha}}\rangle\propto\sum_{j=0}^3 |i^j\tilde{\alpha}\rangle$. This analysis gives useful insight on the optimization  landscape for the open-loop control preparation of complex  superpositions of Fock states using Jaynes-Cummings controls. 

Our numerical results for $\tilde{\alpha}=3$, corresponding to the average photon number $\bar{n}\approx |\tilde{\alpha} |^2=9$ are summarized in Figure \ref{fig:kitten_state_preparation_supplementary}.  We preliminary note that the excitation number of the target state is   not bounded in this case. On the other hand, the Law and Eberly protocol allows to reach only the first    $N$ Fock states in  $N$ preparation steps. Thus, the optimal strategy will project the target state into the Hilbert space spanned by the first $N$ Fock states and can be obtained using Law and Eberly algorithm. This procedure allows also to find the smallest possible infidelity. We choose $N=20$ corresponding to the minimal infidelity ${\cal F}\approx 6\times 10^{-5}$. 

The fidelity as a function of the number of training iterations (or, equivalently, trajectories used during training) for 10 different training runs is shown in panel a. As it should be expected given that the control parameters slide down a rugged learning landscape  the lineshape of the fidelity depends strongly on the initialization. Nevertheless, it displays  robust features in the form of a series of plateaus whose heights do not depend on the intialization. I turns out that  the  states $\hat{\rho}(t_N)$  prepared following strategies obtained in different runs but corresponding to the same fidelity plateau are also approximately equal. In panel b and c, we show the Wigner function and excitation number  distribution $ \langle n|\hat{\rho}(t_N)|n\rangle$ obtained with a representative strategy for each plateau in panel a.
We note that the state is approximately the projection of the target state on a Hilbert space containing the first $N_{\rm th}$  Fock states with the threshold excitation numbers $N_{\rm th}=8,12,16$, cf panel c. Importantly, each of these states are prepared using  different strategies in the different training runs, cf panel d. Indeed, it is easy to construct several different strategies to prepare exactly these states setting $2(N-N_{\rm th})$ controls to zero and choosing  the remaining  $2N_{\rm th}$ controls to solve the Law and Eberly equations (\ref{eq:law_and_eberly}) for $N_{\rm th}$ preparation steps. It can also be shown that such suboptimal strategies correspond to  saddle points of the optimization landscape and that  the curvature of the optimization landscape  in the direction of increasing fidelity is zero, giving rise to a narrow valley. In this way, we can construct many suboptimal strategies for $N$ preparation steps from an optimal strategy with $N_{\rm th}\leq N$ steps. Each of these suboptimal strategies corresponds to a narrow valley in the optimization landscape. These valley can cause the training to stall in a suboptimal solution. In fact, we were not able to recover an optimal solution in any of $10$ training runs each comprising  $1000$ training iterations. Nonetheless, we were able to obtain very good quality solutions. The best solution we have obtained (in one out of 10 runs) allows to prepare the oscillator in the target state
projected onto an Hilbert space with cut off $N_{\rm th}=16$ (third row in panels b and c). This corresponds to the last valley before reaching an optimal solution. We have also run a set of 10 simulation runs comprising also the same number of learning iterations but with a larger number of preparation time steps ($N=28$). In this set of run (not shown), we have reached the fidelity  plateau corresponding to  $N_{\rm th}=16$ more consistently and could even reach the plateau for $N_{\rm th}=20$.

In conclusion, our analysis indicates that the optimization landscape for the preparation of complex superpositions of Fock states using open-loop Jaynes-Cummings controls features a very large number of narrow valleys.  In this setting, it is crucial to have a direct access to the landscape gradient to be able to slowly but steadily slide down the optimization landscape. Very good quality solutions can be consistently obtained.

\begin{figure}
    \centering
    \includegraphics[width=\columnwidth]{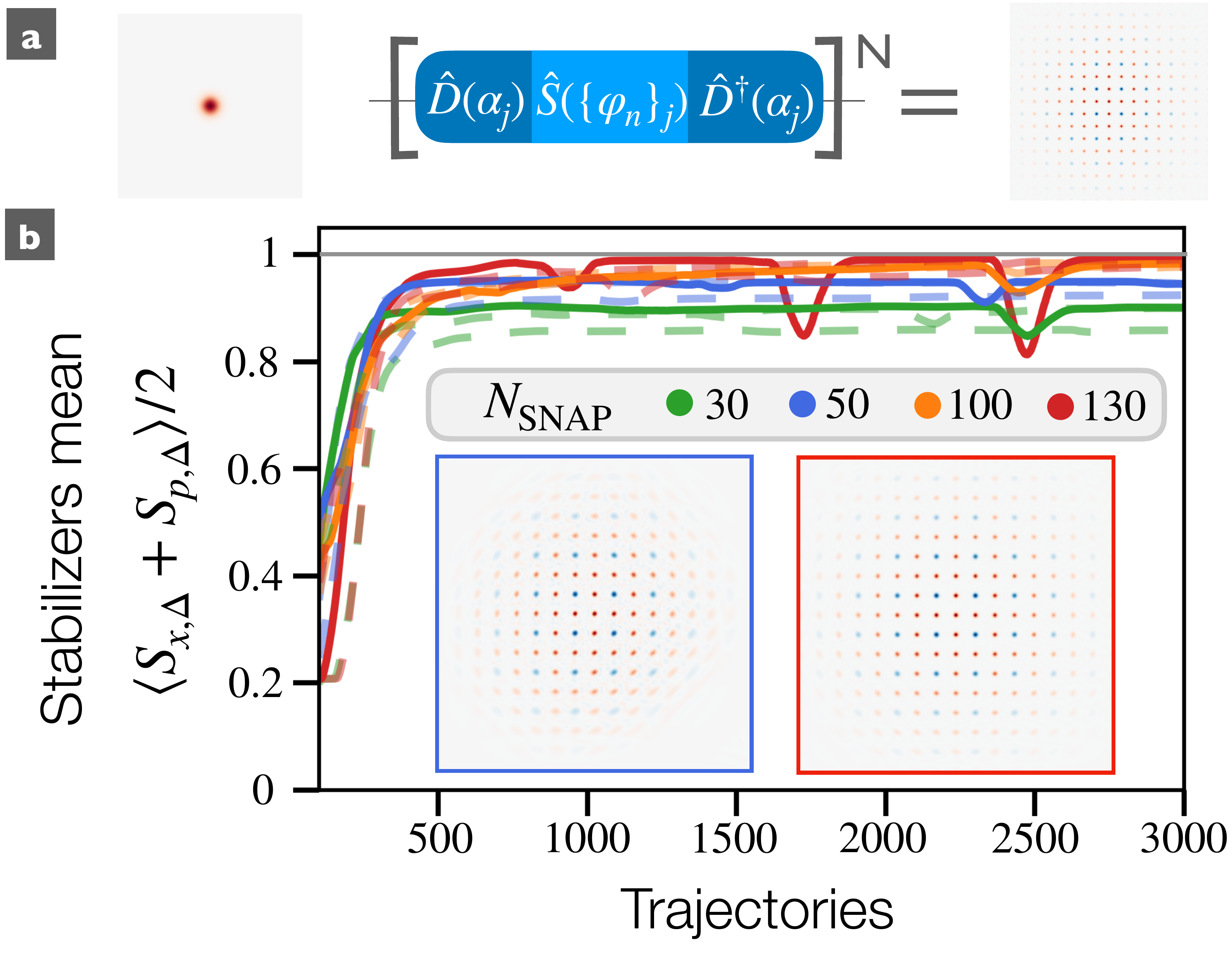}
    \caption{State preparation  (open-loop control) of a grid state with more powerful controls.  a)  Sequence of parametrized controls during a  time-evolution trajectory. Each time step consists of two phase-space displacement gates separated by  a SNAP gate $\hat{S}(\{\varphi_n\})$. The phases $\varphi_n$ for $0\le n<N_{\rm SNAP}$ are predicted by a recurrent NN that is given the time $j$ as an input (for $n\ge N_{\rm SNAP}$ $\varphi_n=0$). Shown are also the Wigner functions of the initial state and of the target grid state.  b) Mean value of the two (finite energy) stabilizers $\hat{S}_{x,\Delta}$ and $\hat{S}_{p,\Delta}$ as a function of the number of trajectories sampled during training (or, equivalently, the number of training iterations) for four values of $N_{\rm SNAP}$ (three runs each; the best run is displayed as a solid line). The target grid state is in the manifold with $\langle\hat{S}_{x,\Delta}\rangle=\langle\hat{S}_{p,\Delta}\rangle=1$. Plotted is the running average over $100$ trajectories.   Also shown is the Wigner function after training for $N_{\rm SNAP}=50$ and $N_{\rm SNAP}=130$. For small $N_{\rm SNAP}$ the tail of the  Wigner function is distorted, while for larger $N_{\rm SNAP}$ it is indistinguishable by bare eye from the target Wigner function. This indicates that the quality of the strategy is limited by the expressivity of the parametrized control sequence. Compared to model-free RL \cite{sivak_model-free_2021}, the model-based approach used here allows to explore a higher-dimensional parameter manifold (larger values of $N_{\rm SNAP}$) and, thus, obtain better quality results for large grid states. Parameters:
     $N=9$ and for the grid state  $\Delta=0.15$ corresponding to $\sqrt{{\rm var} (n)}\approx\bar{n}\approx 1/(2\Delta^2)\approx 22$ (The Hilbert space contains $130$ Fock states).}
    \label{fig:Figure_GKP}
\end{figure}

\section{State preparation with SNAP and displacement gates}
\label{App:grid_state}

 In this Appendix, we report our results for the open-loop control state preparation  of an oscillator state using the universal control scheme based on a set of interleaved \textit{Selective Number-dependent Arbitrary Phase} (SNAP) gates $\hat{S}(\{\varphi_n\}))=\sum_{n}e^{i\varphi_n}|n\rangle \langle n|$  and displacement gates $\hat{D}(\alpha)=\exp[\alpha\hat{a}^\dagger-\alpha^*\hat{a}]$ \cite{Krasnatov_Universal_2015},  cf Fig.~\ref{fig:Figure_GKP}(a). In this case,  the control parameters  are the real and imaginary part of the phase-space displacement $\alpha^j$ together with the phases $\varphi^j_n$ for the first $N_{\rm SNAP}$ Fock states (the remaining phases are set to zero).  This very same optimization  task has been already considered 
by Sivak et al \cite{sivak_model-free_2021} to demonstrate their model-free  optimal control approach. This allows an instructive comparison of our model-based results with results obtained using a cutting-edge model-free approach. In order to facilitate the comparison, we have used a similar controller as in \cite{sivak_model-free_2021}, i.e.  a recurrent NN that takes as input the time step $j$, cf Table \ref{tab:RNN_1}.

The authors of \cite{sivak_model-free_2021} found that the main limiting factor in the way of preparing evermore complex quantum states using their model-free approach is a trade-off introduced by the choice of the hyperparameter $N_{\rm SNAP}$: It is helpful to increase this hyperparameter to improve the expressivity of  the sequence of control gates.  At the same time,  the resulting increased dimensionality of the control parameter space can  make  the training unstable. In order to remain in the regime of efficient training they chose $N_{\rm SNAP}=30$ for $N=9$ preparation steps. Overall, this corresponds to a $288$-dimensional  control parameter space. 
In their numerical experiments, this trade-off became apparent during the  preparation of large finite energy grid states, $|\psi^{\rm GKP}_\Delta\rangle=\hat{E}_\Delta\sum_{j\in \mathbb{Z}}\hat{D}(j\sqrt{\pi})|0_x\rangle$ where $|0_x\rangle$  is a position eigenstate  localized in the origin and $\hat{E}_\Delta=\exp [-\Delta^2 \hat{a}^\dagger \hat{a}]$ is the envelope operator.  These states  have been proposed by Gottesman-Kitaev-Preskill
\cite{Gottesman_encoding_2001} to encode  logical qubit states in an oscillator Hilbert space. They are eigenstates of the stabilizers 
$\hat{S}_{x,\Delta}=\hat{E}_\Delta\hat{D}(\sqrt{\pi})\hat{E}^{-1}_\Delta$ and $\hat{S}_{p,\Delta}=\hat{E}_\Delta\hat{D}(i\sqrt{\pi})\hat{E}^{-1}_\Delta$ with eigenvalue $+1$.
For small $\Delta$, their Wigner function displays a very fine structure in phase space in the form of a large grid of peaks in phase space. The Wigner function  the largest grid state considered in Ref.~\cite{sivak_model-free_2021} (corresponding to $\Delta=0.15$) is shown in Fig.~\ref{fig:Figure_GKP}(a). In that work, they considered as a  figure of merit  the stabilizer mean value $\langle \hat{S}_{x,\Delta}+\hat{S}_{p,\Delta}\rangle/2$ which has optimal value  $\langle \hat{S}_{x,\Delta}+\hat{S}_{p,\Delta}\rangle/2=1$. They   were able to demonstrate a best value of $\langle \hat{S}_{x,\Delta}+\hat{S}_{p,\Delta}\rangle/2\approx 0.93$  out of 6 training runs with $\Delta=0.15$, which  was limited by the expressivity of their control sequence with $N_{\rm SNAP}=30$.

In our GRAPE-type approach, thanks to  the direct access to the gradient of the reward, our training remains efficient for much larger values of $N_{\rm SNAP}$, cf Fig.~\ref{fig:Figure_GKP}(b). More precisely, we have considered  $N_{\rm SNAP}$ up to $130$ with the same number of preparation time-steps ($N=9$) and the same  type of controller  (a recurrent neural network that takes as  input the time-step) as in  Ref.~\cite{sivak_model-free_2021}. This corresponds to  a $1188$-dimensional control space, much larger than the one that could be handled in the model-free approach. In spite of this larger control parameter space, training required only one hundredth of the trajectories. Most importantly, we obtained a better quality solution as reflected by  the best stabilizer value of  $\langle \hat{S}_{x,\Delta}+\hat{S}_{p,\Delta}\rangle/2\approx 0.995$ out of three training runs with $\Delta=0.15$ and $N_{\rm SNAP}=130$.

\begin{figure*}[t]
    \centering
    
    \includegraphics[trim=0cm 7cm 0cm 6cm,width=\textwidth]{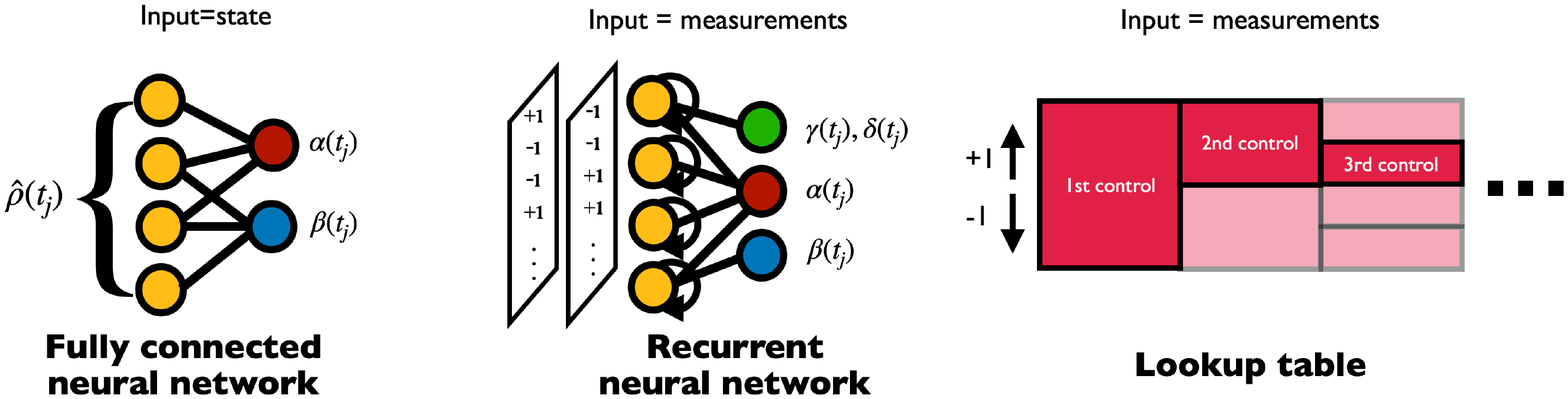}
    \caption{\label{fig:Figure2_supplementary}Sketch of the three alternative types of trainable controls that can be employed in feedback-GRAPE: the first one is a fully connected neural network which receives the density matrix (quantum state) of the system as input and output the controls. The second one is a RNN with GRU cells as recurrent neurons. The third one is a lookup table, with $\sum_{n=0}^{N}2^n$ entries (when feedback is required and when the measurement outcomes are binary, as shown here), and each entry contains the controls that need to be applied after observing a particular measurement sequence.}
   
\end{figure*}

\section{Details on Feedback-GRAPE algorithm and on physical simulations}
\label{appendix:NN_lookup}
As explained in the main text, in the feedback-GRAPE approach presented in this manuscript we can produce the control values (conditioned on previous measurement results) either with the help of a neural network or with the help of a lookup table (containing trainable control values). In this section we present more details on both of these approaches, as implemented for the specific numerical examples shown in the main text.

In our illustrative physical scenario (the feedback-controlled Jaynes-Cummings model), there are four control parameters: $\alpha_j, \beta_j, \gamma_j$ and $\delta_j$. In the most general case, where arbitrary superpositions should be generated, $\alpha_j$ and $\beta_j$ need to be complex. In the scenarios whose results are displayed in the main text, this was not needed due to the nature of the target states. However, we have checked independently that the whole approach works just as well for complex control parameters.

{\bf Neural Network} - We first discuss the case when the controls are computed by means of a neural network. This network can  receive the measurement results so far, $m_1,m_2,
\ldots,m_j$. Alternatively, we can also supply it with the quantum state as input, which has been updated according to the measurement outcomes. Both techniques supply the full information content needed to apply the next control.

For the ``state as input" approach, we defined a fully connected neural network that takes the density matrix of the system as input. Since the density matrix is complex-valued, we chose to split it into its real and imaginary parts and to stack it, in such a way that for a $N_{\rm H}\times N_{\rm H}$ density matrix, the input tensor has shape $[N_{\rm H}\times N_{\rm H}, 2]$. 

The fully connected NN has been employed both for the no-feedback case (pure state preparation), where in principle no such input would be needed (but can still be helpful for convergence), and also for the more interesting feedback cases.

If, on the other hand, we want to supply directly the measurement results, then we employ a recurrent neural network (RNN). For our scenario, its input at each time step is a binary measurement outcome $m_j\in\{-1, +1\}$. When a RNN network is used, due to the probabilistic outcome of the trajectories during a simulation, it is useful to feed batches of multiple randomly sampled trajectories as input to the network. 

As already mentioned, both types of neural networks output real-valued controls $\alpha_j, \beta_j, \gamma_j$ and $\delta_j$ to be applied in the next time step. When complex-valued controls are required, two additional neurons can be added to the output of the neural networks, and they correspond to the imaginary parts of $\alpha_j$ and $\beta_j$. In the main text, we did not use complex controls, because these were not needed for the tasks considered there. 

Our neural networks are implemented using Keras and their hyperparameters are shown for completeness in Tables \ref{tab:NN},  \ref{tab:RNN},  \ref{tab:RNN_1}. \\
\begin{table}[h!]

\centering
\begin{tabular}{|c|c|}

\hline
Parameter&Value\\
\hline
Neurons & [$[N_H\times N_H, 2]$, Flatten, 30, 30, 2 or 4]\\
Batch size & 1\\
Activation & ReLU \\
Initializer & Glorot uniform\\
Initial bias last layer & $\pi$\\
\hline
\end{tabular}
\caption{\label{tab:NN} Hyperparameters of the fully connected neural network.}

\end{table}

\begin{table}[h!]

\centering
\begin{tabular}{|c|c|}

\hline
Parameter&Value\\
\hline
Type RNN cells & GRU\\
Neurons & [30, 2 or 4]\\
Batch size & 10\\
Dropout & 0.2\\
Input shape & [batch\_size, 1, 1]\\
Activation & tanh\\
Recurrent activation & Sigmoid\\
Initializer & Glorot uniform\\
Initial bias last layer & $\pi$\\
\hline
\end{tabular}
\caption{\label{tab:RNN} Hyperparameters  of the Recurrent neural network for Figures 5, 15 and 16.}
\end{table}

\begin{table}[h!]

\centering
\begin{tabular}{|c|c|}

\hline
Parameter&Value\\
\hline
Type RNN cells & GRU\\
Neurons & [30, 30, 30, 2+$N_{\rm SNAP}$]\\
Input shape & [batch\_size, 1, 1]\\
Activation dense layers & Relu\\
Recurrent activation & Sigmoid\\
Initializer & Glorot uniform\\
Initial bias last layer & $0.1$\\
\hline
\end{tabular}
\caption{\label{tab:RNN_1} Hyperparameters  of the Recurrent neural network used for Figures 7 and 12. The batch size is $16$ and $1$ for figure 7 and 12, respectively. The hyperparameter $N_{\rm SNAP}$ is $15$ for Figure 7  and varies from 30 to 130 in Fig.12, cf inset. For figure 12 the input is the time step $j$, $0\leq j<N=9$, while for Figure 7 is the measurement outcome $m_j$.}
\end{table}

{\bf Lookup Table} - Another way to represent the entire feedback-based control strategy is to use a lookup table, which essentially is just a list of optimisable parameters. In the case of feedback, we have to build a lookup table that encodes the structure of a decision tree. For binary measurement outcomes (as used here), this has $\sum_{n=0}^N 2^n$ entries, each of which is the vector of all control parameters, i.e. in our scenario ($\alpha_j$, $\beta_j$, $\gamma_j$, $\delta_j$). Each column of this table represents the $2^j$ possible control parameter vectors at time step $j\in\{0,...N\}$. At $j=0$, we have only one set of numbers, which stand for the (only) possible control vector to apply (not dependent on any previous measurement; in our case reduced to only the entries controlling the first measurement). At step $j=1$, we have two sets of numbers, and we apply the set of controls corresponding to the observed measurement, and so on and so forth. By doing so, we can apply controls conditioned on the ``memory" of all previous measurements, at the cost of keeping an exponentially growing number of entries in the computer's memory. Many of those will likely not be explored at all, if their probabilities are too small. 

In our numerical experiments, we went as far as lookup tables containing about $2^{21} \sim 2\cdot 10^6$ entries, which still was easily handled. The initial condition for the whole table was to set each parameter value to a random number uniformly distributed within $(0,\pi)$.

For figure $8$, we have used a large batch of size $10000$ ($1000$) in the approaches with (without) Monte Carlo sampling. For the approach with random initialization the look-up table entries $\bar{g}\tau_i(\mathbf{m}_{j-1})$  were uniformly distributed within $(-2\pi,2\pi)$. In the approach with smart initialization, the initial entries are weakly randomized: For $\mathbf{m}_{j-1}=(1,\ldots,1)$, we have chosen  $\bar{g}\tau_j(\mathbf{m}_{j-1}) = \pi+z(\mathbf{m}_{j-1})$ with $z(\mathbf{m}_{j-1})$ uniformly distributed in the interval $(0,1)$. For the remaining measurement outcomes we set   $\tau_j(\mathbf{m}_{j-1})=0$.

In several results mentioned in the main text, we use a lookup table ``without memory". This means that there is just one control parameter vector for each step $j$, instead of a tree-type structure with an exponentially growing number of parameters.  Thus, we still optimize the controls but ignore the result of previous measurements. This is used both for the ``non-adaptive" scheme for the purification task in Fig.~\ref{fig:Figure3_b_purification}c) and in figure \ref{fig:Figure3_JCResults}g).

A sketch of all of the three feedback-based strategies discussed here and in the main text  (neural network with state as input, recurrent neural network with measurement sequence as input, and a tree-type lookup table) is shown in Fig.~\ref{fig:Figure2_supplementary}.

In any case, in whatever ways we choose to parametrize our controls, we have a finite number of parameters that need to be learned. In order to do so, the optimizer employed for every example is Adam \cite{kingma2017adam}, and its hyperparameters are shown in table \ref{tab:optimizer}.

\begin{table}[h]
\centering

\begin{threeparttable}
\begin{tabular}{|c|c|}

\hline
Parameter&Value\\
\hline
learning\_rate & 0.01*\\
beta\_1 & 0.9\\
beta\_2 & 0.999\\
epsilon & 1E-7\\
clipnorm & 1\\
clipvalue & 0.5\\
\hline
\end{tabular}
\begin{tablenotes}[para,flushleft]
  \tiny *unless otherwise specified
  \end{tablenotes}
  \end{threeparttable}
\caption{\label{tab:optimizer}Adam hyperparameters}
\end{table}

{\bf Physical Simulations} - In the unitary case, we simply apply the sequence of parametrized unitaries, as explained in the main text. In the case of decay (in the state stabilization scenario), we have solved the master equation for the density matrix during the respective time intervals (where decay is present).
Specifically, we have simulated the weak Markovian coupling of the oscillator to a zero temperature bath via the Lindblad master equation,

\begin{equation}
\dot{\hat \rho} =  \kappa \left( \hat a \hat \rho \hat a^\dagger -\frac{1}{2} \left\{ \hat a^{\dagger} \hat a, \hat \rho
    \right\} \right).
\end{equation}
We discretize this continuous time-evolution applying   the fourth-order Runge-Kutta method.

We chose the Hilbert-space to have a finite dimension $N_H$  with  a  cut-off in the Fock states excitation number. An appropriate choice of the cut-off depends both on the initial and the target state and ranges from $10$ to $130$ in our simulations. 

\section{Further numerical results}
\label{app-further-num-res}
In this subsection we present a few more numerical results to illustrate various options or aspects of the technique.

\begin{figure}[t]
    \centering
    \includegraphics[width=\columnwidth]{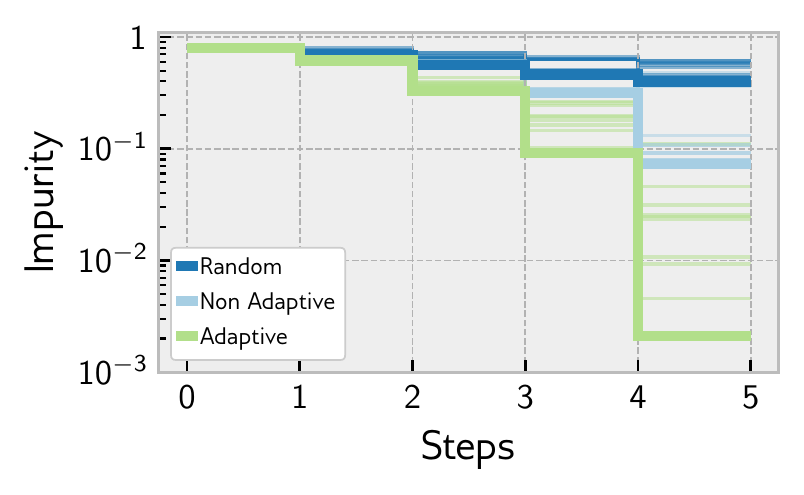}
    \caption{Purification of a thermal state (with $\bar{n}=2$) like in Fig.~\ref{fig:Figure3_b_purification}c) of the main text. Here, the shaded lines show 10 different strategies found by repeated runs of the algorithm, from different random starting points. The thick lines represent the best strategy found.\label{fig:purity_complete_supplementary}}
   
\end{figure}

\subsection{Effect of different initial condition on the training}
\begin{figure*}[t]
    \centering
    
    \includegraphics[width=\textwidth]{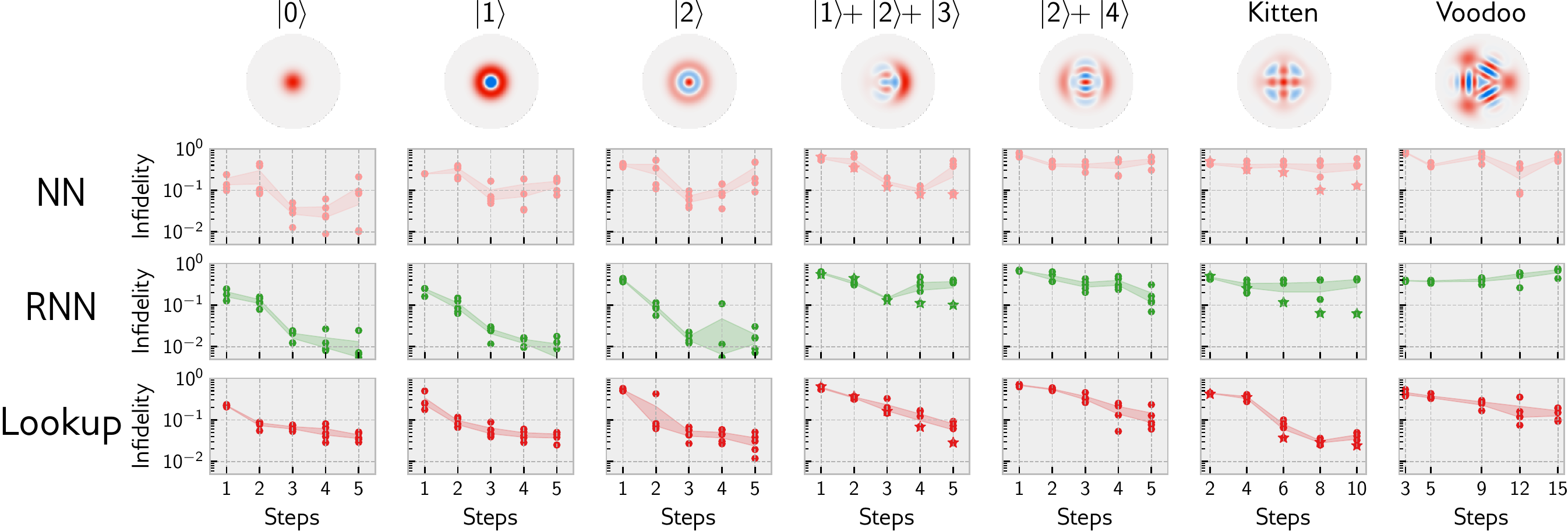}
    \caption{\label{fig:variance_supplementary}State preparation from a thermal state (${\bar n}=1$), employing feedback, like in Fig.~\ref{fig:Figure3_JCResults}). We show the final infidelity as a function of the number of time steps available for the strategy. The columns represents various final target states, while the rows shows the three different approaches to obtain the control parameters (NN: neural network being fed the current quantum state as input; RNN: recurrent neural network obtaining the measurement sequence step-wise; Lookup: a lookup table as defined in the main text). Each dot represents a different training run with different initial condition. For each number of steps, 5 runs are shown.  Their mean and standard deviation   are represented, respectively, as the center and  width of the  shaded area. For the Kitten and the $|1\rangle+|2\rangle+|3\rangle$ state we have added as reference  the data (plotted as stars) shown in Fig.~5. }
\end{figure*}

In order to assess the variability during the training, we show in this subsection how the results of Figs.~\ref{fig:Figure3_b_purification} and \ref{fig:Figure3_JCResults} can change, depending on the choice of different random initial conditions of the algorithm. As a first example, we show in Fig.~\ref{fig:purity_complete_supplementary} an equivalent plot of Fig.~\ref{fig:Figure3_b_purification}c), but in this case we want to differentiate the distinct strategies found by feedback-GRAPE. The majority of the adaptive runs can systematically reach higher purities then the other strategies (Random and Non-Adaptive). Nonetheless, one should be aware of such variability of strategies at the end of the training.\\
To further analyse the variability of training, we focus on the state preparation case from a thermal state, like in Fig.~\ref{fig:Figure3_JCResults}. In Fig.~\ref{fig:variance_supplementary}, the performance for many different target state was evaluated, along with the uncertainty due to different initial condition.
\subsection{Impact of the batch size on generalization}
\begin{figure*}[t]
    \centering
    
    \includegraphics[width=\textwidth]{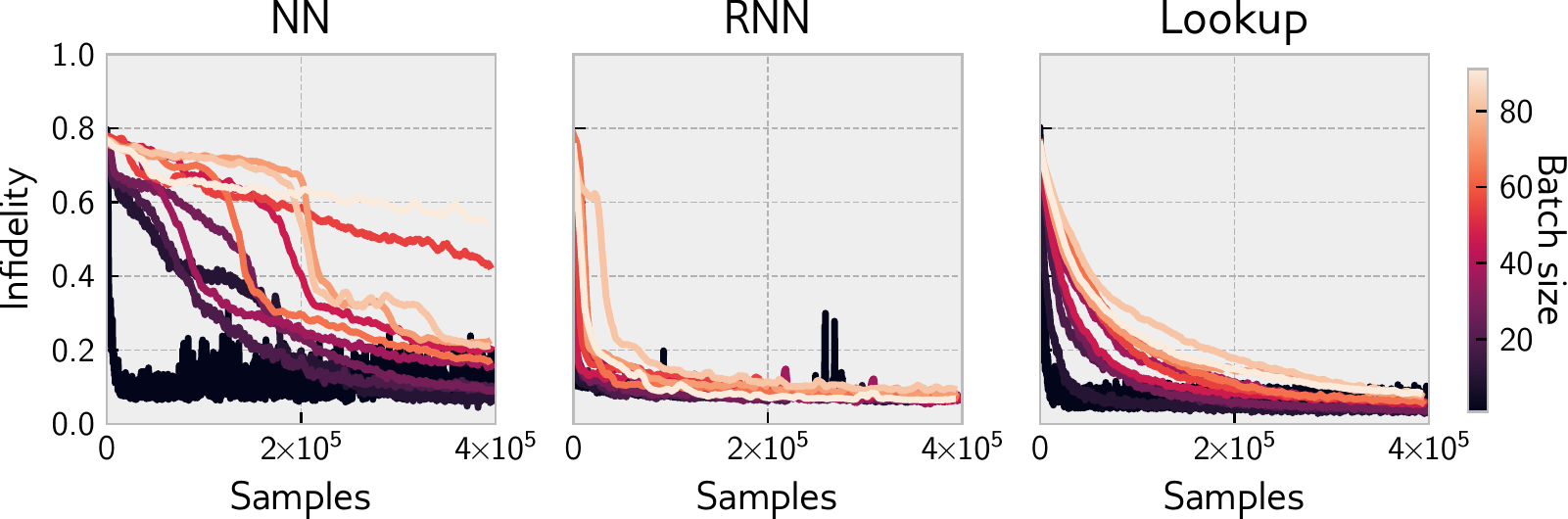}
    \caption{\label{fig:batch_sizes_supplementary}State preparation from a thermal state (here $\bar{n}=1$) and for target state $\psi =( \ket{1}+\ket{2}+\ket{3})/\sqrt{3}$, in 6 steps. Each training was run with a different batch size (encoded in the color), and the performance of the resulting strategy (NN/RNN/Lookup) was evaluated with a larger batch size (here 1000) to suppress statistical noise. The training curves are plotted with a moving average of 50 trajectories in order to suppress fluctuations. The number of trajectories (i.e. batch size $\times$ number of gradient optimization steps) for each curve is the same.}
   
\end{figure*}

A final analysis that we conducted deals with the effect of the batch size during training. We want to analyze both the performance during training and the generalization capabilities of the strategy learned. In order to asses that, we focus on the state preparation case from thermal state (Fig.~\ref{fig:Figure3_JCResults}). We run different trainings with distinct batch sizes (ranging from 1 to 100). For each batch size, we run 5 different training. We then post-select the best one, by computing the average fidelity on a much larger batch size (i.e. 1000). In \ref{fig:batch_sizes_supplementary} we then show the best performing NN/RNN/Lookup table. Interestingly, even though the training is noisier with a lower batch size, it seems that feedback-GRAPE can converge faster and to higher fidelity solution. Also, it seems that lower batch sizes can generalize well to higher ones. The better performance of lower batch sizes could be due to the possibility of the optimizer to escape local minima more efficiently than larger ones.

\section{Detailed analysis of strategies discovered by feedback-GRAPE for the Jaynes-Cummings model scenario}
\label{app-analysis-strategies}
In our work, we chose several different tasks within a Jaynes-Cummings model to illustrate the performance of our approach. Despite being only an illustrative physical example in this context, the model is of sufficient interest as a paradigm for actual feedback control of quantum-optical systems. In this section, we describe some of the insights we were able to extract by closer inspection of the numerical results obtained by feedback-GRAPE, in situations with feedback.

In the main text, we show the decision tree for the purification of a thermal state with initial occupation number $\langle \hat{a}^\dagger \hat{a} \rangle=2$ in four measurements. Here, we want to show how the insight gained by analyzing the decision tree for this special case allows to derive an analytical solution for an optimal purification strategy  valid for  arbitrary temperature and number of measurements.

We start by reviewing the physics for the building block measurement, cf  Eq.~(\ref{eq:POVM_Haroche}). This type of measurement has been originally proposed in \cite{brune_quantum_1990} and has been extensively used in quantum optics experiments with flying Rydberg atoms, e.g. to monitor the occupation number of a cavity in the presence of very small thermal fluctuations \cite{gleyzes_quantum_2007} or to prepare a Fock state starting from an initial coherent state \cite{guerlin_progressive_2007}. After each measurement, the Fock state probability distribution  $P_j(n)$ is updated  by  multiplying it with a sinusoidal mask,
\begin{equation}\label{eq:exclude_ni}
    P_{j+1}(n)\propto P_j(n)\cos^2\left[\gamma_i n+\frac{\delta_i}{2}+\pi(1-m_j)/4\right].
\end{equation}
To better understand the effects of the measurement it is important to keep in mind two key insights: (i) If the measurement strength can be well approximated with a rational multiple of $\pi$, $\gamma_i=\pi p_i/q_i$ where $p_i$ and $q_i$ are co-prime numbers, the denominator $q_i$ represents the period of the mask. Thus, the relative occupations $P(n)/P(n')$ of any pair of Fock states that have the same excitation number modulus $q_i$,   $(n-n')\,{\rm mod}\, q_i=0$, do not change after the measurement. (ii) If the phase $\delta_i$ satisfies either condition  $$\pi \frac{p_i}{q_i}n_i+\frac{\delta_i}{2}=0\,\rm mod\, \pi,\quad{\rm or}\quad  =\pi/2\,\rm mod\, \pi,$$ for an integer $n_i$, one measurement outcome ($m_i=-1$ or $m_i=1$, respectively) rules out the infinite set of Fock states with excitation numbers $n$ satisfying $n\,{\rm mod}\,q_i=n_i$. We note that if $q_i$ is an even number any $\delta_i$ that satisfies the first condition for $n_i\equiv n_{i,-1}$ satisfies also the second condition for $n_i=n_{i,1}\equiv(n_{i,-1}+q_i/2)\,{\rm mod}\,q_i$. In this scenario, each of the two possible measurement outcomes rules out a (different) infinite set of Fock states, $n_{i,\pm 1}\,{\rm mod}\,q_i$ for $m_i=\pm 1$. We note further that there are infinitely many  values of $\delta_i$  satisfying one of the two conditions in Eq.~(\ref{eq:exclude_ni}) for the same $n_i$. All of these  values of $\delta_i$ are rational multiples of $\pi$.

Motivated by the  insights (i) and (ii), we have written an algorithm that identifies  values of $\gamma_i$ and $\delta_i$ that are close to rational multiples of $\pi$ with small denominators  (we allow a deviation of  $1\%$ of $\pi$) and displays these rational values (in units of $\pi$) in the decision tree as shown in Fig.~\ref{fig:Figure3_b_purification}d. By inspecting this decision tree,  one can immediately observe that the NN tends to use measurement strength $\gamma_j$ corresponding to the period $q_j=2^j$ for the $j$-th measurement.  In order to understand this pattern, we  inspect   the phases $\delta_j$ selected by the NN. For the first measurement, the measurement strength is $\gamma_1=\pi/2$ and the phase $\delta_1=0$. This corresponds to $n_{1,-1}=0$  and $n_{1,1}=1$. In other words, the Fock state $0$ ($1$) along with all other even (odd) states are  ruled out by the measurement $m_1=-1$ ($m_1=1$).
Thus, the net effect is that,  irrespective of the measurement outcome, the probability of every second Fock state is set to zero.
Such a measurement extracts exactly 1 bit of information in the large temperature  limit.   For the second measurement, the NN  doubles the period of the sinusoidal mask, $q_2=4$, (independent of the outcome of the first measurement). By inspecting the phases $\delta_2$ chosen adaptively by the NN we find out that they always allow to rule out either of the two most likely states after the  measurement. For example, in the upper branch (corresponding to $m_1=1$) all odd states have been decimated and, thus, the two more likely states are the $0$ and $2$  Fock states. From the tree we see that  $\delta_2=\pi/2$ in this branch. This indeed satisfies the two conditions  in Eq.~(\ref{eq:exclude_ni}) with $n_i=n_{2,-1}=2$ and  $n_i=n_{2,1}=0$, respectively. In other words,  the Fock states with $n\,{\rm mod}\, 4=0$ ($n\,{\rm mod}\, 4=2$)  are ruled out by the measurement outcome $m_2=1$ ($m_2=-1$). Since all odd Fock states had been already ruled out after the first measurement, the overall effect of the first two measurements is to postselect every fourth Fock state, $n\,{\rm mod}\, 4=0$ ($n\,{\rm mod}\, 4=2$) for $m_1=1$ and $m_2=-1$ ($m_1=m_2=1$).   Likewise, the choice of the phase $\delta_2=-\pi/4$ in the lower branch  allows to postselect  every fourth Fock state, now, $n{\rm mod} 4=1$ and $n{\rm mod} 4=3$ for  $m_2=1$ and $m_2=-1$, respectively. This strategy can be easily generalized for any arbitrarily large number  of measurements $J$: the period $q_i$  is doubled  after every measurement, $q_j=2^j$, independent of the measurement outcomes and appropriate adaptive  phases $\delta_j$ are selected to always rule out either of the two most likely states. Such a strategy allows  to postselect the Fock states with $n\,{\rm mod}\, 2^J=n_i$ where $n_i$ depends on the measurement history. More precisely there is a bijective mapping between  $0\leq n_i<2^J-1$ and the $2^J$ possible measurements outcomes. Indeed, a close inspection of the strength $\gamma_i$ and phases $\delta_i$ selected by the NN shows that the NN  adopts this strategy for all four measurements in most (but not all) branches. A notable exception is  the third measurement in the lowest branch (corresponding to $m_1=m_2=-1$). This choice results in an ineffective measurement that does not allow to exclude either of the two most likeliest states. Interestingly, in this case the NN selects for the fourth measurement the measurement settings that were expected (according to the strategy identified above) already for the third measurement. We believe that this sub-optimal strategy corresponds to a local minimum for the gradient ascent. We note that the strategy whose tree is displayed in Fig.~\ref{fig:Figure3_b_purification}(d) has been obtained after selecting the best gradient ascent training run out of 10 runs with different random initializations. A tree without any such suboptimal measurements  could be obtained by performing more gradient ascent runs or, more efficiently, by increasing the temperature of the initial mixed state (which will punish more suboptimal purification strategies).

The same optimal strategy discussed above can be implemented for infinitely many different choices of $\gamma_j$ and $\delta_j$. In particular, different bijective  mappings between the measurement outcomes and the likeliest state $n_j$ after $j$ measurements can be implemented. To find a simple analytical solution for the phases $\gamma_j$ for one of the implementations of the optimal strategy, we  choose $p_j=1$ and, thus, $\gamma_j=\pi/2^j$. In addition, we choose $n_j$ as the  number whose binary representation is $d_{j-1}\ldots d_2d_1$ with $d_i=(1-m_i)/2$, e.g. for $m_1=m_2=-1$ corresponding to $d_1=d_2=1$ we have $n_3=1+2=3$. This mapping is implemented, if the phase $\delta_j$ always allows  to rule out  the Fock state with largest probability (or, equivalently, lowest excitation number among the states that have not yet been decimated by previous measurements) for the measurement outcome $m_j=-1$. With these constraints we find a simple analytical solution for  the phases, $\delta_j=\pi n_{j}/2^j$.

\section{Symmetry of the optimization landscape for the spin state preparation with uncertain parameters}
\label{Appendix_opt_landscape}

In this Appendix we analyze the symmetries of the optimization landscape for the learning of feedback strategies to prepare an ensemble of qubits in the excited state investigated in Section \ref{state_prep_uncertain_model} of the main text.

For $N=2$ the 3D optimization landscape displayed as three cuts in Fig.~\ref{fig:figure_6}(d) is the average $\langle\langle{\cal F}_2\rangle_{\mathbf{m}}\rangle_{g\sim P(g)}$ of the  coupling-dependent  fidelity
\begin{eqnarray}
\langle{\cal F}_2\rangle_{\mathbf{m}}&=&\sin^2[g\tau_0/2]\cos^2[g\tau_1(m_0=-1)/2]\nonumber\\
&&+\cos^2[g\tau_0/2]\sin^2[g\tau_1(m_0=1)/2]
\end{eqnarray}
We note that  this function  has three mirror planes because it is invariant under a sign change  of $\tau_0$, $\tau_1(m_0=1)$, or $\tau_1(m_0=-1)$. In addition in the plane $\tau_1(m_0=-1)=0$, corresponding to the  leftmost cut in Fig.~\ref{fig:figure_6}(d), it can be rewritten as
\begin{eqnarray}
\langle{\cal F}_2\rangle_{\mathbf{m}}&=&\sin^2[g\tau_0/2]+\sin^2[g\tau_1(m_0=1)/2]\nonumber\\
&&-\sin^2[g\tau_0/2]\sin^2[g\tau_1(m_0=1)/2].
\end{eqnarray}
From the above expression, it becomes clear that one can exchange $\tau_0$, $\tau_1(m_0=1)$ without changing the fidelity. These symmetries are present for any value of $g$ and, thus, also for any weighted average over $g$ and, in particular, for the optimization landscape $\langle\langle{\cal F}_2\rangle_{\mathbf{m}}\rangle_{g\sim P(g)}$. Since the optimal solutions lie on the plane $\tau_1(m_0=-1)=0$, there are 8 symmetry-related optimal solutions corresponding to the same coupling-dependent fidelity $\langle{\cal F}\rangle_{\mathbf{m}}$.

This result can be generalized to the case of $N$ measurements. On the hyperplane with $\tau_j(\mathbf{m}_{j-1})=0$ for all $\mathbf{m}_{j-1}\neq(1,\ldots,1)$ the coupling-dependent fidelity is the function
\begin{eqnarray}
\langle{\cal F}_N\rangle_{\mathbf{m}}&=&\sum_{j=0}^{N-1}\sin^2[g\tau_j/2]\prod_{j'=0}^{N-2}\cos^2[g\tau_{j'}/2].
\end{eqnarray}
with $\tau_j=\tau_j(\mathbf{m}_{j-1})$ with $\mathbf{m}_{j-1}=(1,\ldots,1)$. It is easy to show that this function is symmetric under permutation of its $N$ variables $\tau_j$. This leads to $4\times N!$ optimal solutions with the same coupling-dependent fidelity $\langle{\cal F}\rangle_{\mathbf{m}}$.

\bibliography{Feedback_Automatic_Diff}

\begin{thebibliography}{86}%
\makeatletter
\providecommand \@ifxundefined [1]{%
 \@ifx{#1\undefined}
}%
\providecommand \@ifnum [1]{%
 \ifnum #1\expandafter \@firstoftwo
 \else \expandafter \@secondoftwo
 \fi
}%
\providecommand \@ifx [1]{%
 \ifx #1\expandafter \@firstoftwo
 \else \expandafter \@secondoftwo
 \fi
}%
\providecommand \natexlab [1]{#1}%
\providecommand \enquote  [1]{``#1''}%
\providecommand \bibnamefont  [1]{#1}%
\providecommand \bibfnamefont [1]{#1}%
\providecommand \citenamefont [1]{#1}%
\providecommand \href@noop [0]{\@secondoftwo}%
\providecommand \href [0]{\begingroup \@sanitize@url \@href}%
\providecommand \@href[1]{\@@startlink{#1}\@@href}%
\providecommand \@@href[1]{\endgroup#1\@@endlink}%
\providecommand \@sanitize@url [0]{\catcode `\\12\catcode `\$12\catcode
  `\&12\catcode `\#12\catcode `\^12\catcode `\_12\catcode `\%12\relax}%
\providecommand \@@startlink[1]{}%
\providecommand \@@endlink[0]{}%
\providecommand \url  [0]{\begingroup\@sanitize@url \@url }%
\providecommand \@url [1]{\endgroup\@href {#1}{\urlprefix }}%
\providecommand \urlprefix  [0]{URL }%
\providecommand \Eprint [0]{\href }%
\providecommand \doibase [0]{http://dx.doi.org/}%
\providecommand \selectlanguage [0]{\@gobble}%
\providecommand \bibinfo  [0]{\@secondoftwo}%
\providecommand \bibfield  [0]{\@secondoftwo}%
\providecommand \translation [1]{[#1]}%
\providecommand \BibitemOpen [0]{}%
\providecommand \bibitemStop [0]{}%
\providecommand \bibitemNoStop [0]{.\EOS\space}%
\providecommand \EOS [0]{\spacefactor3000\relax}%
\providecommand \BibitemShut  [1]{\csname bibitem#1\endcsname}%
\let\auto@bib@innerbib\@empty
\bibitem [{\citenamefont {Glaser}\ \emph {et~al.}(2015)\citenamefont {Glaser},
  \citenamefont {Boscain}, \citenamefont {Calarco}, \citenamefont {Koch},
  \citenamefont {K{\"o}ckenberger}, \citenamefont {Kosloff}, \citenamefont
  {Kuprov}, \citenamefont {Luy}, \citenamefont {Schirmer}, \citenamefont
  {Schulte-Herbr{\"u}ggen}, \citenamefont {Sugny},\ and\ \citenamefont
  {Wilhelm}}]{glaser_training_2015}%
  \BibitemOpen
  \bibfield  {author} {\bibinfo {author} {\bibfnamefont {S.~J.}\ \bibnamefont
  {Glaser}}, \bibinfo {author} {\bibfnamefont {U.}~\bibnamefont {Boscain}},
  \bibinfo {author} {\bibfnamefont {T.}~\bibnamefont {Calarco}}, \bibinfo
  {author} {\bibfnamefont {C.~P.}\ \bibnamefont {Koch}}, \bibinfo {author}
  {\bibfnamefont {W.}~\bibnamefont {K{\"o}ckenberger}}, \bibinfo {author}
  {\bibfnamefont {R.}~\bibnamefont {Kosloff}}, \bibinfo {author} {\bibfnamefont
  {I.}~\bibnamefont {Kuprov}}, \bibinfo {author} {\bibfnamefont
  {B.}~\bibnamefont {Luy}}, \bibinfo {author} {\bibfnamefont {S.}~\bibnamefont
  {Schirmer}}, \bibinfo {author} {\bibfnamefont {T.}~\bibnamefont
  {Schulte-Herbr{\"u}ggen}}, \bibinfo {author} {\bibfnamefont {D.}~\bibnamefont
  {Sugny}}, \ and\ \bibinfo {author} {\bibfnamefont {F.~K.}\ \bibnamefont
  {Wilhelm}},\ }\href {\doibase 10.1140/epjd/e2015-60464-1} {\bibfield
  {journal} {\bibinfo  {journal} {The European Physical Journal D}\ }\textbf
  {\bibinfo {volume} {69}},\ \bibinfo {pages} {279} (\bibinfo {year}
  {2015})}\BibitemShut {NoStop}%
\bibitem [{\citenamefont {Koch}(2016)}]{koch_controlling_2016}%
  \BibitemOpen
  \bibfield  {author} {\bibinfo {author} {\bibfnamefont {C.~P.}\ \bibnamefont
  {Koch}},\ }\href {\doibase 10.1088/0953-8984/28/21/213001} {\bibfield
  {journal} {\bibinfo  {journal} {Journal of Physics: Condensed Matter}\
  }\textbf {\bibinfo {volume} {28}},\ \bibinfo {pages} {213001} (\bibinfo
  {year} {2016})},\ \bibinfo {note} {publisher: IOP Publishing}\BibitemShut
  {NoStop}%
\bibitem [{\citenamefont {Khaneja}\ \emph {et~al.}(2005)\citenamefont
  {Khaneja}, \citenamefont {Reiss}, \citenamefont {Kehlet}, \citenamefont
  {Schulte-Herbr{\"u}ggen},\ and\ \citenamefont
  {Glaser}}]{khaneja_optimal_2005}%
  \BibitemOpen
  \bibfield  {author} {\bibinfo {author} {\bibfnamefont {N.}~\bibnamefont
  {Khaneja}}, \bibinfo {author} {\bibfnamefont {T.}~\bibnamefont {Reiss}},
  \bibinfo {author} {\bibfnamefont {C.}~\bibnamefont {Kehlet}}, \bibinfo
  {author} {\bibfnamefont {T.}~\bibnamefont {Schulte-Herbr{\"u}ggen}}, \ and\
  \bibinfo {author} {\bibfnamefont {S.~J.}\ \bibnamefont {Glaser}},\ }\href
  {\doibase 10.1016/j.jmr.2004.11.004} {\bibfield  {journal} {\bibinfo
  {journal} {Journal of Magnetic Resonance}\ }\textbf {\bibinfo {volume}
  {172}},\ \bibinfo {pages} {296} (\bibinfo {year} {2005})}\BibitemShut
  {NoStop}%
\bibitem [{\citenamefont {Machnes}\ \emph {et~al.}(2011)\citenamefont
  {Machnes}, \citenamefont {Sander}, \citenamefont {Glaser}, \citenamefont
  {de~Fouqui{\`e}res}, \citenamefont {Gruslys}, \citenamefont {Schirmer},\ and\
  \citenamefont {Schulte-Herbr{\"u}ggen}}]{machnes_comparing_2011}%
  \BibitemOpen
  \bibfield  {author} {\bibinfo {author} {\bibfnamefont {S.}~\bibnamefont
  {Machnes}}, \bibinfo {author} {\bibfnamefont {U.}~\bibnamefont {Sander}},
  \bibinfo {author} {\bibfnamefont {S.~J.}\ \bibnamefont {Glaser}}, \bibinfo
  {author} {\bibfnamefont {P.}~\bibnamefont {de~Fouqui{\`e}res}}, \bibinfo
  {author} {\bibfnamefont {A.}~\bibnamefont {Gruslys}}, \bibinfo {author}
  {\bibfnamefont {S.}~\bibnamefont {Schirmer}}, \ and\ \bibinfo {author}
  {\bibfnamefont {T.}~\bibnamefont {Schulte-Herbr{\"u}ggen}},\ }\href {\doibase
  10.1103/PhysRevA.84.022305} {\bibfield  {journal} {\bibinfo  {journal}
  {Physical Review A}\ }\textbf {\bibinfo {volume} {84}},\ \bibinfo {pages}
  {022305} (\bibinfo {year} {2011})},\ \bibinfo {note} {publisher: American
  Physical Society}\BibitemShut {NoStop}%
\bibitem [{\citenamefont {Dolde}\ \emph {et~al.}(2014)\citenamefont {Dolde},
  \citenamefont {Bergholm}, \citenamefont {Wang}, \citenamefont {Jakobi},
  \citenamefont {Naydenov}, \citenamefont {Pezzagna}, \citenamefont {Meijer},
  \citenamefont {Jelezko}, \citenamefont {Neumann}, \citenamefont
  {Schulte-Herbr{\"u}ggen}, \citenamefont {Biamonte},\ and\ \citenamefont
  {Wrachtrup}}]{dolde_high-fidelity_2014}%
  \BibitemOpen
  \bibfield  {author} {\bibinfo {author} {\bibfnamefont {F.}~\bibnamefont
  {Dolde}}, \bibinfo {author} {\bibfnamefont {V.}~\bibnamefont {Bergholm}},
  \bibinfo {author} {\bibfnamefont {Y.}~\bibnamefont {Wang}}, \bibinfo {author}
  {\bibfnamefont {I.}~\bibnamefont {Jakobi}}, \bibinfo {author} {\bibfnamefont
  {B.}~\bibnamefont {Naydenov}}, \bibinfo {author} {\bibfnamefont
  {S.}~\bibnamefont {Pezzagna}}, \bibinfo {author} {\bibfnamefont
  {J.}~\bibnamefont {Meijer}}, \bibinfo {author} {\bibfnamefont
  {F.}~\bibnamefont {Jelezko}}, \bibinfo {author} {\bibfnamefont
  {P.}~\bibnamefont {Neumann}}, \bibinfo {author} {\bibfnamefont
  {T.}~\bibnamefont {Schulte-Herbr{\"u}ggen}}, \bibinfo {author} {\bibfnamefont
  {J.}~\bibnamefont {Biamonte}}, \ and\ \bibinfo {author} {\bibfnamefont
  {J.}~\bibnamefont {Wrachtrup}},\ }\href {\doibase 10.1038/ncomms4371}
  {\bibfield  {journal} {\bibinfo  {journal} {Nature Communications}\ }\textbf
  {\bibinfo {volume} {5}},\ \bibinfo {pages} {3371} (\bibinfo {year} {2014})},\
  \bibinfo {note} {number: 1 Publisher: Nature Publishing Group}\BibitemShut
  {NoStop}%
\bibitem [{\citenamefont {Yang}\ \emph {et~al.}(2019)\citenamefont {Yang},
  \citenamefont {Chan}, \citenamefont {Harper}, \citenamefont {Huang},
  \citenamefont {Evans}, \citenamefont {Hwang}, \citenamefont {Hensen},
  \citenamefont {Laucht}, \citenamefont {Tanttu}, \citenamefont {Hudson},
  \citenamefont {Flammia}, \citenamefont {Itoh}, \citenamefont {Morello},
  \citenamefont {Bartlett},\ and\ \citenamefont {Dzurak}}]{yang_silicon_2019}%
  \BibitemOpen
  \bibfield  {author} {\bibinfo {author} {\bibfnamefont {C.~H.}\ \bibnamefont
  {Yang}}, \bibinfo {author} {\bibfnamefont {K.~W.}\ \bibnamefont {Chan}},
  \bibinfo {author} {\bibfnamefont {R.}~\bibnamefont {Harper}}, \bibinfo
  {author} {\bibfnamefont {W.}~\bibnamefont {Huang}}, \bibinfo {author}
  {\bibfnamefont {T.}~\bibnamefont {Evans}}, \bibinfo {author} {\bibfnamefont
  {J.~C.~C.}\ \bibnamefont {Hwang}}, \bibinfo {author} {\bibfnamefont
  {B.}~\bibnamefont {Hensen}}, \bibinfo {author} {\bibfnamefont
  {A.}~\bibnamefont {Laucht}}, \bibinfo {author} {\bibfnamefont
  {T.}~\bibnamefont {Tanttu}}, \bibinfo {author} {\bibfnamefont {F.~E.}\
  \bibnamefont {Hudson}}, \bibinfo {author} {\bibfnamefont {S.~T.}\
  \bibnamefont {Flammia}}, \bibinfo {author} {\bibfnamefont {K.~M.}\
  \bibnamefont {Itoh}}, \bibinfo {author} {\bibfnamefont {A.}~\bibnamefont
  {Morello}}, \bibinfo {author} {\bibfnamefont {S.~D.}\ \bibnamefont
  {Bartlett}}, \ and\ \bibinfo {author} {\bibfnamefont {A.~S.}\ \bibnamefont
  {Dzurak}},\ }\href {\doibase 10.1038/s41928-019-0234-1} {\bibfield  {journal}
  {\bibinfo  {journal} {Nature Electronics}\ }\textbf {\bibinfo {volume} {2}},\
  \bibinfo {pages} {151} (\bibinfo {year} {2019})},\ \bibinfo {note} {number: 4
  Publisher: Nature Publishing Group}\BibitemShut {NoStop}%
\bibitem [{\citenamefont {Schulte-Herbr{\"u}ggen}\ \emph
  {et~al.}(2005)\citenamefont {Schulte-Herbr{\"u}ggen}, \citenamefont
  {Sp{\"o}rl}, \citenamefont {Khaneja},\ and\ \citenamefont
  {Glaser}}]{schulte-herbruggen_optimal_2005}%
  \BibitemOpen
  \bibfield  {author} {\bibinfo {author} {\bibfnamefont {T.}~\bibnamefont
  {Schulte-Herbr{\"u}ggen}}, \bibinfo {author} {\bibfnamefont {A.}~\bibnamefont
  {Sp{\"o}rl}}, \bibinfo {author} {\bibfnamefont {N.}~\bibnamefont {Khaneja}},
  \ and\ \bibinfo {author} {\bibfnamefont {S.~J.}\ \bibnamefont {Glaser}},\
  }\href {\doibase 10.1103/PhysRevA.72.042331} {\bibfield  {journal} {\bibinfo
  {journal} {Physical Review A}\ }\textbf {\bibinfo {volume} {72}},\ \bibinfo
  {pages} {042331} (\bibinfo {year} {2005})},\ \bibinfo {note} {publisher:
  American Physical Society}\BibitemShut {NoStop}%
\bibitem [{\citenamefont {Sp{\"o}rl}\ \emph {et~al.}(2007)\citenamefont
  {Sp{\"o}rl}, \citenamefont {Schulte-Herbr{\"u}ggen}, \citenamefont {Glaser},
  \citenamefont {Bergholm}, \citenamefont {Storcz}, \citenamefont {Ferber},\
  and\ \citenamefont {Wilhelm}}]{sporl_optimal_2007}%
  \BibitemOpen
  \bibfield  {author} {\bibinfo {author} {\bibfnamefont {A.}~\bibnamefont
  {Sp{\"o}rl}}, \bibinfo {author} {\bibfnamefont {T.}~\bibnamefont
  {Schulte-Herbr{\"u}ggen}}, \bibinfo {author} {\bibfnamefont {S.~J.}\
  \bibnamefont {Glaser}}, \bibinfo {author} {\bibfnamefont {V.}~\bibnamefont
  {Bergholm}}, \bibinfo {author} {\bibfnamefont {M.~J.}\ \bibnamefont
  {Storcz}}, \bibinfo {author} {\bibfnamefont {J.}~\bibnamefont {Ferber}}, \
  and\ \bibinfo {author} {\bibfnamefont {F.~K.}\ \bibnamefont {Wilhelm}},\
  }\href {\doibase 10.1103/PhysRevA.75.012302} {\bibfield  {journal} {\bibinfo
  {journal} {Physical Review A}\ }\textbf {\bibinfo {volume} {75}},\ \bibinfo
  {pages} {012302} (\bibinfo {year} {2007})},\ \bibinfo {note} {publisher:
  American Physical Society}\BibitemShut {NoStop}%
\bibitem [{\citenamefont {Heeres}\ \emph {et~al.}(2017)\citenamefont {Heeres},
  \citenamefont {Reinhold}, \citenamefont {Ofek}, \citenamefont {Frunzio},
  \citenamefont {Jiang}, \citenamefont {Devoret},\ and\ \citenamefont
  {Schoelkopf}}]{heeres_implementing_2017}%
  \BibitemOpen
  \bibfield  {author} {\bibinfo {author} {\bibfnamefont {R.~W.}\ \bibnamefont
  {Heeres}}, \bibinfo {author} {\bibfnamefont {P.}~\bibnamefont {Reinhold}},
  \bibinfo {author} {\bibfnamefont {N.}~\bibnamefont {Ofek}}, \bibinfo {author}
  {\bibfnamefont {L.}~\bibnamefont {Frunzio}}, \bibinfo {author} {\bibfnamefont
  {L.}~\bibnamefont {Jiang}}, \bibinfo {author} {\bibfnamefont {M.~H.}\
  \bibnamefont {Devoret}}, \ and\ \bibinfo {author} {\bibfnamefont {R.~J.}\
  \bibnamefont {Schoelkopf}},\ }\href {\doibase 10.1038/s41467-017-00045-1}
  {\bibfield  {journal} {\bibinfo  {journal} {Nature Communications}\ }\textbf
  {\bibinfo {volume} {8}},\ \bibinfo {pages} {94} (\bibinfo {year} {2017})},\
  \bibinfo {note} {number: 1 Publisher: Nature Publishing Group}\BibitemShut
  {NoStop}%
\bibitem [{\citenamefont {Fisher}\ \emph {et~al.}(2010)\citenamefont {Fisher},
  \citenamefont {Helmer}, \citenamefont {Glaser}, \citenamefont {Marquardt},\
  and\ \citenamefont {Schulte-Herbr{\"u}ggen}}]{fisher_optimal_2010}%
  \BibitemOpen
  \bibfield  {author} {\bibinfo {author} {\bibfnamefont {R.}~\bibnamefont
  {Fisher}}, \bibinfo {author} {\bibfnamefont {F.}~\bibnamefont {Helmer}},
  \bibinfo {author} {\bibfnamefont {S.~J.}\ \bibnamefont {Glaser}}, \bibinfo
  {author} {\bibfnamefont {F.}~\bibnamefont {Marquardt}}, \ and\ \bibinfo
  {author} {\bibfnamefont {T.}~\bibnamefont {Schulte-Herbr{\"u}ggen}},\ }\href
  {\doibase 10.1103/PhysRevB.81.085328} {\bibfield  {journal} {\bibinfo
  {journal} {Physical Review B}\ }\textbf {\bibinfo {volume} {81}},\ \bibinfo
  {pages} {085328} (\bibinfo {year} {2010})},\ \bibinfo {note} {publisher:
  American Physical Society}\BibitemShut {NoStop}%
\bibitem [{\citenamefont {Schulte-Herbr{\"u}ggen}\ \emph
  {et~al.}(2011)\citenamefont {Schulte-Herbr{\"u}ggen}, \citenamefont
  {Sp{\"o}rl}, \citenamefont {Khaneja},\ and\ \citenamefont
  {Glaser}}]{schulte-herbruggen_optimal_2011}%
  \BibitemOpen
  \bibfield  {author} {\bibinfo {author} {\bibfnamefont {T.}~\bibnamefont
  {Schulte-Herbr{\"u}ggen}}, \bibinfo {author} {\bibfnamefont {A.}~\bibnamefont
  {Sp{\"o}rl}}, \bibinfo {author} {\bibfnamefont {N.}~\bibnamefont {Khaneja}},
  \ and\ \bibinfo {author} {\bibfnamefont {S.~J.}\ \bibnamefont {Glaser}},\
  }\href {\doibase 10.1088/0953-4075/44/15/154013} {\bibfield  {journal}
  {\bibinfo  {journal} {Journal of Physics B: Atomic, Molecular and Optical
  Physics}\ }\textbf {\bibinfo {volume} {44}},\ \bibinfo {pages} {154013}
  (\bibinfo {year} {2011})}\BibitemShut {NoStop}%
\bibitem [{\citenamefont {Boutin}\ \emph {et~al.}(2017)\citenamefont {Boutin},
  \citenamefont {Andersen}, \citenamefont {Venkatraman}, \citenamefont
  {Ferris},\ and\ \citenamefont {Blais}}]{boutin_resonator_2017}%
  \BibitemOpen
  \bibfield  {author} {\bibinfo {author} {\bibfnamefont {S.}~\bibnamefont
  {Boutin}}, \bibinfo {author} {\bibfnamefont {C.~K.}\ \bibnamefont
  {Andersen}}, \bibinfo {author} {\bibfnamefont {J.}~\bibnamefont
  {Venkatraman}}, \bibinfo {author} {\bibfnamefont {A.~J.}\ \bibnamefont
  {Ferris}}, \ and\ \bibinfo {author} {\bibfnamefont {A.}~\bibnamefont
  {Blais}},\ }\href {\doibase 10.1103/PhysRevA.96.042315} {\bibfield  {journal}
  {\bibinfo  {journal} {Physical Review A}\ }\textbf {\bibinfo {volume} {96}},\
  \bibinfo {pages} {042315} (\bibinfo {year} {2017})}\BibitemShut {NoStop}%
\bibitem [{\citenamefont {Egger}\ and\ \citenamefont
  {Wilhelm}(2014)}]{egger_adaptive_2014}%
  \BibitemOpen
  \bibfield  {author} {\bibinfo {author} {\bibfnamefont {D.~J.}\ \bibnamefont
  {Egger}}\ and\ \bibinfo {author} {\bibfnamefont {F.~K.}\ \bibnamefont
  {Wilhelm}},\ }\href {\doibase 10.1103/PhysRevLett.112.240503} {\bibfield
  {journal} {\bibinfo  {journal} {Physical Review Letters}\ }\textbf {\bibinfo
  {volume} {112}},\ \bibinfo {pages} {240503} (\bibinfo {year} {2014})},\
  \bibinfo {note} {publisher: American Physical Society}\BibitemShut {NoStop}%
\bibitem [{\citenamefont {Wu}\ \emph {et~al.}(2018)\citenamefont {Wu},
  \citenamefont {Chu}, \citenamefont {Owens},\ and\ \citenamefont
  {Rabitz}}]{wu_data-driven_2018}%
  \BibitemOpen
  \bibfield  {author} {\bibinfo {author} {\bibfnamefont {R.-B.}\ \bibnamefont
  {Wu}}, \bibinfo {author} {\bibfnamefont {B.}~\bibnamefont {Chu}}, \bibinfo
  {author} {\bibfnamefont {D.~H.}\ \bibnamefont {Owens}}, \ and\ \bibinfo
  {author} {\bibfnamefont {H.}~\bibnamefont {Rabitz}},\ }\href {\doibase
  10.1103/PhysRevA.97.042122} {\bibfield  {journal} {\bibinfo  {journal}
  {Physical Review A}\ }\textbf {\bibinfo {volume} {97}},\ \bibinfo {pages}
  {042122} (\bibinfo {year} {2018})}\BibitemShut {NoStop}%
\bibitem [{\citenamefont {Wu}\ \emph {et~al.}(2019)\citenamefont {Wu},
  \citenamefont {Ding}, \citenamefont {Dong},\ and\ \citenamefont
  {Wang}}]{wu_learning_2019}%
  \BibitemOpen
  \bibfield  {author} {\bibinfo {author} {\bibfnamefont {R.-B.}\ \bibnamefont
  {Wu}}, \bibinfo {author} {\bibfnamefont {H.}~\bibnamefont {Ding}}, \bibinfo
  {author} {\bibfnamefont {D.}~\bibnamefont {Dong}}, \ and\ \bibinfo {author}
  {\bibfnamefont {X.}~\bibnamefont {Wang}},\ }\href {\doibase
  10.1103/PhysRevA.99.042327} {\bibfield  {journal} {\bibinfo  {journal}
  {Physical Review A}\ }\textbf {\bibinfo {volume} {99}},\ \bibinfo {pages}
  {042327} (\bibinfo {year} {2019})}\BibitemShut {NoStop}%
\bibitem [{\citenamefont {Ding}\ \emph {et~al.}(2021)\citenamefont {Ding},
  \citenamefont {Chu}, \citenamefont {Qi},\ and\ \citenamefont
  {Wu}}]{ding_collaborative_2021}%
  \BibitemOpen
  \bibfield  {author} {\bibinfo {author} {\bibfnamefont {H.-J.}\ \bibnamefont
  {Ding}}, \bibinfo {author} {\bibfnamefont {B.}~\bibnamefont {Chu}}, \bibinfo
  {author} {\bibfnamefont {B.}~\bibnamefont {Qi}}, \ and\ \bibinfo {author}
  {\bibfnamefont {R.-B.}\ \bibnamefont {Wu}},\ }\href {\doibase
  10.1103/PhysRevApplied.16.014056} {\bibfield  {journal} {\bibinfo  {journal}
  {Physical Review Applied}\ }\textbf {\bibinfo {volume} {16}},\ \bibinfo
  {pages} {014056} (\bibinfo {year} {2021})},\ \bibinfo {note} {publisher:
  American Physical Society}\BibitemShut {NoStop}%
\bibitem [{\citenamefont {de~Fouquieres}\ \emph {et~al.}(2011)\citenamefont
  {de~Fouquieres}, \citenamefont {Schirmer}, \citenamefont {Glaser},\ and\
  \citenamefont {Kuprov}}]{de_fouquieres_second_2011}%
  \BibitemOpen
  \bibfield  {author} {\bibinfo {author} {\bibfnamefont {P.}~\bibnamefont
  {de~Fouquieres}}, \bibinfo {author} {\bibfnamefont {S.~G.}\ \bibnamefont
  {Schirmer}}, \bibinfo {author} {\bibfnamefont {S.~J.}\ \bibnamefont
  {Glaser}}, \ and\ \bibinfo {author} {\bibfnamefont {I.}~\bibnamefont
  {Kuprov}},\ }\href {\doibase 10.1016/j.jmr.2011.07.023} {\bibfield  {journal}
  {\bibinfo  {journal} {Journal of Magnetic Resonance}\ }\textbf {\bibinfo
  {volume} {212}},\ \bibinfo {pages} {412} (\bibinfo {year}
  {2011})}\BibitemShut {NoStop}%
\bibitem [{\citenamefont {Machnes}\ \emph {et~al.}(2018)\citenamefont
  {Machnes}, \citenamefont {Ass{\'e}mat}, \citenamefont {Tannor},\ and\
  \citenamefont {Wilhelm}}]{machnes_tunable_2018}%
  \BibitemOpen
  \bibfield  {author} {\bibinfo {author} {\bibfnamefont {S.}~\bibnamefont
  {Machnes}}, \bibinfo {author} {\bibfnamefont {E.}~\bibnamefont
  {Ass{\'e}mat}}, \bibinfo {author} {\bibfnamefont {D.}~\bibnamefont {Tannor}},
  \ and\ \bibinfo {author} {\bibfnamefont {F.~K.}\ \bibnamefont {Wilhelm}},\
  }\href {\doibase 10.1103/PhysRevLett.120.150401} {\bibfield  {journal}
  {\bibinfo  {journal} {Physical Review Letters}\ }\textbf {\bibinfo {volume}
  {120}},\ \bibinfo {pages} {150401} (\bibinfo {year} {2018})},\ \bibinfo
  {note} {publisher: American Physical Society}\BibitemShut {NoStop}%
\bibitem [{\citenamefont {Zhang}\ \emph {et~al.}(2017)\citenamefont {Zhang},
  \citenamefont {Liu}, \citenamefont {Wu}, \citenamefont {Jacobs},\ and\
  \citenamefont {Nori}}]{zhang_quantum_2017}%
  \BibitemOpen
  \bibfield  {author} {\bibinfo {author} {\bibfnamefont {J.}~\bibnamefont
  {Zhang}}, \bibinfo {author} {\bibfnamefont {Y.-x.}\ \bibnamefont {Liu}},
  \bibinfo {author} {\bibfnamefont {R.-B.}\ \bibnamefont {Wu}}, \bibinfo
  {author} {\bibfnamefont {K.}~\bibnamefont {Jacobs}}, \ and\ \bibinfo {author}
  {\bibfnamefont {F.}~\bibnamefont {Nori}},\ }\href {\doibase
  10.1016/j.physrep.2017.02.003} {\bibfield  {journal} {\bibinfo  {journal}
  {Physics Reports}\ }\bibinfo {series} {Quantum feedback: theory, experiments,
  and applications},\ \textbf {\bibinfo {volume} {679}},\ \bibinfo {pages} {1}
  (\bibinfo {year} {2017})}\BibitemShut {NoStop}%
\bibitem [{\citenamefont {Handel}\ \emph {et~al.}(2005)\citenamefont {Handel},
  \citenamefont {Stockton},\ and\ \citenamefont
  {Mabuchi}}]{handel_modelling_2005}%
  \BibitemOpen
  \bibfield  {author} {\bibinfo {author} {\bibfnamefont {R.~v.}\ \bibnamefont
  {Handel}}, \bibinfo {author} {\bibfnamefont {J.~K.}\ \bibnamefont
  {Stockton}}, \ and\ \bibinfo {author} {\bibfnamefont {H.}~\bibnamefont
  {Mabuchi}},\ }\href {\doibase 10.1088/1464-4266/7/10/001} {\bibfield
  {journal} {\bibinfo  {journal} {Journal of Optics B: Quantum and
  Semiclassical Optics}\ }\textbf {\bibinfo {volume} {7}},\ \bibinfo {pages}
  {S179} (\bibinfo {year} {2005})},\ \bibinfo {note} {publisher: IOP
  Publishing}\BibitemShut {NoStop}%
\bibitem [{\citenamefont {Sayrin}\ \emph {et~al.}(2011)\citenamefont {Sayrin},
  \citenamefont {Dotsenko}, \citenamefont {Zhou}, \citenamefont {Peaudecerf},
  \citenamefont {Rybarczyk}, \citenamefont {Gleyzes}, \citenamefont {Rouchon},
  \citenamefont {Mirrahimi}, \citenamefont {Amini}, \citenamefont {Brune},
  \citenamefont {Raimond},\ and\ \citenamefont
  {Haroche}}]{sayrin_real-time_2011}%
  \BibitemOpen
  \bibfield  {author} {\bibinfo {author} {\bibfnamefont {C.}~\bibnamefont
  {Sayrin}}, \bibinfo {author} {\bibfnamefont {I.}~\bibnamefont {Dotsenko}},
  \bibinfo {author} {\bibfnamefont {X.}~\bibnamefont {Zhou}}, \bibinfo {author}
  {\bibfnamefont {B.}~\bibnamefont {Peaudecerf}}, \bibinfo {author}
  {\bibfnamefont {T.}~\bibnamefont {Rybarczyk}}, \bibinfo {author}
  {\bibfnamefont {S.}~\bibnamefont {Gleyzes}}, \bibinfo {author} {\bibfnamefont
  {P.}~\bibnamefont {Rouchon}}, \bibinfo {author} {\bibfnamefont
  {M.}~\bibnamefont {Mirrahimi}}, \bibinfo {author} {\bibfnamefont
  {H.}~\bibnamefont {Amini}}, \bibinfo {author} {\bibfnamefont
  {M.}~\bibnamefont {Brune}}, \bibinfo {author} {\bibfnamefont {J.-M.}\
  \bibnamefont {Raimond}}, \ and\ \bibinfo {author} {\bibfnamefont
  {S.}~\bibnamefont {Haroche}},\ }\href {\doibase 10.1038/nature10376}
  {\bibfield  {journal} {\bibinfo  {journal} {Nature}\ }\textbf {\bibinfo
  {volume} {477}},\ \bibinfo {pages} {73} (\bibinfo {year} {2011})},\ \bibinfo
  {note} {arXiv: 1107.4027}\BibitemShut {NoStop}%
\bibitem [{\citenamefont {Vijay}\ \emph {et~al.}(2012)\citenamefont {Vijay},
  \citenamefont {Macklin}, \citenamefont {Slichter}, \citenamefont {Weber},
  \citenamefont {Murch}, \citenamefont {Naik}, \citenamefont {Korotkov},\ and\
  \citenamefont {Siddiqi}}]{vijay_stabilizing_2012}%
  \BibitemOpen
  \bibfield  {author} {\bibinfo {author} {\bibfnamefont {R.}~\bibnamefont
  {Vijay}}, \bibinfo {author} {\bibfnamefont {C.}~\bibnamefont {Macklin}},
  \bibinfo {author} {\bibfnamefont {D.~H.}\ \bibnamefont {Slichter}}, \bibinfo
  {author} {\bibfnamefont {S.~J.}\ \bibnamefont {Weber}}, \bibinfo {author}
  {\bibfnamefont {K.~W.}\ \bibnamefont {Murch}}, \bibinfo {author}
  {\bibfnamefont {R.}~\bibnamefont {Naik}}, \bibinfo {author} {\bibfnamefont
  {A.~N.}\ \bibnamefont {Korotkov}}, \ and\ \bibinfo {author} {\bibfnamefont
  {I.}~\bibnamefont {Siddiqi}},\ }\href {\doibase 10.1038/nature11505}
  {\bibfield  {journal} {\bibinfo  {journal} {Nature}\ }\textbf {\bibinfo
  {volume} {490}},\ \bibinfo {pages} {77} (\bibinfo {year} {2012})},\ \bibinfo
  {note} {number: 7418 Publisher: Nature Publishing Group}\BibitemShut
  {NoStop}%
\bibitem [{\citenamefont {Hirose}\ and\ \citenamefont
  {Cappellaro}(2016)}]{hirose_coherent_2016}%
  \BibitemOpen
  \bibfield  {author} {\bibinfo {author} {\bibfnamefont {M.}~\bibnamefont
  {Hirose}}\ and\ \bibinfo {author} {\bibfnamefont {P.}~\bibnamefont
  {Cappellaro}},\ }\href {\doibase 10.1038/nature17404} {\bibfield  {journal}
  {\bibinfo  {journal} {Nature}\ }\textbf {\bibinfo {volume} {532}},\ \bibinfo
  {pages} {77} (\bibinfo {year} {2016})},\ \bibinfo {note} {number: 7597
  Publisher: Nature Publishing Group}\BibitemShut {NoStop}%
\bibitem [{\citenamefont {Andersen}\ \emph {et~al.}(2019)\citenamefont
  {Andersen}, \citenamefont {Remm}, \citenamefont {Lazar}, \citenamefont
  {Krinner}, \citenamefont {Heinsoo}, \citenamefont {Besse}, \citenamefont
  {Gabureac}, \citenamefont {Wallraff},\ and\ \citenamefont
  {Eichler}}]{andersen_entanglement_2019}%
  \BibitemOpen
  \bibfield  {author} {\bibinfo {author} {\bibfnamefont {C.~K.}\ \bibnamefont
  {Andersen}}, \bibinfo {author} {\bibfnamefont {A.}~\bibnamefont {Remm}},
  \bibinfo {author} {\bibfnamefont {S.}~\bibnamefont {Lazar}}, \bibinfo
  {author} {\bibfnamefont {S.}~\bibnamefont {Krinner}}, \bibinfo {author}
  {\bibfnamefont {J.}~\bibnamefont {Heinsoo}}, \bibinfo {author} {\bibfnamefont
  {J.-C.}\ \bibnamefont {Besse}}, \bibinfo {author} {\bibfnamefont
  {M.}~\bibnamefont {Gabureac}}, \bibinfo {author} {\bibfnamefont
  {A.}~\bibnamefont {Wallraff}}, \ and\ \bibinfo {author} {\bibfnamefont
  {C.}~\bibnamefont {Eichler}},\ }\href {\doibase 10.1038/s41534-019-0185-4}
  {\bibfield  {journal} {\bibinfo  {journal} {npj Quantum Information}\
  }\textbf {\bibinfo {volume} {5}},\ \bibinfo {pages} {1} (\bibinfo {year}
  {2019})},\ \bibinfo {note} {number: 1 Publisher: Nature Publishing
  Group}\BibitemShut {NoStop}%
\bibitem [{\citenamefont {Hentschel}\ and\ \citenamefont
  {Sanders}(2011)}]{hentschel_efficient_2011}%
  \BibitemOpen
  \bibfield  {author} {\bibinfo {author} {\bibfnamefont {A.}~\bibnamefont
  {Hentschel}}\ and\ \bibinfo {author} {\bibfnamefont {B.~C.}\ \bibnamefont
  {Sanders}},\ }\href {\doibase 10.1103/PhysRevLett.107.233601} {\bibfield
  {journal} {\bibinfo  {journal} {Physical Review Letters}\ }\textbf {\bibinfo
  {volume} {107}},\ \bibinfo {pages} {233601} (\bibinfo {year}
  {2011})}\BibitemShut {NoStop}%
\bibitem [{\citenamefont {Ahn}\ \emph {et~al.}(2002)\citenamefont {Ahn},
  \citenamefont {Doherty},\ and\ \citenamefont
  {Landahl}}]{ahn_continuous_2002}%
  \BibitemOpen
  \bibfield  {author} {\bibinfo {author} {\bibfnamefont {C.}~\bibnamefont
  {Ahn}}, \bibinfo {author} {\bibfnamefont {A.~C.}\ \bibnamefont {Doherty}}, \
  and\ \bibinfo {author} {\bibfnamefont {A.~J.}\ \bibnamefont {Landahl}},\
  }\href {\doibase 10.1103/PhysRevA.65.042301} {\bibfield  {journal} {\bibinfo
  {journal} {Physical Review A}\ }\textbf {\bibinfo {volume} {65}},\ \bibinfo
  {pages} {042301} (\bibinfo {year} {2002})},\ \bibinfo {note} {publisher:
  American Physical Society}\BibitemShut {NoStop}%
\bibitem [{\citenamefont {Cramer}\ \emph {et~al.}(2016)\citenamefont {Cramer},
  \citenamefont {Kalb}, \citenamefont {Rol}, \citenamefont {Hensen},
  \citenamefont {Blok}, \citenamefont {Markham}, \citenamefont {Twitchen},
  \citenamefont {Hanson},\ and\ \citenamefont
  {Taminiau}}]{cramer_repeated_2016}%
  \BibitemOpen
  \bibfield  {author} {\bibinfo {author} {\bibfnamefont {J.}~\bibnamefont
  {Cramer}}, \bibinfo {author} {\bibfnamefont {N.}~\bibnamefont {Kalb}},
  \bibinfo {author} {\bibfnamefont {M.~A.}\ \bibnamefont {Rol}}, \bibinfo
  {author} {\bibfnamefont {B.}~\bibnamefont {Hensen}}, \bibinfo {author}
  {\bibfnamefont {M.~S.}\ \bibnamefont {Blok}}, \bibinfo {author}
  {\bibfnamefont {M.}~\bibnamefont {Markham}}, \bibinfo {author} {\bibfnamefont
  {D.~J.}\ \bibnamefont {Twitchen}}, \bibinfo {author} {\bibfnamefont
  {R.}~\bibnamefont {Hanson}}, \ and\ \bibinfo {author} {\bibfnamefont {T.~H.}\
  \bibnamefont {Taminiau}},\ }\href {\doibase 10.1038/ncomms11526} {\bibfield
  {journal} {\bibinfo  {journal} {Nature Communications}\ }\textbf {\bibinfo
  {volume} {7}},\ \bibinfo {pages} {11526} (\bibinfo {year} {2016})},\ \bibinfo
  {note} {number: 1 Publisher: Nature Publishing Group}\BibitemShut {NoStop}%
\bibitem [{\citenamefont {Ryan-Anderson}\ \emph {et~al.}(2021)\citenamefont
  {Ryan-Anderson}, \citenamefont {Bohnet}, \citenamefont {Lee}, \citenamefont
  {Gresh}, \citenamefont {Hankin}, \citenamefont {Gaebler}, \citenamefont
  {Francois}, \citenamefont {Chernoguzov}, \citenamefont {Lucchetti},
  \citenamefont {Brown}, \citenamefont {Gatterman}, \citenamefont {Halit},
  \citenamefont {Gilmore}, \citenamefont {Gerber}, \citenamefont {Neyenhuis},
  \citenamefont {Hayes},\ and\ \citenamefont
  {Stutz}}]{ryan-anderson_realization_2021}%
  \BibitemOpen
  \bibfield  {author} {\bibinfo {author} {\bibfnamefont {C.}~\bibnamefont
  {Ryan-Anderson}}, \bibinfo {author} {\bibfnamefont {J.~G.}\ \bibnamefont
  {Bohnet}}, \bibinfo {author} {\bibfnamefont {K.}~\bibnamefont {Lee}},
  \bibinfo {author} {\bibfnamefont {D.}~\bibnamefont {Gresh}}, \bibinfo
  {author} {\bibfnamefont {A.}~\bibnamefont {Hankin}}, \bibinfo {author}
  {\bibfnamefont {J.~P.}\ \bibnamefont {Gaebler}}, \bibinfo {author}
  {\bibfnamefont {D.}~\bibnamefont {Francois}}, \bibinfo {author}
  {\bibfnamefont {A.}~\bibnamefont {Chernoguzov}}, \bibinfo {author}
  {\bibfnamefont {D.}~\bibnamefont {Lucchetti}}, \bibinfo {author}
  {\bibfnamefont {N.~C.}\ \bibnamefont {Brown}}, \bibinfo {author}
  {\bibfnamefont {T.~M.}\ \bibnamefont {Gatterman}}, \bibinfo {author}
  {\bibfnamefont {S.~K.}\ \bibnamefont {Halit}}, \bibinfo {author}
  {\bibfnamefont {K.}~\bibnamefont {Gilmore}}, \bibinfo {author} {\bibfnamefont
  {J.~A.}\ \bibnamefont {Gerber}}, \bibinfo {author} {\bibfnamefont
  {B.}~\bibnamefont {Neyenhuis}}, \bibinfo {author} {\bibfnamefont
  {D.}~\bibnamefont {Hayes}}, \ and\ \bibinfo {author} {\bibfnamefont {R.~P.}\
  \bibnamefont {Stutz}},\ }\href {\doibase 10.1103/PhysRevX.11.041058}
  {\bibfield  {journal} {\bibinfo  {journal} {Physical Review X}\ }\textbf
  {\bibinfo {volume} {11}},\ \bibinfo {pages} {041058} (\bibinfo {year}
  {2021})},\ \bibinfo {note} {publisher: American Physical Society}\BibitemShut
  {NoStop}%
\bibitem [{\citenamefont {Krinner}\ \emph {et~al.}(2021)\citenamefont
  {Krinner}, \citenamefont {Lacroix}, \citenamefont {Remm}, \citenamefont
  {Di~Paolo}, \citenamefont {Genois}, \citenamefont {Leroux}, \citenamefont
  {Hellings}, \citenamefont {Lazar}, \citenamefont {Swiadek}, \citenamefont
  {Herrmann}, \citenamefont {Norris}, \citenamefont {Andersen}, \citenamefont
  {M{\"u}ller}, \citenamefont {Blais}, \citenamefont {Eichler},\ and\
  \citenamefont {Wallraff}}]{krinner_realizing_2021}%
  \BibitemOpen
  \bibfield  {author} {\bibinfo {author} {\bibfnamefont {S.}~\bibnamefont
  {Krinner}}, \bibinfo {author} {\bibfnamefont {N.}~\bibnamefont {Lacroix}},
  \bibinfo {author} {\bibfnamefont {A.}~\bibnamefont {Remm}}, \bibinfo {author}
  {\bibfnamefont {A.}~\bibnamefont {Di~Paolo}}, \bibinfo {author}
  {\bibfnamefont {E.}~\bibnamefont {Genois}}, \bibinfo {author} {\bibfnamefont
  {C.}~\bibnamefont {Leroux}}, \bibinfo {author} {\bibfnamefont
  {C.}~\bibnamefont {Hellings}}, \bibinfo {author} {\bibfnamefont
  {S.}~\bibnamefont {Lazar}}, \bibinfo {author} {\bibfnamefont
  {F.}~\bibnamefont {Swiadek}}, \bibinfo {author} {\bibfnamefont
  {J.}~\bibnamefont {Herrmann}}, \bibinfo {author} {\bibfnamefont {G.~J.}\
  \bibnamefont {Norris}}, \bibinfo {author} {\bibfnamefont {C.~K.}\
  \bibnamefont {Andersen}}, \bibinfo {author} {\bibfnamefont {M.}~\bibnamefont
  {M{\"u}ller}}, \bibinfo {author} {\bibfnamefont {A.}~\bibnamefont {Blais}},
  \bibinfo {author} {\bibfnamefont {C.}~\bibnamefont {Eichler}}, \ and\
  \bibinfo {author} {\bibfnamefont {A.}~\bibnamefont {Wallraff}},\ }\href
  {http://arxiv.org/abs/2112.03708} {\bibfield  {journal} {\bibinfo  {journal}
  {arXiv:2112.03708 [cond-mat, physics:quant-ph]}\ } (\bibinfo {year}
  {2021})},\ \bibinfo {note} {arXiv: 2112.03708}\BibitemShut {NoStop}%
\bibitem [{\citenamefont {Sutton}\ and\ \citenamefont
  {Barto}(2018)}]{sutton_reinforcement_2018}%
  \BibitemOpen
  \bibfield  {author} {\bibinfo {author} {\bibfnamefont {R.~S.}\ \bibnamefont
  {Sutton}}\ and\ \bibinfo {author} {\bibfnamefont {A.~G.}\ \bibnamefont
  {Barto}},\ }\href@noop {} {\emph {\bibinfo {title} {Reinforcement {Learning},
  second edition: {An} {Introduction}}}}\ (\bibinfo  {publisher} {MIT Press},\
  \bibinfo {year} {2018})\BibitemShut {NoStop}%
\bibitem [{\citenamefont {Bukov}\ \emph {et~al.}(2018)\citenamefont {Bukov},
  \citenamefont {Day}, \citenamefont {Sels}, \citenamefont {Weinberg},
  \citenamefont {Polkovnikov},\ and\ \citenamefont
  {Mehta}}]{bukov_reinforcement_2018}%
  \BibitemOpen
  \bibfield  {author} {\bibinfo {author} {\bibfnamefont {M.}~\bibnamefont
  {Bukov}}, \bibinfo {author} {\bibfnamefont {A.~G.~R.}\ \bibnamefont {Day}},
  \bibinfo {author} {\bibfnamefont {D.}~\bibnamefont {Sels}}, \bibinfo {author}
  {\bibfnamefont {P.}~\bibnamefont {Weinberg}}, \bibinfo {author}
  {\bibfnamefont {A.}~\bibnamefont {Polkovnikov}}, \ and\ \bibinfo {author}
  {\bibfnamefont {P.}~\bibnamefont {Mehta}},\ }\href {\doibase
  10.1103/PhysRevX.8.031086} {\bibfield  {journal} {\bibinfo  {journal}
  {Physical Review X}\ }\textbf {\bibinfo {volume} {8}},\ \bibinfo {pages}
  {031086} (\bibinfo {year} {2018})},\ \bibinfo {note} {arXiv:
  1705.00565}\BibitemShut {NoStop}%
\bibitem [{\citenamefont {August}\ and\ \citenamefont
  {Hern{\'a}ndez-Lobato}(2018)}]{august_taking_2018}%
  \BibitemOpen
  \bibfield  {author} {\bibinfo {author} {\bibfnamefont {M.}~\bibnamefont
  {August}}\ and\ \bibinfo {author} {\bibfnamefont {J.~M.}\ \bibnamefont
  {Hern{\'a}ndez-Lobato}},\ }in\ \href {\doibase 10.1007/978-3-030-02465-9_43}
  {\emph {\bibinfo {booktitle} {High {Performance} {Computing}}}},\ \bibinfo
  {series and number} {Lecture {Notes} in {Computer} {Science}},\ \bibinfo
  {editor} {edited by\ \bibinfo {editor} {\bibfnamefont {R.}~\bibnamefont
  {Yokota}}, \bibinfo {editor} {\bibfnamefont {M.}~\bibnamefont {Weiland}},
  \bibinfo {editor} {\bibfnamefont {J.}~\bibnamefont {Shalf}}, \ and\ \bibinfo
  {editor} {\bibfnamefont {S.}~\bibnamefont {Alam}}}\ (\bibinfo  {publisher}
  {Springer International Publishing},\ \bibinfo {address} {Cham},\ \bibinfo
  {year} {2018})\ pp.\ \bibinfo {pages} {591--613}\BibitemShut {NoStop}%
\bibitem [{\citenamefont {Niu}\ \emph {et~al.}(2019)\citenamefont {Niu},
  \citenamefont {Boixo}, \citenamefont {Smelyanskiy},\ and\ \citenamefont
  {Neven}}]{niu_universal_2019}%
  \BibitemOpen
  \bibfield  {author} {\bibinfo {author} {\bibfnamefont {M.~Y.}\ \bibnamefont
  {Niu}}, \bibinfo {author} {\bibfnamefont {S.}~\bibnamefont {Boixo}}, \bibinfo
  {author} {\bibfnamefont {V.~N.}\ \bibnamefont {Smelyanskiy}}, \ and\ \bibinfo
  {author} {\bibfnamefont {H.}~\bibnamefont {Neven}},\ }\href {\doibase
  10.1038/s41534-019-0141-3} {\bibfield  {journal} {\bibinfo  {journal} {npj
  Quantum Information}\ }\textbf {\bibinfo {volume} {5}},\ \bibinfo {pages} {1}
  (\bibinfo {year} {2019})},\ \bibinfo {note} {publisher: Nature Publishing
  Group tex.copyright: 2019 The Author(s)}\BibitemShut {NoStop}%
\bibitem [{\citenamefont {Porotti}\ \emph {et~al.}(2019)\citenamefont
  {Porotti}, \citenamefont {Tamascelli}, \citenamefont {Restelli},\ and\
  \citenamefont {Prati}}]{porotti_coherent_2019}%
  \BibitemOpen
  \bibfield  {author} {\bibinfo {author} {\bibfnamefont {R.}~\bibnamefont
  {Porotti}}, \bibinfo {author} {\bibfnamefont {D.}~\bibnamefont {Tamascelli}},
  \bibinfo {author} {\bibfnamefont {M.}~\bibnamefont {Restelli}}, \ and\
  \bibinfo {author} {\bibfnamefont {E.}~\bibnamefont {Prati}},\ }\href
  {\doibase 10.1038/s42005-019-0169-x} {\bibfield  {journal} {\bibinfo
  {journal} {Communications On Physics}\ }\textbf {\bibinfo {volume} {2}},\
  \bibinfo {pages} {1} (\bibinfo {year} {2019})},\ \bibinfo {note} {publisher:
  Nature Publishing Group tex.copyright: 2019 The Author(s)}\BibitemShut
  {NoStop}%
\bibitem [{\citenamefont {Baum}\ \emph {et~al.}(2021)\citenamefont {Baum},
  \citenamefont {Amico}, \citenamefont {Howell}, \citenamefont {Hush},
  \citenamefont {Liuzzi}, \citenamefont {Mundada}, \citenamefont {Merkh},
  \citenamefont {Carvalho},\ and\ \citenamefont
  {Biercuk}}]{baum_experimental_2021}%
  \BibitemOpen
  \bibfield  {author} {\bibinfo {author} {\bibfnamefont {Y.}~\bibnamefont
  {Baum}}, \bibinfo {author} {\bibfnamefont {M.}~\bibnamefont {Amico}},
  \bibinfo {author} {\bibfnamefont {S.}~\bibnamefont {Howell}}, \bibinfo
  {author} {\bibfnamefont {M.}~\bibnamefont {Hush}}, \bibinfo {author}
  {\bibfnamefont {M.}~\bibnamefont {Liuzzi}}, \bibinfo {author} {\bibfnamefont
  {P.}~\bibnamefont {Mundada}}, \bibinfo {author} {\bibfnamefont
  {T.}~\bibnamefont {Merkh}}, \bibinfo {author} {\bibfnamefont {A.~R.}\
  \bibnamefont {Carvalho}}, \ and\ \bibinfo {author} {\bibfnamefont {M.~J.}\
  \bibnamefont {Biercuk}},\ }\href {\doibase 10.1103/PRXQuantum.2.040324}
  {\bibfield  {journal} {\bibinfo  {journal} {PRX Quantum}\ }\textbf {\bibinfo
  {volume} {2}},\ \bibinfo {pages} {040324} (\bibinfo {year}
  {2021})}\BibitemShut {NoStop}%
\bibitem [{\citenamefont {F{\"o}sel}\ \emph {et~al.}(2018)\citenamefont
  {F{\"o}sel}, \citenamefont {Tighineanu}, \citenamefont {Weiss},\ and\
  \citenamefont {Marquardt}}]{fosel_reinforcement_2018}%
  \BibitemOpen
  \bibfield  {author} {\bibinfo {author} {\bibfnamefont {T.}~\bibnamefont
  {F{\"o}sel}}, \bibinfo {author} {\bibfnamefont {P.}~\bibnamefont
  {Tighineanu}}, \bibinfo {author} {\bibfnamefont {T.}~\bibnamefont {Weiss}}, \
  and\ \bibinfo {author} {\bibfnamefont {F.}~\bibnamefont {Marquardt}},\ }\href
  {\doibase 10.1103/PhysRevX.8.031084} {\bibfield  {journal} {\bibinfo
  {journal} {Physical Review X}\ }\textbf {\bibinfo {volume} {8}},\ \bibinfo
  {pages} {031084} (\bibinfo {year} {2018})},\ \bibinfo {note} {publisher:
  American Physical Society}\BibitemShut {NoStop}%
\bibitem [{\citenamefont {Borah}\ \emph {et~al.}(2021)\citenamefont {Borah},
  \citenamefont {Sarma}, \citenamefont {Kewming}, \citenamefont {Milburn},\
  and\ \citenamefont {Twamley}}]{borah_measurement-based_2021}%
  \BibitemOpen
  \bibfield  {author} {\bibinfo {author} {\bibfnamefont {S.}~\bibnamefont
  {Borah}}, \bibinfo {author} {\bibfnamefont {B.}~\bibnamefont {Sarma}},
  \bibinfo {author} {\bibfnamefont {M.}~\bibnamefont {Kewming}}, \bibinfo
  {author} {\bibfnamefont {G.~J.}\ \bibnamefont {Milburn}}, \ and\ \bibinfo
  {author} {\bibfnamefont {J.}~\bibnamefont {Twamley}},\ }\href {\doibase
  10.1103/PhysRevLett.127.190403} {\bibfield  {journal} {\bibinfo  {journal}
  {Physical Review Letters}\ }\textbf {\bibinfo {volume} {127}},\ \bibinfo
  {pages} {190403} (\bibinfo {year} {2021})}\BibitemShut {NoStop}%
\bibitem [{\citenamefont {Sivak}\ \emph {et~al.}(2022)\citenamefont {Sivak},
  \citenamefont {Eickbusch}, \citenamefont {Liu}, \citenamefont {Royer},
  \citenamefont {Tsioutsios},\ and\ \citenamefont
  {Devoret}}]{sivak_model-free_2021}%
  \BibitemOpen
  \bibfield  {author} {\bibinfo {author} {\bibfnamefont {V.~V.}\ \bibnamefont
  {Sivak}}, \bibinfo {author} {\bibfnamefont {A.}~\bibnamefont {Eickbusch}},
  \bibinfo {author} {\bibfnamefont {H.}~\bibnamefont {Liu}}, \bibinfo {author}
  {\bibfnamefont {B.}~\bibnamefont {Royer}}, \bibinfo {author} {\bibfnamefont
  {I.}~\bibnamefont {Tsioutsios}}, \ and\ \bibinfo {author} {\bibfnamefont
  {M.~H.}\ \bibnamefont {Devoret}},\ }\href {\doibase
  10.1103/PhysRevX.12.011059} {\bibfield  {journal} {\bibinfo  {journal} {Phys.
  Rev. X}\ }\textbf {\bibinfo {volume} {12}},\ \bibinfo {pages} {011059}
  (\bibinfo {year} {2022})}\BibitemShut {NoStop}%
\bibitem [{\citenamefont {Porotti}\ \emph {et~al.}(2021)\citenamefont
  {Porotti}, \citenamefont {Essig}, \citenamefont {Huard},\ and\ \citenamefont
  {Marquardt}}]{porotti_deep_2021}%
  \BibitemOpen
  \bibfield  {author} {\bibinfo {author} {\bibfnamefont {R.}~\bibnamefont
  {Porotti}}, \bibinfo {author} {\bibfnamefont {A.}~\bibnamefont {Essig}},
  \bibinfo {author} {\bibfnamefont {B.}~\bibnamefont {Huard}}, \ and\ \bibinfo
  {author} {\bibfnamefont {F.}~\bibnamefont {Marquardt}},\ }\href
  {http://arxiv.org/abs/2107.08816} {\bibfield  {journal} {\bibinfo  {journal}
  {arXiv:2107.08816 [quant-ph]}\ } (\bibinfo {year} {2021})}\BibitemShut
  {NoStop}%
\bibitem [{\citenamefont {Reuer}\ \emph {et~al.}(2022)\citenamefont {Reuer},
  \citenamefont {Landgraf}, \citenamefont {F{\"o}sel}, \citenamefont
  {O'Sullivan}, \citenamefont {Beltr{\'a}n}, \citenamefont {Akin},
  \citenamefont {Norris}, \citenamefont {Remm}, \citenamefont {Kerschbaum},
  \citenamefont {Besse} \emph {et~al.}}]{reuer2022realizing}%
  \BibitemOpen
  \bibfield  {author} {\bibinfo {author} {\bibfnamefont {K.}~\bibnamefont
  {Reuer}}, \bibinfo {author} {\bibfnamefont {J.}~\bibnamefont {Landgraf}},
  \bibinfo {author} {\bibfnamefont {T.}~\bibnamefont {F{\"o}sel}}, \bibinfo
  {author} {\bibfnamefont {J.}~\bibnamefont {O'Sullivan}}, \bibinfo {author}
  {\bibfnamefont {L.}~\bibnamefont {Beltr{\'a}n}}, \bibinfo {author}
  {\bibfnamefont {A.}~\bibnamefont {Akin}}, \bibinfo {author} {\bibfnamefont
  {G.~J.}\ \bibnamefont {Norris}}, \bibinfo {author} {\bibfnamefont
  {A.}~\bibnamefont {Remm}}, \bibinfo {author} {\bibfnamefont {M.}~\bibnamefont
  {Kerschbaum}}, \bibinfo {author} {\bibfnamefont {J.-C.}\ \bibnamefont
  {Besse}},  \emph {et~al.},\ }\href@noop {} {\bibfield  {journal} {\bibinfo
  {journal} {arXiv preprint arXiv:2210.16715}\ } (\bibinfo {year}
  {2022})}\BibitemShut {NoStop}%
\bibitem [{\citenamefont {Sivak}\ \emph {et~al.}(2023)\citenamefont {Sivak},
  \citenamefont {Eickbusch}, \citenamefont {Royer} \emph
  {et~al.}}]{sivak2023breakeven}%
  \BibitemOpen
  \bibfield  {author} {\bibinfo {author} {\bibfnamefont {V.}~\bibnamefont
  {Sivak}}, \bibinfo {author} {\bibfnamefont {A.}~\bibnamefont {Eickbusch}},
  \bibinfo {author} {\bibfnamefont {B.}~\bibnamefont {Royer}},  \emph
  {et~al.},\ }\href@noop {} {\bibfield  {journal} {\bibinfo  {journal}
  {Nature}\ ,\ \bibinfo {pages} {50–55}} (\bibinfo {year}
  {2023})}\BibitemShut {NoStop}%
\bibitem [{\citenamefont {Moerland}\ \emph {et~al.}(2022)\citenamefont
  {Moerland}, \citenamefont {Broekens}, \citenamefont {Plaat},\ and\
  \citenamefont {Jonker}}]{moerland_model-based_2022}%
  \BibitemOpen
  \bibfield  {author} {\bibinfo {author} {\bibfnamefont {T.~M.}\ \bibnamefont
  {Moerland}}, \bibinfo {author} {\bibfnamefont {J.}~\bibnamefont {Broekens}},
  \bibinfo {author} {\bibfnamefont {A.}~\bibnamefont {Plaat}}, \ and\ \bibinfo
  {author} {\bibfnamefont {C.~M.}\ \bibnamefont {Jonker}},\ }\href {\doibase
  10.48550/arXiv.2006.16712} {\enquote {\bibinfo {title} {Model-based
  {Reinforcement} {Learning}: {A} {Survey}},}\ } (\bibinfo {year} {2022}),\
  \bibinfo {note} {arXiv:2006.16712 [cs, stat]}\BibitemShut {NoStop}%
\bibitem [{\citenamefont {Leung}\ \emph {et~al.}(2017)\citenamefont {Leung},
  \citenamefont {Abdelhafez}, \citenamefont {Koch},\ and\ \citenamefont
  {Schuster}}]{leung_speedup_2017}%
  \BibitemOpen
  \bibfield  {author} {\bibinfo {author} {\bibfnamefont {N.}~\bibnamefont
  {Leung}}, \bibinfo {author} {\bibfnamefont {M.}~\bibnamefont {Abdelhafez}},
  \bibinfo {author} {\bibfnamefont {J.}~\bibnamefont {Koch}}, \ and\ \bibinfo
  {author} {\bibfnamefont {D.}~\bibnamefont {Schuster}},\ }\href {\doibase
  10.1103/PhysRevA.95.042318} {\bibfield  {journal} {\bibinfo  {journal}
  {Physical Review A}\ }\textbf {\bibinfo {volume} {95}},\ \bibinfo {pages}
  {042318} (\bibinfo {year} {2017})},\ \bibinfo {note} {publisher: American
  Physical Society}\BibitemShut {NoStop}%
\bibitem [{\citenamefont {Abdelhafez}\ \emph {et~al.}(2020)\citenamefont
  {Abdelhafez}, \citenamefont {Baker}, \citenamefont {Gyenis}, \citenamefont
  {Mundada}, \citenamefont {Houck}, \citenamefont {Schuster},\ and\
  \citenamefont {Koch}}]{abdelhafez_universal_2020}%
  \BibitemOpen
  \bibfield  {author} {\bibinfo {author} {\bibfnamefont {M.}~\bibnamefont
  {Abdelhafez}}, \bibinfo {author} {\bibfnamefont {B.}~\bibnamefont {Baker}},
  \bibinfo {author} {\bibfnamefont {A.}~\bibnamefont {Gyenis}}, \bibinfo
  {author} {\bibfnamefont {P.}~\bibnamefont {Mundada}}, \bibinfo {author}
  {\bibfnamefont {A.~A.}\ \bibnamefont {Houck}}, \bibinfo {author}
  {\bibfnamefont {D.}~\bibnamefont {Schuster}}, \ and\ \bibinfo {author}
  {\bibfnamefont {J.}~\bibnamefont {Koch}},\ }\href {\doibase
  10.1103/PhysRevA.101.022321} {\bibfield  {journal} {\bibinfo  {journal}
  {Physical Review A}\ }\textbf {\bibinfo {volume} {101}},\ \bibinfo {pages}
  {022321} (\bibinfo {year} {2020})},\ \bibinfo {note} {publisher: American
  Physical Society}\BibitemShut {NoStop}%
\bibitem [{\citenamefont {Sch{\"a}fer}\ \emph {et~al.}(2020)\citenamefont
  {Sch{\"a}fer}, \citenamefont {Kloc}, \citenamefont {Bruder},\ and\
  \citenamefont {L{\"o}rch}}]{schafer_differentiable_2020}%
  \BibitemOpen
  \bibfield  {author} {\bibinfo {author} {\bibfnamefont {F.}~\bibnamefont
  {Sch{\"a}fer}}, \bibinfo {author} {\bibfnamefont {M.}~\bibnamefont {Kloc}},
  \bibinfo {author} {\bibfnamefont {C.}~\bibnamefont {Bruder}}, \ and\ \bibinfo
  {author} {\bibfnamefont {N.}~\bibnamefont {L{\"o}rch}},\ }\href {\doibase
  10.1088/2632-2153/ab9802} {\bibfield  {journal} {\bibinfo  {journal} {Machine
  Learning: Science and Technology}\ }\textbf {\bibinfo {volume} {1}},\
  \bibinfo {pages} {035009} (\bibinfo {year} {2020})},\ \bibinfo {note}
  {publisher: IOP Publishing}\BibitemShut {NoStop}%
\bibitem [{\citenamefont {Liao}\ \emph {et~al.}(2019)\citenamefont {Liao},
  \citenamefont {Liu}, \citenamefont {Wang},\ and\ \citenamefont
  {Xiang}}]{liao_differentiable_2019}%
  \BibitemOpen
  \bibfield  {author} {\bibinfo {author} {\bibfnamefont {H.-J.}\ \bibnamefont
  {Liao}}, \bibinfo {author} {\bibfnamefont {J.-G.}\ \bibnamefont {Liu}},
  \bibinfo {author} {\bibfnamefont {L.}~\bibnamefont {Wang}}, \ and\ \bibinfo
  {author} {\bibfnamefont {T.}~\bibnamefont {Xiang}},\ }\href {\doibase
  10.1103/PhysRevX.9.031041} {\bibfield  {journal} {\bibinfo  {journal}
  {Physical Review X}\ }\textbf {\bibinfo {volume} {9}},\ \bibinfo {pages}
  {031041} (\bibinfo {year} {2019})}\BibitemShut {NoStop}%
\bibitem [{\citenamefont {Coopmans}\ \emph {et~al.}(2021)\citenamefont
  {Coopmans}, \citenamefont {Luo}, \citenamefont {Kells}, \citenamefont
  {Clark},\ and\ \citenamefont {Carrasquilla}}]{coopmans_protocol_2021}%
  \BibitemOpen
  \bibfield  {author} {\bibinfo {author} {\bibfnamefont {L.}~\bibnamefont
  {Coopmans}}, \bibinfo {author} {\bibfnamefont {D.}~\bibnamefont {Luo}},
  \bibinfo {author} {\bibfnamefont {G.}~\bibnamefont {Kells}}, \bibinfo
  {author} {\bibfnamefont {B.~K.}\ \bibnamefont {Clark}}, \ and\ \bibinfo
  {author} {\bibfnamefont {J.}~\bibnamefont {Carrasquilla}},\ }\href {\doibase
  10.1103/PRXQuantum.2.020332} {\bibfield  {journal} {\bibinfo  {journal} {PRX
  Quantum}\ }\textbf {\bibinfo {volume} {2}},\ \bibinfo {pages} {020332}
  (\bibinfo {year} {2021})},\ \bibinfo {note} {publisher: American Physical
  Society}\BibitemShut {NoStop}%
\bibitem [{\citenamefont {Abdelhafez}\ \emph {et~al.}(2019)\citenamefont
  {Abdelhafez}, \citenamefont {Schuster},\ and\ \citenamefont
  {Koch}}]{abdelhafez_gradient-based_2019}%
  \BibitemOpen
  \bibfield  {author} {\bibinfo {author} {\bibfnamefont {M.}~\bibnamefont
  {Abdelhafez}}, \bibinfo {author} {\bibfnamefont {D.~I.}\ \bibnamefont
  {Schuster}}, \ and\ \bibinfo {author} {\bibfnamefont {J.}~\bibnamefont
  {Koch}},\ }\href {\doibase 10.1103/PhysRevA.99.052327} {\bibfield  {journal}
  {\bibinfo  {journal} {Physical Review A}\ }\textbf {\bibinfo {volume} {99}},\
  \bibinfo {pages} {052327} (\bibinfo {year} {2019})}\BibitemShut {NoStop}%
\bibitem [{\citenamefont {Wiseman}\ and\ \citenamefont
  {Milburn}(2009)}]{wiseman_quantum_2009}%
  \BibitemOpen
  \bibfield  {author} {\bibinfo {author} {\bibfnamefont {H.~M.}\ \bibnamefont
  {Wiseman}}\ and\ \bibinfo {author} {\bibfnamefont {G.~J.}\ \bibnamefont
  {Milburn}},\ }\href {\doibase 10.1017/CBO9780511813948} {\emph {\bibinfo
  {title} {Quantum {Measurement} and {Control}}}}\ (\bibinfo  {publisher}
  {Cambridge University Press},\ \bibinfo {address} {Cambridge},\ \bibinfo
  {year} {2009})\BibitemShut {NoStop}%
\bibitem [{\citenamefont {Sch{\"a}fer}\ \emph {et~al.}(2021)\citenamefont
  {Sch{\"a}fer}, \citenamefont {Sekatski}, \citenamefont {Koppenh{\"o}fer},
  \citenamefont {Bruder},\ and\ \citenamefont {Kloc}}]{schafer_control_2021}%
  \BibitemOpen
  \bibfield  {author} {\bibinfo {author} {\bibfnamefont {F.}~\bibnamefont
  {Sch{\"a}fer}}, \bibinfo {author} {\bibfnamefont {P.}~\bibnamefont
  {Sekatski}}, \bibinfo {author} {\bibfnamefont {M.}~\bibnamefont
  {Koppenh{\"o}fer}}, \bibinfo {author} {\bibfnamefont {C.}~\bibnamefont
  {Bruder}}, \ and\ \bibinfo {author} {\bibfnamefont {M.}~\bibnamefont
  {Kloc}},\ }\href {\doibase 10.1088/2632-2153/abec22} {\bibfield  {journal}
  {\bibinfo  {journal} {Machine Learning: Science and Technology}\ }\textbf
  {\bibinfo {volume} {2}},\ \bibinfo {pages} {035004} (\bibinfo {year}
  {2021})},\ \bibinfo {note} {publisher: IOP Publishing}\BibitemShut {NoStop}%
\bibitem [{\citenamefont {Doherty}\ and\ \citenamefont
  {Jacobs}(1999)}]{doherty_feedback_1999}%
  \BibitemOpen
  \bibfield  {author} {\bibinfo {author} {\bibfnamefont {A.~C.}\ \bibnamefont
  {Doherty}}\ and\ \bibinfo {author} {\bibfnamefont {K.}~\bibnamefont
  {Jacobs}},\ }\href {\doibase 10.1103/PhysRevA.60.2700} {\bibfield  {journal}
  {\bibinfo  {journal} {Physical Review A: Atomic, Molecular, and Optical
  Physics}\ }\textbf {\bibinfo {volume} {60}},\ \bibinfo {pages} {2700}
  (\bibinfo {year} {1999})},\ \bibinfo {note} {publisher: American Physical
  Society}\BibitemShut {NoStop}%
\bibitem [{\citenamefont {Blais}\ \emph {et~al.}(2021)\citenamefont {Blais},
  \citenamefont {Grimsmo}, \citenamefont {Girvin},\ and\ \citenamefont
  {Wallraff}}]{blais_circuit_2021}%
  \BibitemOpen
  \bibfield  {author} {\bibinfo {author} {\bibfnamefont {A.}~\bibnamefont
  {Blais}}, \bibinfo {author} {\bibfnamefont {A.~L.}\ \bibnamefont {Grimsmo}},
  \bibinfo {author} {\bibfnamefont {S.~M.}\ \bibnamefont {Girvin}}, \ and\
  \bibinfo {author} {\bibfnamefont {A.}~\bibnamefont {Wallraff}},\ }\href
  {\doibase 10.1103/RevModPhys.93.025005} {\bibfield  {journal} {\bibinfo
  {journal} {Reviews of Modern Physics}\ }\textbf {\bibinfo {volume} {93}},\
  \bibinfo {pages} {025005} (\bibinfo {year} {2021})},\ \bibinfo {note}
  {publisher: American Physical Society}\BibitemShut {NoStop}%
\bibitem [{\citenamefont {Chen}\ \emph {et~al.}(2018)\citenamefont {Chen},
  \citenamefont {Rubanova}, \citenamefont {Bettencourt},\ and\ \citenamefont
  {Duvenaud}}]{chen_neural_2018}%
  \BibitemOpen
  \bibfield  {author} {\bibinfo {author} {\bibfnamefont {T.~Q.}\ \bibnamefont
  {Chen}}, \bibinfo {author} {\bibfnamefont {Y.}~\bibnamefont {Rubanova}},
  \bibinfo {author} {\bibfnamefont {J.}~\bibnamefont {Bettencourt}}, \ and\
  \bibinfo {author} {\bibfnamefont {D.~K.}\ \bibnamefont {Duvenaud}},\ }in\
  \href@noop {} {\emph {\bibinfo {booktitle} {32nd {Conference} on {Neural}
  {Information} {Processing} {Systems}}}}\ (\bibinfo {address} {Montreal},\
  \bibinfo {year} {2018})\ p.~\bibinfo {pages} {13}\BibitemShut {NoStop}%
\bibitem [{\citenamefont {Abadi}\ \emph {et~al.}(2015)\citenamefont {Abadi},
  \citenamefont {Agarwal}, \citenamefont {Barham}, \citenamefont {Brevdo},
  \citenamefont {Chen}, \citenamefont {Citro}, \citenamefont {Corrado},
  \citenamefont {Davis}, \citenamefont {Dean}, \citenamefont {Devin},
  \citenamefont {Ghemawat}, \citenamefont {Goodfellow}, \citenamefont {Harp},
  \citenamefont {Irving}, \citenamefont {Isard}, \citenamefont {Jia},
  \citenamefont {Jozefowicz}, \citenamefont {Kaiser}, \citenamefont {Kudlur},
  \citenamefont {Levenberg}, \citenamefont {Man{\'e}}, \citenamefont {Monga},
  \citenamefont {Moore}, \citenamefont {Murray}, \citenamefont {Olah},
  \citenamefont {Schuster}, \citenamefont {Shlens}, \citenamefont {Steiner},
  \citenamefont {Sutskever}, \citenamefont {Talwar}, \citenamefont {Tucker},
  \citenamefont {Vanhoucke}, \citenamefont {Vasudevan}, \citenamefont
  {Vi{\'e}gas}, \citenamefont {Vinyals}, \citenamefont {Warden}, \citenamefont
  {Wattenberg}, \citenamefont {Wicke}, \citenamefont {Yu},\ and\ \citenamefont
  {Zheng}}]{tensorflow2015-whitepaper}%
  \BibitemOpen
  \bibfield  {author} {\bibinfo {author} {\bibfnamefont {M.}~\bibnamefont
  {Abadi}}, \bibinfo {author} {\bibfnamefont {A.}~\bibnamefont {Agarwal}},
  \bibinfo {author} {\bibfnamefont {P.}~\bibnamefont {Barham}}, \bibinfo
  {author} {\bibfnamefont {E.}~\bibnamefont {Brevdo}}, \bibinfo {author}
  {\bibfnamefont {Z.}~\bibnamefont {Chen}}, \bibinfo {author} {\bibfnamefont
  {C.}~\bibnamefont {Citro}}, \bibinfo {author} {\bibfnamefont {G.~S.}\
  \bibnamefont {Corrado}}, \bibinfo {author} {\bibfnamefont {A.}~\bibnamefont
  {Davis}}, \bibinfo {author} {\bibfnamefont {J.}~\bibnamefont {Dean}},
  \bibinfo {author} {\bibfnamefont {M.}~\bibnamefont {Devin}}, \bibinfo
  {author} {\bibfnamefont {S.}~\bibnamefont {Ghemawat}}, \bibinfo {author}
  {\bibfnamefont {I.}~\bibnamefont {Goodfellow}}, \bibinfo {author}
  {\bibfnamefont {A.}~\bibnamefont {Harp}}, \bibinfo {author} {\bibfnamefont
  {G.}~\bibnamefont {Irving}}, \bibinfo {author} {\bibfnamefont
  {M.}~\bibnamefont {Isard}}, \bibinfo {author} {\bibfnamefont
  {Y.}~\bibnamefont {Jia}}, \bibinfo {author} {\bibfnamefont {R.}~\bibnamefont
  {Jozefowicz}}, \bibinfo {author} {\bibfnamefont {L.}~\bibnamefont {Kaiser}},
  \bibinfo {author} {\bibfnamefont {M.}~\bibnamefont {Kudlur}}, \bibinfo
  {author} {\bibfnamefont {J.}~\bibnamefont {Levenberg}}, \bibinfo {author}
  {\bibfnamefont {D.}~\bibnamefont {Man{\'e}}}, \bibinfo {author}
  {\bibfnamefont {R.}~\bibnamefont {Monga}}, \bibinfo {author} {\bibfnamefont
  {S.}~\bibnamefont {Moore}}, \bibinfo {author} {\bibfnamefont
  {D.}~\bibnamefont {Murray}}, \bibinfo {author} {\bibfnamefont
  {C.}~\bibnamefont {Olah}}, \bibinfo {author} {\bibfnamefont {M.}~\bibnamefont
  {Schuster}}, \bibinfo {author} {\bibfnamefont {J.}~\bibnamefont {Shlens}},
  \bibinfo {author} {\bibfnamefont {B.}~\bibnamefont {Steiner}}, \bibinfo
  {author} {\bibfnamefont {I.}~\bibnamefont {Sutskever}}, \bibinfo {author}
  {\bibfnamefont {K.}~\bibnamefont {Talwar}}, \bibinfo {author} {\bibfnamefont
  {P.}~\bibnamefont {Tucker}}, \bibinfo {author} {\bibfnamefont
  {V.}~\bibnamefont {Vanhoucke}}, \bibinfo {author} {\bibfnamefont
  {V.}~\bibnamefont {Vasudevan}}, \bibinfo {author} {\bibfnamefont
  {F.}~\bibnamefont {Vi{\'e}gas}}, \bibinfo {author} {\bibfnamefont
  {O.}~\bibnamefont {Vinyals}}, \bibinfo {author} {\bibfnamefont
  {P.}~\bibnamefont {Warden}}, \bibinfo {author} {\bibfnamefont
  {M.}~\bibnamefont {Wattenberg}}, \bibinfo {author} {\bibfnamefont
  {M.}~\bibnamefont {Wicke}}, \bibinfo {author} {\bibfnamefont
  {Y.}~\bibnamefont {Yu}}, \ and\ \bibinfo {author} {\bibfnamefont
  {X.}~\bibnamefont {Zheng}},\ }\href {https://www.tensorflow.org/} {\enquote
  {\bibinfo {title} {{TensorFlow}: {Large}-scale machine learning on
  heterogeneous systems},}\ } (\bibinfo {year} {2015})\BibitemShut {NoStop}%
\bibitem [{\citenamefont {Hochreiter}\ and\ \citenamefont
  {Schmidhuber}(1997)}]{hochreiter_long_1997}%
  \BibitemOpen
  \bibfield  {author} {\bibinfo {author} {\bibfnamefont {S.}~\bibnamefont
  {Hochreiter}}\ and\ \bibinfo {author} {\bibfnamefont {J.}~\bibnamefont
  {Schmidhuber}},\ }\href {\doibase 10.1162/neco.1997.9.8.1735} {\bibfield
  {journal} {\bibinfo  {journal} {Neural Computation}\ }\textbf {\bibinfo
  {volume} {9}},\ \bibinfo {pages} {1735} (\bibinfo {year} {1997})}\BibitemShut
  {NoStop}%
\bibitem [{\citenamefont {Jaynes}\ and\ \citenamefont
  {Cummings}(1963)}]{jaynes_comparison_1963}%
  \BibitemOpen
  \bibfield  {author} {\bibinfo {author} {\bibfnamefont {E.}~\bibnamefont
  {Jaynes}}\ and\ \bibinfo {author} {\bibfnamefont {F.}~\bibnamefont
  {Cummings}},\ }\href {\doibase 10.1109/PROC.1963.1664} {\bibfield  {journal}
  {\bibinfo  {journal} {Proceedings of the IEEE}\ }\textbf {\bibinfo {volume}
  {51}},\ \bibinfo {pages} {89} (\bibinfo {year} {1963})},\ \bibinfo {note}
  {conference Name: Proceedings of the IEEE}\BibitemShut {NoStop}%
\bibitem [{\citenamefont {Shore}\ and\ \citenamefont
  {Knight}(1993)}]{shore_jaynes-cummings_1993}%
  \BibitemOpen
  \bibfield  {author} {\bibinfo {author} {\bibfnamefont {B.~W.}\ \bibnamefont
  {Shore}}\ and\ \bibinfo {author} {\bibfnamefont {P.~L.}\ \bibnamefont
  {Knight}},\ }\href {\doibase 10.1080/09500349314551321} {\bibfield  {journal}
  {\bibinfo  {journal} {Journal of Modern Optics}\ }\textbf {\bibinfo {volume}
  {40}},\ \bibinfo {pages} {1195} (\bibinfo {year} {1993})},\ \bibinfo {note}
  {publisher: Taylor \& Francis \_eprint:
  https://doi.org/10.1080/09500349314551321}\BibitemShut {NoStop}%
\bibitem [{\citenamefont {Law}\ and\ \citenamefont
  {Eberly}(1996)}]{law_arbitrary_1996}%
  \BibitemOpen
  \bibfield  {author} {\bibinfo {author} {\bibfnamefont {C.~K.}\ \bibnamefont
  {Law}}\ and\ \bibinfo {author} {\bibfnamefont {J.~H.}\ \bibnamefont
  {Eberly}},\ }\href {\doibase 10.1103/PhysRevLett.76.1055} {\bibfield
  {journal} {\bibinfo  {journal} {Physical Review Letters}\ }\textbf {\bibinfo
  {volume} {76}},\ \bibinfo {pages} {1055} (\bibinfo {year}
  {1996})}\BibitemShut {NoStop}%
\bibitem [{\citenamefont {Hofheinz}\ \emph {et~al.}(2009)\citenamefont
  {Hofheinz}, \citenamefont {Wang}, \citenamefont {Ansmann}, \citenamefont
  {Bialczak}, \citenamefont {Lucero}, \citenamefont {Neeley}, \citenamefont
  {O'Connell}, \citenamefont {Sank}, \citenamefont {Wenner}, \citenamefont
  {Martinis},\ and\ \citenamefont {Cleland}}]{hofheinz_synthesizing_2009}%
  \BibitemOpen
  \bibfield  {author} {\bibinfo {author} {\bibfnamefont {M.}~\bibnamefont
  {Hofheinz}}, \bibinfo {author} {\bibfnamefont {H.}~\bibnamefont {Wang}},
  \bibinfo {author} {\bibfnamefont {M.}~\bibnamefont {Ansmann}}, \bibinfo
  {author} {\bibfnamefont {R.~C.}\ \bibnamefont {Bialczak}}, \bibinfo {author}
  {\bibfnamefont {E.}~\bibnamefont {Lucero}}, \bibinfo {author} {\bibfnamefont
  {M.}~\bibnamefont {Neeley}}, \bibinfo {author} {\bibfnamefont {A.~D.}\
  \bibnamefont {O'Connell}}, \bibinfo {author} {\bibfnamefont {D.}~\bibnamefont
  {Sank}}, \bibinfo {author} {\bibfnamefont {J.}~\bibnamefont {Wenner}},
  \bibinfo {author} {\bibfnamefont {J.~M.}\ \bibnamefont {Martinis}}, \ and\
  \bibinfo {author} {\bibfnamefont {A.~N.}\ \bibnamefont {Cleland}},\ }\href
  {\doibase 10.1038/nature08005} {\bibfield  {journal} {\bibinfo  {journal}
  {Nature}\ }\textbf {\bibinfo {volume} {459}},\ \bibinfo {pages} {546}
  (\bibinfo {year} {2009})},\ \bibinfo {note} {tex.copyright: 2009 Macmillan
  Publishers Limited. All rights reserved}\BibitemShut {NoStop}%
\bibitem [{\citenamefont {Schulman}\ \emph
  {et~al.}(2017{\natexlab{a}})\citenamefont {Schulman}, \citenamefont {Wolski},
  \citenamefont {Dhariwal}, \citenamefont {Radford},\ and\ \citenamefont
  {Klimov}}]{PPO}%
  \BibitemOpen
  \bibfield  {author} {\bibinfo {author} {\bibfnamefont {J.}~\bibnamefont
  {Schulman}}, \bibinfo {author} {\bibfnamefont {F.}~\bibnamefont {Wolski}},
  \bibinfo {author} {\bibfnamefont {P.}~\bibnamefont {Dhariwal}}, \bibinfo
  {author} {\bibfnamefont {A.}~\bibnamefont {Radford}}, \ and\ \bibinfo
  {author} {\bibfnamefont {O.}~\bibnamefont {Klimov}},\ }\href@noop {}
  {\enquote {\bibinfo {title} {Proximal policy optimization algorithms},}\ }
  (\bibinfo {year} {2017}{\natexlab{a}}),\ \bibinfo {note} {arXiv: 1707.06347
  [cs.LG]}\BibitemShut {NoStop}%
\bibitem [{\citenamefont {Gottesman}\ \emph {et~al.}(2001)\citenamefont
  {Gottesman}, \citenamefont {Kitaev},\ and\ \citenamefont
  {Preskill}}]{Gottesman_encoding_2001}%
  \BibitemOpen
  \bibfield  {author} {\bibinfo {author} {\bibfnamefont {D.}~\bibnamefont
  {Gottesman}}, \bibinfo {author} {\bibfnamefont {A.}~\bibnamefont {Kitaev}}, \
  and\ \bibinfo {author} {\bibfnamefont {J.}~\bibnamefont {Preskill}},\ }\href
  {\doibase 10.1103/PhysRevA.64.012310} {\bibfield  {journal} {\bibinfo
  {journal} {Phys. Rev. A}\ }\textbf {\bibinfo {volume} {64}},\ \bibinfo
  {pages} {012310} (\bibinfo {year} {2001})}\BibitemShut {NoStop}%
\bibitem [{\citenamefont {Brune}\ \emph {et~al.}(1990)\citenamefont {Brune},
  \citenamefont {Haroche}, \citenamefont {Lefevre}, \citenamefont {Raimond},\
  and\ \citenamefont {Zagury}}]{brune_quantum_1990}%
  \BibitemOpen
  \bibfield  {author} {\bibinfo {author} {\bibfnamefont {M.}~\bibnamefont
  {Brune}}, \bibinfo {author} {\bibfnamefont {S.}~\bibnamefont {Haroche}},
  \bibinfo {author} {\bibfnamefont {V.}~\bibnamefont {Lefevre}}, \bibinfo
  {author} {\bibfnamefont {J.~M.}\ \bibnamefont {Raimond}}, \ and\ \bibinfo
  {author} {\bibfnamefont {N.}~\bibnamefont {Zagury}},\ }\href {\doibase
  10.1103/PhysRevLett.65.976} {\bibfield  {journal} {\bibinfo  {journal}
  {Physical Review Letters}\ }\textbf {\bibinfo {volume} {65}},\ \bibinfo
  {pages} {976} (\bibinfo {year} {1990})}\BibitemShut {NoStop}%
\bibitem [{\citenamefont {Brune}\ \emph {et~al.}(1996)\citenamefont {Brune},
  \citenamefont {Hagley}, \citenamefont {Dreyer}, \citenamefont {Ma{\^i}tre},
  \citenamefont {Maali}, \citenamefont {Wunderlich}, \citenamefont {Raimond},\
  and\ \citenamefont {Haroche}}]{brune_observing_1996}%
  \BibitemOpen
  \bibfield  {author} {\bibinfo {author} {\bibfnamefont {M.}~\bibnamefont
  {Brune}}, \bibinfo {author} {\bibfnamefont {E.}~\bibnamefont {Hagley}},
  \bibinfo {author} {\bibfnamefont {J.}~\bibnamefont {Dreyer}}, \bibinfo
  {author} {\bibfnamefont {X.}~\bibnamefont {Ma{\^i}tre}}, \bibinfo {author}
  {\bibfnamefont {A.}~\bibnamefont {Maali}}, \bibinfo {author} {\bibfnamefont
  {C.}~\bibnamefont {Wunderlich}}, \bibinfo {author} {\bibfnamefont {J.~M.}\
  \bibnamefont {Raimond}}, \ and\ \bibinfo {author} {\bibfnamefont
  {S.}~\bibnamefont {Haroche}},\ }\href {\doibase 10.1103/PhysRevLett.77.4887}
  {\bibfield  {journal} {\bibinfo  {journal} {Physical Review Letters}\
  }\textbf {\bibinfo {volume} {77}},\ \bibinfo {pages} {4887} (\bibinfo {year}
  {1996})},\ \bibinfo {note} {publisher: American Physical Society}\BibitemShut
  {NoStop}%
\bibitem [{\citenamefont {Guerlin}\ \emph {et~al.}(2007)\citenamefont
  {Guerlin}, \citenamefont {Bernu}, \citenamefont {Del{\'e}glise},
  \citenamefont {Sayrin}, \citenamefont {Gleyzes}, \citenamefont {Kuhr},
  \citenamefont {Brune}, \citenamefont {Raimond},\ and\ \citenamefont
  {Haroche}}]{guerlin_progressive_2007}%
  \BibitemOpen
  \bibfield  {author} {\bibinfo {author} {\bibfnamefont {C.}~\bibnamefont
  {Guerlin}}, \bibinfo {author} {\bibfnamefont {J.}~\bibnamefont {Bernu}},
  \bibinfo {author} {\bibfnamefont {S.}~\bibnamefont {Del{\'e}glise}}, \bibinfo
  {author} {\bibfnamefont {C.}~\bibnamefont {Sayrin}}, \bibinfo {author}
  {\bibfnamefont {S.}~\bibnamefont {Gleyzes}}, \bibinfo {author} {\bibfnamefont
  {S.}~\bibnamefont {Kuhr}}, \bibinfo {author} {\bibfnamefont {M.}~\bibnamefont
  {Brune}}, \bibinfo {author} {\bibfnamefont {J.-M.}\ \bibnamefont {Raimond}},
  \ and\ \bibinfo {author} {\bibfnamefont {S.}~\bibnamefont {Haroche}},\ }\href
  {\doibase 10.1038/nature06057} {\bibfield  {journal} {\bibinfo  {journal}
  {Nature}\ }\textbf {\bibinfo {volume} {448}},\ \bibinfo {pages} {889}
  (\bibinfo {year} {2007})},\ \bibinfo {note} {tex.copyright: 2007 Nature
  Publishing Group}\BibitemShut {NoStop}%
\bibitem [{\citenamefont {Gleyzes}\ \emph {et~al.}(2007)\citenamefont
  {Gleyzes}, \citenamefont {Kuhr}, \citenamefont {Guerlin}, \citenamefont
  {Bernu}, \citenamefont {Del{\'e}glise}, \citenamefont {Busk~Hoff},
  \citenamefont {Brune}, \citenamefont {Raimond},\ and\ \citenamefont
  {Haroche}}]{gleyzes_quantum_2007}%
  \BibitemOpen
  \bibfield  {author} {\bibinfo {author} {\bibfnamefont {S.}~\bibnamefont
  {Gleyzes}}, \bibinfo {author} {\bibfnamefont {S.}~\bibnamefont {Kuhr}},
  \bibinfo {author} {\bibfnamefont {C.}~\bibnamefont {Guerlin}}, \bibinfo
  {author} {\bibfnamefont {J.}~\bibnamefont {Bernu}}, \bibinfo {author}
  {\bibfnamefont {S.}~\bibnamefont {Del{\'e}glise}}, \bibinfo {author}
  {\bibfnamefont {U.}~\bibnamefont {Busk~Hoff}}, \bibinfo {author}
  {\bibfnamefont {M.}~\bibnamefont {Brune}}, \bibinfo {author} {\bibfnamefont
  {J.-M.}\ \bibnamefont {Raimond}}, \ and\ \bibinfo {author} {\bibfnamefont
  {S.}~\bibnamefont {Haroche}},\ }\href {\doibase 10.1038/nature05589}
  {\bibfield  {journal} {\bibinfo  {journal} {Nature}\ }\textbf {\bibinfo
  {volume} {446}},\ \bibinfo {pages} {297} (\bibinfo {year} {2007})},\ \bibinfo
  {note} {tex.copyright: 2007 Nature Publishing Group}\BibitemShut {NoStop}%
\bibitem [{\citenamefont {Krastanov}\ \emph {et~al.}(2015)\citenamefont
  {Krastanov}, \citenamefont {Albert}, \citenamefont {Shen}, \citenamefont
  {Zou}, \citenamefont {Heeres}, \citenamefont {Vlastakis}, \citenamefont
  {Schoelkopf},\ and\ \citenamefont {Jiang}}]{Krasnatov_Universal_2015}%
  \BibitemOpen
  \bibfield  {author} {\bibinfo {author} {\bibfnamefont {S.}~\bibnamefont
  {Krastanov}}, \bibinfo {author} {\bibfnamefont {V.~V.}\ \bibnamefont
  {Albert}}, \bibinfo {author} {\bibfnamefont {C.}~\bibnamefont {Shen}},
  \bibinfo {author} {\bibfnamefont {C.-L.}\ \bibnamefont {Zou}}, \bibinfo
  {author} {\bibfnamefont {R.~W.}\ \bibnamefont {Heeres}}, \bibinfo {author}
  {\bibfnamefont {B.}~\bibnamefont {Vlastakis}}, \bibinfo {author}
  {\bibfnamefont {R.~J.}\ \bibnamefont {Schoelkopf}}, \ and\ \bibinfo {author}
  {\bibfnamefont {L.}~\bibnamefont {Jiang}},\ }\href {\doibase
  10.1103/PhysRevA.92.040303} {\bibfield  {journal} {\bibinfo  {journal} {Phys.
  Rev. A}\ }\textbf {\bibinfo {volume} {92}},\ \bibinfo {pages} {040303}
  (\bibinfo {year} {2015})}\BibitemShut {NoStop}%
\bibitem [{\citenamefont {Degrave}\ \emph {et~al.}(2022)\citenamefont
  {Degrave}, \citenamefont {Felici}, \citenamefont {Buchli}, \citenamefont
  {Neunert}, \citenamefont {Tracey}, \citenamefont {Carpanese}, \citenamefont
  {Ewalds}, \citenamefont {Hafner}, \citenamefont {Abdolmaleki}, \citenamefont
  {de~las Casas}, \citenamefont {Donner}, \citenamefont {Fritz}, \citenamefont
  {Galperti}, \citenamefont {Huber}, \citenamefont {Keeling}, \citenamefont
  {Tsimpoukelli}, \citenamefont {Kay}, \citenamefont {Merle}, \citenamefont
  {Moret}, \citenamefont {Noury}, \citenamefont {Pesamosca}, \citenamefont
  {Pfau}, \citenamefont {Sauter}, \citenamefont {Sommariva}, \citenamefont
  {Coda}, \citenamefont {Duval}, \citenamefont {Fasoli}, \citenamefont {Kohli},
  \citenamefont {Kavukcuoglu}, \citenamefont {Hassabis},\ and\ \citenamefont
  {Riedmiller}}]{degrave_magnetic_2022}%
  \BibitemOpen
  \bibfield  {author} {\bibinfo {author} {\bibfnamefont {J.}~\bibnamefont
  {Degrave}}, \bibinfo {author} {\bibfnamefont {F.}~\bibnamefont {Felici}},
  \bibinfo {author} {\bibfnamefont {J.}~\bibnamefont {Buchli}}, \bibinfo
  {author} {\bibfnamefont {M.}~\bibnamefont {Neunert}}, \bibinfo {author}
  {\bibfnamefont {B.}~\bibnamefont {Tracey}}, \bibinfo {author} {\bibfnamefont
  {F.}~\bibnamefont {Carpanese}}, \bibinfo {author} {\bibfnamefont
  {T.}~\bibnamefont {Ewalds}}, \bibinfo {author} {\bibfnamefont
  {R.}~\bibnamefont {Hafner}}, \bibinfo {author} {\bibfnamefont
  {A.}~\bibnamefont {Abdolmaleki}}, \bibinfo {author} {\bibfnamefont
  {D.}~\bibnamefont {de~las Casas}}, \bibinfo {author} {\bibfnamefont
  {C.}~\bibnamefont {Donner}}, \bibinfo {author} {\bibfnamefont
  {L.}~\bibnamefont {Fritz}}, \bibinfo {author} {\bibfnamefont
  {C.}~\bibnamefont {Galperti}}, \bibinfo {author} {\bibfnamefont
  {A.}~\bibnamefont {Huber}}, \bibinfo {author} {\bibfnamefont
  {J.}~\bibnamefont {Keeling}}, \bibinfo {author} {\bibfnamefont
  {M.}~\bibnamefont {Tsimpoukelli}}, \bibinfo {author} {\bibfnamefont
  {J.}~\bibnamefont {Kay}}, \bibinfo {author} {\bibfnamefont {A.}~\bibnamefont
  {Merle}}, \bibinfo {author} {\bibfnamefont {J.-M.}\ \bibnamefont {Moret}},
  \bibinfo {author} {\bibfnamefont {S.}~\bibnamefont {Noury}}, \bibinfo
  {author} {\bibfnamefont {F.}~\bibnamefont {Pesamosca}}, \bibinfo {author}
  {\bibfnamefont {D.}~\bibnamefont {Pfau}}, \bibinfo {author} {\bibfnamefont
  {O.}~\bibnamefont {Sauter}}, \bibinfo {author} {\bibfnamefont
  {C.}~\bibnamefont {Sommariva}}, \bibinfo {author} {\bibfnamefont
  {S.}~\bibnamefont {Coda}}, \bibinfo {author} {\bibfnamefont {B.}~\bibnamefont
  {Duval}}, \bibinfo {author} {\bibfnamefont {A.}~\bibnamefont {Fasoli}},
  \bibinfo {author} {\bibfnamefont {P.}~\bibnamefont {Kohli}}, \bibinfo
  {author} {\bibfnamefont {K.}~\bibnamefont {Kavukcuoglu}}, \bibinfo {author}
  {\bibfnamefont {D.}~\bibnamefont {Hassabis}}, \ and\ \bibinfo {author}
  {\bibfnamefont {M.}~\bibnamefont {Riedmiller}},\ }\href {\doibase
  10.1038/s41586-021-04301-9} {\bibfield  {journal} {\bibinfo  {journal}
  {Nature}\ }\textbf {\bibinfo {volume} {602}},\ \bibinfo {pages} {414}
  (\bibinfo {year} {2022})}\BibitemShut {NoStop}%
\bibitem [{\citenamefont {Xie}\ \emph {et~al.}(2022)\citenamefont {Xie},
  \citenamefont {Lin}, \citenamefont {Li}, \citenamefont {Li},\ and\
  \citenamefont {Ye}}]{xie_pretraining_2022}%
  \BibitemOpen
  \bibfield  {author} {\bibinfo {author} {\bibfnamefont {Z.}~\bibnamefont
  {Xie}}, \bibinfo {author} {\bibfnamefont {Z.}~\bibnamefont {Lin}}, \bibinfo
  {author} {\bibfnamefont {J.}~\bibnamefont {Li}}, \bibinfo {author}
  {\bibfnamefont {S.}~\bibnamefont {Li}}, \ and\ \bibinfo {author}
  {\bibfnamefont {D.}~\bibnamefont {Ye}},\ }\href
  {http://arxiv.org/abs/2211.03959} {{\enquote {\bibinfo
  {title} {Pretraining in {Deep} {Reinforcement} {Learning}: {A} {Survey}},}\
  }} (\bibinfo {year} {2022}),\ \bibinfo {note} {arXiv:2211.03959
  [cs]}\BibitemShut {NoStop}%
\bibitem [{\citenamefont {Yao}\ \emph {et~al.}(2021)\citenamefont {Yao},
  \citenamefont {Lin},\ and\ \citenamefont {Bukov}}]{Yao_reinforcement_2021}%
  \BibitemOpen
  \bibfield  {author} {\bibinfo {author} {\bibfnamefont {J.}~\bibnamefont
  {Yao}}, \bibinfo {author} {\bibfnamefont {L.}~\bibnamefont {Lin}}, \ and\
  \bibinfo {author} {\bibfnamefont {M.}~\bibnamefont {Bukov}},\ }\href
  {\doibase 10.1103/PhysRevX.11.031070} {\bibfield  {journal} {\bibinfo
  {journal} {Phys. Rev. X}\ }\textbf {\bibinfo {volume} {11}},\ \bibinfo
  {pages} {031070} (\bibinfo {year} {2021})}\BibitemShut {NoStop}%
\bibitem [{\citenamefont {Wierichs}\ \emph {et~al.}(2020)\citenamefont
  {Wierichs}, \citenamefont {Gogolin},\ and\ \citenamefont
  {Kastoryano}}]{Wierichs_avoiding_2020}%
  \BibitemOpen
  \bibfield  {author} {\bibinfo {author} {\bibfnamefont {D.}~\bibnamefont
  {Wierichs}}, \bibinfo {author} {\bibfnamefont {C.}~\bibnamefont {Gogolin}}, \
  and\ \bibinfo {author} {\bibfnamefont {M.}~\bibnamefont {Kastoryano}},\
  }\href {\doibase 10.1103/PhysRevResearch.2.043246} {\bibfield  {journal}
  {\bibinfo  {journal} {Phys. Rev. Res.}\ }\textbf {\bibinfo {volume} {2}},\
  \bibinfo {pages} {043246} (\bibinfo {year} {2020})}\BibitemShut {NoStop}%
\bibitem [{\citenamefont {McClean}\ \emph {et~al.}(2018)\citenamefont
  {McClean}, \citenamefont {Boixo}, \citenamefont {Smelyanskiy}, \citenamefont
  {Babbush},\ and\ \citenamefont {Neven}}]{mcclean_barren_2018}%
  \BibitemOpen
  \bibfield  {author} {\bibinfo {author} {\bibfnamefont {J.~R.}\ \bibnamefont
  {McClean}}, \bibinfo {author} {\bibfnamefont {S.}~\bibnamefont {Boixo}},
  \bibinfo {author} {\bibfnamefont {V.~N.}\ \bibnamefont {Smelyanskiy}},
  \bibinfo {author} {\bibfnamefont {R.}~\bibnamefont {Babbush}}, \ and\
  \bibinfo {author} {\bibfnamefont {H.}~\bibnamefont {Neven}},\ }\href
  {\doibase 10.1038/s41467-018-07090-4} {\bibfield  {journal} {\bibinfo
  {journal} {Nature Communications}\ }\textbf {\bibinfo {volume} {9}},\
  \bibinfo {pages} {4812} (\bibinfo {year} {2018})},\ \bibinfo {note} {number:
  1 Publisher: Nature Publishing Group}\BibitemShut {NoStop}%
\bibitem [{\citenamefont {Grant}\ \emph {et~al.}(2019)\citenamefont {Grant},
  \citenamefont {Wossnig}, \citenamefont {Ostaszewski},\ and\ \citenamefont
  {Benedetti}}]{grant_initialization_2019}%
  \BibitemOpen
  \bibfield  {author} {\bibinfo {author} {\bibfnamefont {E.}~\bibnamefont
  {Grant}}, \bibinfo {author} {\bibfnamefont {L.}~\bibnamefont {Wossnig}},
  \bibinfo {author} {\bibfnamefont {M.}~\bibnamefont {Ostaszewski}}, \ and\
  \bibinfo {author} {\bibfnamefont {M.}~\bibnamefont {Benedetti}},\ }\href
  {\doibase 10.22331/q-2019-12-09-214} {\bibfield  {journal} {\bibinfo
  {journal} {Quantum}\ }\textbf {\bibinfo {volume} {3}},\ \bibinfo {pages}
  {214} (\bibinfo {year} {2019})},\ \bibinfo {note} {publisher: Verein zur
  F{\"o}rderung des Open Access Publizierens in den
  Quantenwissenschaften}\BibitemShut {NoStop}%
\bibitem [{\citenamefont {Volkoff}\ and\ \citenamefont
  {Coles}(2021)}]{volkoff_large_2021}%
  \BibitemOpen
  \bibfield  {author} {\bibinfo {author} {\bibfnamefont {T.}~\bibnamefont
  {Volkoff}}\ and\ \bibinfo {author} {\bibfnamefont {P.~J.}\ \bibnamefont
  {Coles}},\ }\href {\doibase 10.1088/2058-9565/abd891} {\bibfield  {journal}
  {\bibinfo  {journal} {Quantum Science and Technology}\ }\textbf {\bibinfo
  {volume} {6}},\ \bibinfo {pages} {025008} (\bibinfo {year} {2021})},\
  \bibinfo {note} {publisher: IOP Publishing}\BibitemShut {NoStop}%
\bibitem [{\citenamefont {Holmes}\ \emph {et~al.}(2022)\citenamefont {Holmes},
  \citenamefont {Sharma}, \citenamefont {Cerezo},\ and\ \citenamefont
  {Coles}}]{holmes_connecting_2022}%
  \BibitemOpen
  \bibfield  {author} {\bibinfo {author} {\bibfnamefont {Z.}~\bibnamefont
  {Holmes}}, \bibinfo {author} {\bibfnamefont {K.}~\bibnamefont {Sharma}},
  \bibinfo {author} {\bibfnamefont {M.}~\bibnamefont {Cerezo}}, \ and\ \bibinfo
  {author} {\bibfnamefont {P.~J.}\ \bibnamefont {Coles}},\ }\href {\doibase
  10.1103/PRXQuantum.3.010313} {\bibfield  {journal} {\bibinfo  {journal} {PRX
  Quantum}\ }\textbf {\bibinfo {volume} {3}},\ \bibinfo {pages} {010313}
  (\bibinfo {year} {2022})}\BibitemShut {NoStop}%
\bibitem [{\citenamefont {Khatri}\ \emph {et~al.}(2019)\citenamefont {Khatri},
  \citenamefont {LaRose}, \citenamefont {Poremba}, \citenamefont {Cincio},
  \citenamefont {Sornborger},\ and\ \citenamefont
  {Coles}}]{khatri_quantum-assisted_2019}%
  \BibitemOpen
  \bibfield  {author} {\bibinfo {author} {\bibfnamefont {S.}~\bibnamefont
  {Khatri}}, \bibinfo {author} {\bibfnamefont {R.}~\bibnamefont {LaRose}},
  \bibinfo {author} {\bibfnamefont {A.}~\bibnamefont {Poremba}}, \bibinfo
  {author} {\bibfnamefont {L.}~\bibnamefont {Cincio}}, \bibinfo {author}
  {\bibfnamefont {A.~T.}\ \bibnamefont {Sornborger}}, \ and\ \bibinfo {author}
  {\bibfnamefont {P.~J.}\ \bibnamefont {Coles}},\ }\href {\doibase
  10.22331/q-2019-05-13-140} {\bibfield  {journal} {\bibinfo  {journal}
  {Quantum}\ }\textbf {\bibinfo {volume} {3}},\ \bibinfo {pages} {140}
  (\bibinfo {year} {2019})},\ \bibinfo {note} {publisher: Verein zur
  F{\"o}rderung des Open Access Publizierens in den
  Quantenwissenschaften}\BibitemShut {NoStop}%
\bibitem [{\citenamefont {Cerezo}\ \emph {et~al.}(2021)\citenamefont {Cerezo},
  \citenamefont {Sone}, \citenamefont {Volkoff}, \citenamefont {Cincio},\ and\
  \citenamefont {Coles}}]{cerezo_cost_2021}%
  \BibitemOpen
  \bibfield  {author} {\bibinfo {author} {\bibfnamefont {M.}~\bibnamefont
  {Cerezo}}, \bibinfo {author} {\bibfnamefont {A.}~\bibnamefont {Sone}},
  \bibinfo {author} {\bibfnamefont {T.}~\bibnamefont {Volkoff}}, \bibinfo
  {author} {\bibfnamefont {L.}~\bibnamefont {Cincio}}, \ and\ \bibinfo {author}
  {\bibfnamefont {P.~J.}\ \bibnamefont {Coles}},\ }\href {\doibase
  10.1038/s41467-021-21728-w} {\bibfield  {journal} {\bibinfo  {journal}
  {Nature Communications}\ }\textbf {\bibinfo {volume} {12}},\ \bibinfo {pages}
  {1791} (\bibinfo {year} {2021})},\ \bibinfo {note} {number: 1 Publisher:
  Nature Publishing Group}\BibitemShut {NoStop}%
\bibitem [{\citenamefont {Wiersema}\ \emph {et~al.}(2021)\citenamefont
  {Wiersema}, \citenamefont {Zhou}, \citenamefont {Carrasquilla},\ and\
  \citenamefont {Kim}}]{wiersema_measurement-induced_2021}%
  \BibitemOpen
  \bibfield  {author} {\bibinfo {author} {\bibfnamefont {R.}~\bibnamefont
  {Wiersema}}, \bibinfo {author} {\bibfnamefont {C.}~\bibnamefont {Zhou}},
  \bibinfo {author} {\bibfnamefont {J.~F.}\ \bibnamefont {Carrasquilla}}, \
  and\ \bibinfo {author} {\bibfnamefont {Y.~B.}\ \bibnamefont {Kim}},\ }\href
  {\doibase 10.48550/arXiv.2111.08035} {\enquote {\bibinfo {title}
  {Measurement-induced entanglement phase transitions in variational quantum
  circuits},}\ } (\bibinfo {year} {2021}),\ \bibinfo {note} {arXiv:2111.08035
  [cond-mat, physics:quant-ph]}\BibitemShut {NoStop}%
\bibitem [{\citenamefont {Mousavian}\ \emph {et~al.}(2019)\citenamefont
  {Mousavian}, \citenamefont {Toshev}, \citenamefont {Fiser}, \citenamefont
  {Kosecka}, \citenamefont {Wahid},\ and\ \citenamefont
  {Davidson}}]{mousavian_visual_2019}%
  \BibitemOpen
  \bibfield  {author} {\bibinfo {author} {\bibfnamefont {A.}~\bibnamefont
  {Mousavian}}, \bibinfo {author} {\bibfnamefont {A.}~\bibnamefont {Toshev}},
  \bibinfo {author} {\bibfnamefont {M.}~\bibnamefont {Fiser}}, \bibinfo
  {author} {\bibfnamefont {J.}~\bibnamefont {Kosecka}}, \bibinfo {author}
  {\bibfnamefont {A.}~\bibnamefont {Wahid}}, \ and\ \bibinfo {author}
  {\bibfnamefont {J.}~\bibnamefont {Davidson}},\ }\href
  {http://arxiv.org/abs/1805.06066} {\bibfield  {journal} {\bibinfo  {journal}
  {arXiv:1805.06066 [cs]}\ } (\bibinfo {year} {2019})},\ \bibinfo {note}
  {arXiv: 1805.06066}\BibitemShut {NoStop}%
\bibitem [{\citenamefont {Kim}\ \emph {et~al.}(2021)\citenamefont {Kim},
  \citenamefont {Lee}, \citenamefont {Kim}, \citenamefont {Ryu}, \citenamefont
  {Lee},\ and\ \citenamefont {Zhang}}]{kim_goal-aware_2021}%
  \BibitemOpen
  \bibfield  {author} {\bibinfo {author} {\bibfnamefont {K.}~\bibnamefont
  {Kim}}, \bibinfo {author} {\bibfnamefont {M.~W.}\ \bibnamefont {Lee}},
  \bibinfo {author} {\bibfnamefont {Y.}~\bibnamefont {Kim}}, \bibinfo {author}
  {\bibfnamefont {J.-H.}\ \bibnamefont {Ryu}}, \bibinfo {author} {\bibfnamefont
  {M.}~\bibnamefont {Lee}}, \ and\ \bibinfo {author} {\bibfnamefont {B.-T.}\
  \bibnamefont {Zhang}},\ }\href {http://arxiv.org/abs/2110.12985} {\bibfield
  {journal} {\bibinfo  {journal} {arXiv:2110.12985 [cs]}\ } (\bibinfo {year}
  {2021})},\ \bibinfo {note} {arXiv: 2110.12985}\BibitemShut {NoStop}%
\bibitem [{\citenamefont {Self}\ \emph {et~al.}(2021)\citenamefont {Self},
  \citenamefont {Khosla}, \citenamefont {Smith}, \citenamefont {Sauvage},
  \citenamefont {Haynes}, \citenamefont {Knolle}, \citenamefont {Mintert},\
  and\ \citenamefont {Kim}}]{self_variational_2021}%
  \BibitemOpen
  \bibfield  {author} {\bibinfo {author} {\bibfnamefont {C.~N.}\ \bibnamefont
  {Self}}, \bibinfo {author} {\bibfnamefont {K.~E.}\ \bibnamefont {Khosla}},
  \bibinfo {author} {\bibfnamefont {A.~W.~R.}\ \bibnamefont {Smith}}, \bibinfo
  {author} {\bibfnamefont {F.}~\bibnamefont {Sauvage}}, \bibinfo {author}
  {\bibfnamefont {P.~D.}\ \bibnamefont {Haynes}}, \bibinfo {author}
  {\bibfnamefont {J.}~\bibnamefont {Knolle}}, \bibinfo {author} {\bibfnamefont
  {F.}~\bibnamefont {Mintert}}, \ and\ \bibinfo {author} {\bibfnamefont
  {M.~S.}\ \bibnamefont {Kim}},\ }\href {\doibase 10.1038/s41534-021-00452-9}
  {\bibfield  {journal} {\bibinfo  {journal} {npj Quantum Information}\
  }\textbf {\bibinfo {volume} {7}},\ \bibinfo {pages} {1} (\bibinfo {year}
  {2021})},\ \bibinfo {note} {number: 1 Publisher: Nature Publishing
  Group}\BibitemShut {NoStop}%
\bibitem [{\citenamefont {Hill}\ \emph {et~al.}(2018)\citenamefont {Hill},
  \citenamefont {Raffin}, \citenamefont {Ernestus}, \citenamefont {Gleave},
  \citenamefont {Kanervisto}, \citenamefont {Traore}, \citenamefont {Dhariwal},
  \citenamefont {Hesse}, \citenamefont {Klimov}, \citenamefont {Nichol},
  \citenamefont {Plappert}, \citenamefont {Radford}, \citenamefont {Schulman},
  \citenamefont {Sidor},\ and\ \citenamefont {Wu}}]{stable-baselines}%
  \BibitemOpen
  \bibfield  {author} {\bibinfo {author} {\bibfnamefont {A.}~\bibnamefont
  {Hill}}, \bibinfo {author} {\bibfnamefont {A.}~\bibnamefont {Raffin}},
  \bibinfo {author} {\bibfnamefont {M.}~\bibnamefont {Ernestus}}, \bibinfo
  {author} {\bibfnamefont {A.}~\bibnamefont {Gleave}}, \bibinfo {author}
  {\bibfnamefont {A.}~\bibnamefont {Kanervisto}}, \bibinfo {author}
  {\bibfnamefont {R.}~\bibnamefont {Traore}}, \bibinfo {author} {\bibfnamefont
  {P.}~\bibnamefont {Dhariwal}}, \bibinfo {author} {\bibfnamefont
  {C.}~\bibnamefont {Hesse}}, \bibinfo {author} {\bibfnamefont
  {O.}~\bibnamefont {Klimov}}, \bibinfo {author} {\bibfnamefont
  {A.}~\bibnamefont {Nichol}}, \bibinfo {author} {\bibfnamefont
  {M.}~\bibnamefont {Plappert}}, \bibinfo {author} {\bibfnamefont
  {A.}~\bibnamefont {Radford}}, \bibinfo {author} {\bibfnamefont
  {J.}~\bibnamefont {Schulman}}, \bibinfo {author} {\bibfnamefont
  {S.}~\bibnamefont {Sidor}}, \ and\ \bibinfo {author} {\bibfnamefont
  {Y.}~\bibnamefont {Wu}},\ }\href {https://github.com/hill-a/stable-baselines}
  {\enquote {\bibinfo {title} {Stable baselines},}\ } (\bibinfo {year}
  {2018})\BibitemShut {NoStop}%
\bibitem [{\citenamefont {Mnih}\ \emph {et~al.}(2016)\citenamefont {Mnih},
  \citenamefont {Badia}, \citenamefont {Mirza}, \citenamefont {Graves},
  \citenamefont {Lillicrap}, \citenamefont {Harley}, \citenamefont {Silver},\
  and\ \citenamefont {Kavukcuoglu}}]{A2C}%
  \BibitemOpen
  \bibfield  {author} {\bibinfo {author} {\bibfnamefont {V.}~\bibnamefont
  {Mnih}}, \bibinfo {author} {\bibfnamefont {A.~P.}\ \bibnamefont {Badia}},
  \bibinfo {author} {\bibfnamefont {M.}~\bibnamefont {Mirza}}, \bibinfo
  {author} {\bibfnamefont {A.}~\bibnamefont {Graves}}, \bibinfo {author}
  {\bibfnamefont {T.~P.}\ \bibnamefont {Lillicrap}}, \bibinfo {author}
  {\bibfnamefont {T.}~\bibnamefont {Harley}}, \bibinfo {author} {\bibfnamefont
  {D.}~\bibnamefont {Silver}}, \ and\ \bibinfo {author} {\bibfnamefont
  {K.}~\bibnamefont {Kavukcuoglu}},\ }\href@noop {} {\enquote {\bibinfo {title}
  {Asynchronous methods for deep reinforcement learning},}\ } (\bibinfo {year}
  {2016}),\ \bibinfo {note} {arXiv: 1602.01783 [cs.LG]}\BibitemShut {NoStop}%
\bibitem [{\citenamefont {Andrychowicz}\ \emph {et~al.}(2018)\citenamefont
  {Andrychowicz}, \citenamefont {Wolski}, \citenamefont {Ray}, \citenamefont
  {Schneider}, \citenamefont {Fong}, \citenamefont {Welinder}, \citenamefont
  {McGrew}, \citenamefont {Tobin}, \citenamefont {Abbeel},\ and\ \citenamefont
  {Zaremba}}]{HER}%
  \BibitemOpen
  \bibfield  {author} {\bibinfo {author} {\bibfnamefont {M.}~\bibnamefont
  {Andrychowicz}}, \bibinfo {author} {\bibfnamefont {F.}~\bibnamefont
  {Wolski}}, \bibinfo {author} {\bibfnamefont {A.}~\bibnamefont {Ray}},
  \bibinfo {author} {\bibfnamefont {J.}~\bibnamefont {Schneider}}, \bibinfo
  {author} {\bibfnamefont {R.}~\bibnamefont {Fong}}, \bibinfo {author}
  {\bibfnamefont {P.}~\bibnamefont {Welinder}}, \bibinfo {author}
  {\bibfnamefont {B.}~\bibnamefont {McGrew}}, \bibinfo {author} {\bibfnamefont
  {J.}~\bibnamefont {Tobin}}, \bibinfo {author} {\bibfnamefont
  {P.}~\bibnamefont {Abbeel}}, \ and\ \bibinfo {author} {\bibfnamefont
  {W.}~\bibnamefont {Zaremba}},\ }\href@noop {} {\enquote {\bibinfo {title}
  {Hindsight experience replay},}\ } (\bibinfo {year} {2018}),\ \bibinfo {note}
  {arXiv: 1707.01495 [cs.LG]}\BibitemShut {NoStop}%
\bibitem [{\citenamefont {Schulman}\ \emph
  {et~al.}(2017{\natexlab{b}})\citenamefont {Schulman}, \citenamefont {Levine},
  \citenamefont {Moritz}, \citenamefont {Jordan},\ and\ \citenamefont
  {Abbeel}}]{TRPO}%
  \BibitemOpen
  \bibfield  {author} {\bibinfo {author} {\bibfnamefont {J.}~\bibnamefont
  {Schulman}}, \bibinfo {author} {\bibfnamefont {S.}~\bibnamefont {Levine}},
  \bibinfo {author} {\bibfnamefont {P.}~\bibnamefont {Moritz}}, \bibinfo
  {author} {\bibfnamefont {M.~I.}\ \bibnamefont {Jordan}}, \ and\ \bibinfo
  {author} {\bibfnamefont {P.}~\bibnamefont {Abbeel}},\ }\href@noop {}
  {\enquote {\bibinfo {title} {Trust region policy optimization},}\ } (\bibinfo
  {year} {2017}{\natexlab{b}}),\ \bibinfo {note} {arXiv: 1502.05477
  [cs.LG]}\BibitemShut {NoStop}%
\bibitem [{\citenamefont {Lillicrap}\ \emph {et~al.}(2019)\citenamefont
  {Lillicrap}, \citenamefont {Hunt}, \citenamefont {Pritzel}, \citenamefont
  {Heess}, \citenamefont {Erez}, \citenamefont {Tassa}, \citenamefont
  {Silver},\ and\ \citenamefont {Wierstra}}]{DDPG}%
  \BibitemOpen
  \bibfield  {author} {\bibinfo {author} {\bibfnamefont {T.~P.}\ \bibnamefont
  {Lillicrap}}, \bibinfo {author} {\bibfnamefont {J.~J.}\ \bibnamefont {Hunt}},
  \bibinfo {author} {\bibfnamefont {A.}~\bibnamefont {Pritzel}}, \bibinfo
  {author} {\bibfnamefont {N.}~\bibnamefont {Heess}}, \bibinfo {author}
  {\bibfnamefont {T.}~\bibnamefont {Erez}}, \bibinfo {author} {\bibfnamefont
  {Y.}~\bibnamefont {Tassa}}, \bibinfo {author} {\bibfnamefont
  {D.}~\bibnamefont {Silver}}, \ and\ \bibinfo {author} {\bibfnamefont
  {D.}~\bibnamefont {Wierstra}},\ }\href@noop {} {\enquote {\bibinfo {title}
  {Continuous control with deep reinforcement learning},}\ } (\bibinfo {year}
  {2019}),\ \bibinfo {note} {arXiv: 1509.02971 [cs.LG]}\BibitemShut {NoStop}%
\bibitem [{\citenamefont {Kingma}\ and\ \citenamefont
  {Ba}(2017)}]{kingma2017adam}%
  \BibitemOpen
  \bibfield  {author} {\bibinfo {author} {\bibfnamefont {D.~P.}\ \bibnamefont
  {Kingma}}\ and\ \bibinfo {author} {\bibfnamefont {J.}~\bibnamefont {Ba}},\
  }\href@noop {} {\enquote {\bibinfo {title} {Adam: {A} method for stochastic
  optimization},}\ } (\bibinfo {year} {2017}),\ \bibinfo {note} {arXiv:
  1412.6980 [cs.LG]}\BibitemShut {NoStop}%
\end{thebibliography}%

\end{document}